\documentclass[11pt]{article}

\usepackage[dvips]{graphicx}
\usepackage[english]{babel}
\usepackage{color}

\newcommand{\lsim}{\raisebox{-4pt}{$
\,\stackrel{\textstyle <}{\sim}\,$}}
\newcommand{\gsim}{\raisebox{-4pt}{$
\,\stackrel{\textstyle >}{\sim}\,$}}

\newcommand{\sms}{\mskip 1.5mu}
\newcommand{\smb}{\mskip -1.5mu}

\newcommand{\mev}{\,\mbox{MeV}}
\newcommand{\gev}{\,\mbox{GeV}}
\newcommand{\nb}{\,\mbox{nb}}

\newcommand{\half}{{\textstyle\frac{1}{2}}}

\newcommand{\xb}{x_{\smb B}}

\newcommand{\re}{\mathrm{Re}\,}
\newcommand{\im}{\mathrm{Im}\,}

\begin{document}

\begin{flushright}
DESY-05-094 \\
JLAB-THY-05-375 \\
hep-ph/0506171 \\
\end{flushright}

\begin{center}
\vskip 4.0\baselineskip
\textbf{\LARGE Exclusive channels in semi-inclusive production \\[0.5em]
  of pions and kaons}
\vskip 4.0\baselineskip
M.~Diehl, W.~Kugler \\[0.5\baselineskip]
\textit{Deutsches Elektronen-Synchroton DESY, 22603 Hamburg, Germany}
\\[1.5\baselineskip] 
A.~Sch{\"a}fer \\[0.5\baselineskip]
\textit{Institut f{\"u}r Theoretische Physik, Universit{\"a}t
  Regensburg, 93040 Regensburg, Germany}
\\[1.5\baselineskip] 
C.~Weiss \\[0.5\baselineskip]
\textit{Theory Group, Jefferson Lab, Newport News, VA 23606, U.S.A.}
\\[1.5\baselineskip] 
\vskip 4.0\baselineskip
\textbf{Abstract}\\[0.5\baselineskip]
\parbox{0.9\textwidth}{We investigate the role of exclusive channels
in semi-inclusive electroproduction of pions and kaons.  Using the QCD
factorization theorem for hard exclusive processes we evaluate the
cross sections for exclusive pseudoscalar and vector meson production
in terms of generalized parton distributions and meson distribution
amplitudes.  We investigate the uncertainties arising from the
modeling of the nonperturbative input quantities.  Combining these
results with available experimental data, we compare the cross
sections for exclusive channels to that obtained from quark
fragmentation in semi-inclusive deep inelastic scattering.  We find
that $\rho^0$ production is the only exclusive channel with
significant contributions to semi-inclusive pion production at large
$z$ and moderate $Q^2$.  The corresponding contribution to kaon
production from the decay of exclusively produced $\phi$ and
$K^{\ast}$ is rather small.}
\end{center}

\vfill
%
%

\newpage
\tableofcontents
\newpage

\section{Introduction}
\label{sec:intro}

Electroproduction processes in the Bjorken regime probe the partonic
structure of the nucleon.  Inclusive deep inelastic scattering (DIS)
provides extensive information about the sum of quark and antiquark
distributions in the nucleon, and allows one to determine the gluon
distribution from the observed scaling violations.  More detailed
information can be obtained from scattering experiments in which one
or more hadrons are observed in the final state.  There are two basic
classes of such experiments.  The first one is semi-inclusive deep
inelastic scattering, in which one observes a single hadron, carrying
a fraction $z$ of the photon energy in the target rest frame, in a
final state of large average multiplicity.  A QCD factorization
theorem states that the semi-inclusive cross section in the Bjorken
limit can be expressed in terms of distribution functions for quarks,
antiquarks and gluons in the target and of the corresponding
fragmentation functions into the observed hadron.  This allows one to
tag the active parton via its fragmentation properties and has
recently been used for a flavor decomposition of polarized quark and
antiquark distributions in the semi-inclusive production of pions and
kaons \cite{Airapetian:2004zf}.  In addition, measurements of
azimuthal asymmetries in semi-inclusive pion and kaon production, such
as the Collins and Sivers asymmetries for a transversely polarized
target \cite{Airapetian:2004tw}, provide interesting information about
the distribution of the spin and transverse momentum carried by quarks
and antiquarks in the nucleon.

The second class are exclusive scattering processes, in which the
final state is a recoiling baryon $B$ together with a single meson or
a few-meson system carrying nearly the full photon energy in the
target rest frame.  A QCD factorization theorem states that in the
Bjorken limit, and for longitudinal photon polarization, the
amplitudes of such processes are calculable in terms of the light-cone
distribution amplitude of the produced meson and generalized parton
distributions (GPDs) for the $p\to B$ transition.  Generalized parton
distributions provide a wealth of information on the parton structure
of the nucleon, in particular about the spatial distribution of
partons in the transverse plane and about quark orbital angular
momentum, see e.g.\ the reviews \cite{Ji:1998pc,Goeke:2001tz,%
Burkardt:2002hr,Diehl:2003ny,Belitsky:2005qn}.

The Bjorken limit for semi-inclusive electroproduction implies a large
average multiplicity of the produced hadronic system.  In practice,
however, there can be situations in which individual exclusive
channels play an important role.  In fixed-target experiments the
limited photon energy restricts the phase space for quark
fragmentation, in particular at large $z$, and for relatively low
photon virtuality $Q^2$ the suppression of individual exclusive
channels due to the faster drop of the exclusive cross sections with
$Q^2$ may not yet be effective.  In particular, phenomenological
studies suggest that a large contribution to $\pi^\pm$ production
comes from exclusive $\rho^0$ production, with subsequent decay
$\rho^0 \rightarrow \pi^+ \pi^-$ \cite{Szczurek:2000mj,Bino}.
Fortunately, the cross sections for exclusive $\rho^0$
electroproduction has been measured by the HERMES and CLAS
experiments, including the cross section ratio for longitudinal and
transverse photons
\cite{Airapetian:2000ni,Borissov:2001fq,Tytgat:2000th,%
Rakness:2000th,Hadjidakis:2004zm}.  It is natural to ask whether the
strange vector mesons $\phi$ and $K^\ast$ play an equally important
role in semi-inclusive kaon production and whether other exclusive
channels may be important, too.  A quantitative answer to these
questions will help to delineate the limits of the kinematic region
where semi-inclusive data can be analyzed using the factorization
theorem.

In this paper we investigate the role of exclusive channels in the
semi-inclusive production of pions and kaons on the basis of the
factorization theorem for hard exclusive processes.  Our investigation
consists of two parts.  Firstly, we evaluate the longitudinal cross
section for the exclusive production of pseudoscalar and vector mesons
in the leading-twist approximation and at leading order in the strong
coupling, focusing on $\pi$, $K$ and $\rho$, $\phi$, $K^*$.  We
explore uncertainties in the obtained cross sections, in particular
those due to the generalized parton distributions, which are still
largely unknown and need to be modeled.  Such uncertainties will
persist if higher-order and higher-twist corrections are included.
Seen from a different perspective, they indicate to which extent
exclusive meson production is sensitive to GPDs and thus interesting
in its own right.  It is known that exclusive meson production cross
sections at moderate $Q^2$ are affected by substantial power
corrections.  For the production of $\rho^0$, $\phi$ and $\pi^+$ there
is data or preliminary data, which we will use to assess the
quantitative validity of our calculated cross sections.  Secondly, we
evaluate the contribution of these exclusive channels to
semi-inclusive production of $\pi$ and $K$, and compare them with the
results obtained from leading-twist quark fragmentation.  For the
exclusive meson production cross sections we rely on experimental data
where possible, and only use our leading-twist calculation to estimate
the \emph{ratio} of cross sections for measured and unmeasured
channels.

The paper is organized as follows.  In Sects.~\ref{sec:excl-mesons}
and \ref{sec:gpd-models} we summarize the leading-twist description of
exclusive meson production and describe the models for the GPDs used
in our investigation.  An analysis of pseudoscalar and vector meson
production channels is then given in Sects.~\ref{sec:pseudo-results}
and \ref{sec:vector}.  In Sect.~\ref{sec:data-compare} we briefly
discuss the limitations of our leading-order results and compare them
with experimental data.  The contribution of exclusive channels to
semi-inclusive meson production is compared with leading-twist quark
fragmentation in Sect.~\ref{sec:exclu-inclu}, and in
Sect.~\ref{sec:summary} we summarize our main results.  Some important
technical details are given in two appendices.


\section{Exclusive meson production in the leading-twist approximation}
\label{sec:excl-mesons}

Let us consider the exclusive electroproduction process
\begin{equation}
  \label{excl-reac}
e(k) \; + \; p(p) \;\; \to \;\; e(k') \; + \; M(q') \; + \; B(p') ,
\end{equation}
where $M$ is a meson and $B$ a baryon, and where four-momenta are
indicated in parentheses.  Throughout this work we assume beam and
target to be unpolarized.  We write $q= k-k'$, $\Delta= p'-p$, and
use the standard kinematic variables
\begin{equation}
t  = \Delta^2 , \qquad
Q^2= -q^2 , \qquad
W^2= (p+q)^2 , \qquad
\xb = Q^2 /(2pq) , \qquad
y = (pq) /(pk) .
\end{equation}
We respectively write $m_p$, $m_B$, $m_M$ for the masses of the
proton, the baryon $B$, and the meson $M$.  With Hand's convention
\cite{Hand:1963bb} for the virtual photon flux, the electroproduction
cross section is given by
\begin{equation}
\label{electroproduction}
\frac{d\sigma(ep\rightarrow eMB)}{dQ^2\,d\xb\,dt}=
\frac{\alpha_{em}}{2\pi}
\frac{y^2}{1-\epsilon}
\frac{1-\xb}{\xb}
\frac{1}{Q^2}
\left[ \frac{d\sigma_T}{dt}+\epsilon\frac{d\sigma_L}{dt} \right]
\end{equation}
in terms of the cross sections $d\sigma_T/dt$ and $d\sigma_L/dt$ of
the $\gamma^* p\to M B$ subprocess for transverse and longitudinal
$\gamma^*$, where
\begin{equation}
\epsilon=\frac{1-y -(y \xb m_p /Q)^2}{1-y+y^2/2 +(y \xb m_p /Q)^2}
\end{equation}
is the ratio of longitudinal to transverse photon flux.

\begin{figure}
\begin{center}
\leavevmode
\includegraphics[width=0.8\textwidth]{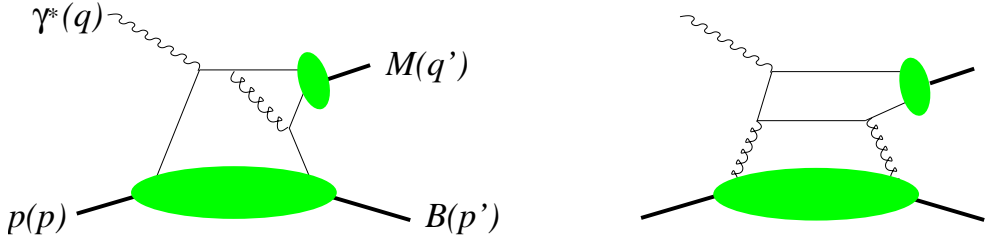}
\end{center}
\caption{\label{fig:meson-graphs} Example graphs for exclusive meson
production in the generalized Bjorken limit.  The large blob denotes
the $p\to B$ transition GPD and the small one the meson distribution
amplitude.}
\end{figure}

In the generalized Bjorken limit of large $Q^2$ at fixed $\xb$ and
fixed $t$, the process amplitude factorizes into a hard-scattering
kernel convoluted with generalized parton distributions for the $p\to
B$ transition and with the distribution amplitude of the meson
\cite{Collins:1996fb}.  Example graphs are shown in
Fig.~\ref{fig:meson-graphs}.  In this limit the longitudinal cross
section can be written as
\begin{equation}
  \label{H-combinations}
\frac{d\sigma_L}{dt}=
\frac{\alpha_{em}}{Q^6}
\frac{\xb^2}{1-\xb} \Bigg\{
(1-\xi^2) |\mathcal{H}|^2 
   - \Bigg[ \frac{2\xi (m_B^2 - m_p^2) + t}{(m_B+m_p)^2} 
   + \xi^2 \Bigg]\, |\mathcal{E}|^2
   - \Bigg[ \xi + \frac{m_B-m_p}{m_B+m_p} \Bigg]\,
   2\xi\, \re( \mathcal{E}^* \mathcal{H} ) \Bigg\}
\end{equation}
for vector mesons, and as
\begin{equation}
  \label{H-tilde-combinations}
\frac{d\sigma_L}{dt}=
\frac{\alpha_{em}}{Q^6}
\frac{\xb^2}{1-\xb} \Bigg\{
 (1-\xi^2) |\tilde\mathcal{H}|^2 
   + \frac{(m_B-m_p)^2 - t}{(m_B+m_p)^2}\,
     \xi^2 |\tilde\mathcal{E}|^2
   - \Bigg[ \xi + \frac{m_B-m_p}{m_B+m_p} \Bigg]\,
   2\xi\, \re( \tilde\mathcal{E}^* \tilde\mathcal{H} ) \Bigg\}
\end{equation}
for pseudoscalar mesons $M$.  The transverse cross section
$d\sigma_T/dt$ is power suppressed by $1/Q^2$ compared with
$d\sigma_L/dt$.  Here we have in addition used the skewness variable
\begin{equation}
  \label{xi_definition}
\xi = \frac{(p-p') (q+q')}{(p+p') (q+q')} \approx \frac{\xb}{2-\xb} ,
\end{equation}
where the approximation holds in the generalized Bjorken limit.  Note
that the prefactor in (\ref{H-combinations}) and
(\ref{H-tilde-combinations}) can be rewritten as $\xb^2 /(1-\xb) =
4\xi^2 /(1-\xi^2)$.  The quantities $\mathcal{H}$, $\mathcal{E}$ and
$\tilde\mathcal{H}$, $\tilde\mathcal{E}$ are specific for each
channel.  Throughout this work we will take their leading order
approximations in $\alpha_s$.  To be specific, we have
\begin{eqnarray}
  \label{script-H}
\mathcal{H}_{K^{*+}\Lambda}(\xi,t) 
&=& \frac{4\pi\alpha_s}{27} f_{K^*}
\Bigg[ \int_0^1 dz\, \frac{1}{z(1-z)}\, \phi_{K^{*+}}(z) 
  \int_{-1}^1 dx\, \frac{2H_{p\to \Lambda}(x,\xi,t) 
      + H_{p\to \Lambda}(-x,\xi,t)}{\xi-x-i\varepsilon} 
\nonumber \\
 && \hspace{3.4em}
{}- \int_0^1 dz\, \frac{2z-1}{z(1-z)}\, \phi_{K^{*+}}(z)
  \int_{-1}^1 dx\, \frac{2H_{p\to \Lambda}(x,\xi,t)
      - H_{p\to \Lambda}(-x,\xi,t)}{\xi-x-i\varepsilon}
\Bigg] ,
\nonumber \\
\tilde\mathcal{H}_{K^{+}\Lambda}(\xi,t)
&=& \frac{4\pi\alpha_s}{27} f_{K}
\Bigg[ \int_0^1 dz\, \frac{1}{z(1-z)}\, \phi_{K^+}(z)
  \int_{-1}^1 dx\, \frac{2\tilde{H}_{p\to \Lambda}(x,\xi,t) 
      + \tilde{H}_{p\to \Lambda}(-x,\xi,t)}{\xi-x-i\varepsilon} 
\nonumber \\
 && \hspace{3.3em}
{}- \int_0^1 dz\, \frac{2z-1}{z(1-z)}\, \phi_{K^+}(z)
  \int_{-1}^1 dx\, \frac{2\tilde{H}_{p\to \Lambda}(x,\xi,t)
      - \tilde{H}_{p\to \Lambda}(-x,\xi,t)}{\xi-x-i\varepsilon}
\Bigg]
\end{eqnarray}
for $\gamma^* p\to K^{*+} \Lambda$ and $\gamma^* p\to K^{+} \Lambda$,
respectively, with analogous expressions for $\mathcal{E}_{K^{*+}
\Lambda}$ and $\tilde\mathcal{E}_{K^{+} \Lambda}$.  The 
GPDs for the $p\to\Lambda$ transition are defined as
\begin{eqnarray}
\lefteqn{
\frac{1}{2} \int \frac{d z^-}{2\pi}\, e^{ix P^+ z^-}
  \langle \Lambda |\, \bar{s}(-\half z)\, \gamma^+ u(\half z) 
  \,|p \rangle \Big|_{z^+=0,\, z_T=0}
} \hspace{3em}
\nonumber \\
&=& \frac{1}{2P^+} \left[
  H_{p\to \Lambda}(x,\xi,t)\, \bar{u} \gamma^+ u +
  E_{p\to \Lambda}(x,\xi,t)\, \bar{u} 
     \frac{i \sigma^{+\alpha} \Delta_\alpha}{m_{\Lambda} + m_p} u
  \, \right] ,
\nonumber \\
\lefteqn{
\frac{1}{2} \int \frac{d z^-}{2\pi}\, e^{ix P^+ z^-}
  \langle \Lambda |\, 
     \bar{s}(-\half z)\, \gamma^+ \gamma_5\, u(\half z)
  \,|p \rangle \Big|_{z^+=0,\, z_T=0}
} \hspace{3em}
\nonumber \\
&=& \frac{1}{2P^+} \left[
  \tilde{H}_{p\to \Lambda}(x,\xi,t)\, 
         \bar{u} \gamma^+ \gamma_5 u +
  \tilde{E}_{p\to \Lambda}(x,\xi,t)\, \bar{u}
         \frac{\gamma_5 \Delta^+}{m_{\Lambda} + m_p} u
  \, \right] ,
\end{eqnarray}
where we use light-cone coordinates $v^\pm= (v^0 \pm v^3) /\sqrt{2}$
and $v_T =(v^1,v^2)$ for a four-vector $v$ and assume light-cone gauge
$A^+ =0$.  For brevity we have not displayed the momentum and
polarization dependence of the baryon spinors on the right-hand sides.
GPDs for other transitions are defined in full analogy.  The integrals
over meson distribution amplitudes in (\ref{script-H}) can be
expressed as
\begin{equation}
\int_0^1 dz\, \frac{1}{z(1-z)}\, \phi(z)
  = 6 \Big[ 1 + \sum_{n=1}^\infty a_{2n} \Big] , \qquad\qquad
\int_0^1 dz\, \frac{2z-1}{z(1-z)}\, \phi(z)
  = 6 \sum_{n=1}^\infty a_{2n-1}
\end{equation}
through their coefficients in the expansion
\begin{equation}
  \label{Gegenbauer}
\phi(z) = 6z(1-z) \Big[ 1 + \sum_{n=1}^\infty a_n\, C_n^{3/2}(2z-1)
  \Big]
\end{equation}
on Gegenbauer polynomials, where $z$ is the light-cone momentum
fraction carried by the quark in the meson.  Note that odd Gegenbauer
coefficients $a_{2n-1}$ describe an asymmetry in the momentum
distribution of the quark and antiquark in the meson.  They can be
nonzero for $K$ and $K^*$ due to flavor SU(3) breaking.  In
(\ref{script-H}) to (\ref{Gegenbauer}) we have not displayed the
logarithmic dependence on the renormalization scale in $\alpha_s$ and
on the factorization scale in the GPDs and the distribution
amplitudes.

\begin{table}
\caption{\label{tab:GPDs} Combinations of proton GPDs to be used for
  various channels $\gamma p\to MB$ at the place of $2H_{p\to
  \Lambda}(x,\xi,t) + H_{p\to \Lambda}(-x,\xi,t)$ or
  $2\tilde{H}_{p\to \Lambda}(x,\xi,t) + \tilde{H}_{p\to
  \Lambda}(-x,\xi,t)$ in (\protect\ref{script-H}).  All distributions
  are to be evaluated at arguments $x$, $\xi$, $t$, with $H^q$,
  $\tilde{H}^q$ and $H^g$ as defined in \protect\cite{Diehl:2003ny}
  and $H^{\bar{q}}$, $\tilde{H}^{\bar{q}}$ given
  above~(\protect\ref{forward-limits}).}
$$
\renewcommand{\arraystretch}{1.7}
\begin{array}{ll} \hline\hline
\rho^+ n & 2 [H^u - H^d] - [H^{\bar{u}} - H^{\bar{d}}\sms] \\
\rho^0\sms p & \frac{1}{\sqrt{2}} \Big(\, [2H^u + H^d] 
         + [2H^{\bar{u}} + H^{\bar{d}}\sms] 
         + \frac{9}{4}\, x^{-1} H^g \,\Big) \\
\omega\sms p & \frac{1}{\sqrt{2}} \Big(\, [2H^u - H^d] 
         + [2H^{\bar{u}} - H^{\bar{d}}\sms] 
         + \frac{3}{4}\, x^{-1} H^g \,\Big) \\
K^{*+}\sms\Lambda & - \frac{1}{\sqrt{6}} 
          \Big(\, 2\sms [2H^u - H^d - H^s]
          - [2H^{\bar{u}} - H^{\bar{d}} - H^{\bar{s}}\sms] \,\Big) \\
K^{*+}\sms\Sigma^0 & - \frac{1}{\sqrt{2}} \Big(\, 2\sms [H^d - H^s] 
                     - [H^{\bar{d}} - H^{\bar{s}}\sms] \,\Big) \\
K^{*0}\sms\Sigma^+ & [H^d - H^s] 
                   + [H^{\bar{d}} - H^{\bar{s}}\sms] \\
\phi\sms p & -\Big(\, [H^s + H^{\bar{s}}\sms]
             + \frac{3}{4}\, x^{-1} H^g \,\Big)
\rule[-0.8em]{0pt}{1em} \\
\hline
\pi^+ n & 2 [\tilde{H}^u - \tilde{H}^d] 
          + [\tilde{H}^{\bar{u}} - \tilde{H}^{\bar{d}}\sms] \\
\pi^0 p & \frac{1}{\sqrt{2}}  \Big(\, [2\tilde{H}^u + \tilde{H}^d] 
        - [2\tilde{H}^{\bar{u}} + \tilde{H}^{\bar{d}}\sms] \,\Big) \\
K^+ \Lambda & - \frac{1}{\sqrt{6}} \Big(\,
            2\sms [2\tilde{H}^u - \tilde{H}^d - \tilde{H}^s]
            + [2\tilde{H}^{\bar{u}} 
            - \tilde{H}^{\bar{d}} - \tilde{H}^{\bar{s}}\sms] \,\Big) \\
K^+ \Sigma^0 & - \frac{1}{\sqrt{2}} 
           \Big(\, 2\sms [\tilde{H}^d - \tilde{H}^s]
           + [\tilde{H}^{\bar{d}} - \tilde{H}^{\bar{s}}\sms] \,\Big) \\
K^0\sms\Sigma^+ & [\tilde{H}^d - \tilde{H}^s] 
                - [\tilde{H}^{\bar{d}} - \tilde{H}^{\bar{s}}\sms] \\
\hline\hline
\end{array}
\renewcommand{\arraystretch}{1}
$$
\end{table}

Using flavor SU(3) symmetry one can relate the transition GPDs from
the proton to a hyperon to the distributions $H^q(x,\xi,t)$ for quark
flavor $q$ in the proton \cite{Goeke:2001tz,Belitsky:2005qn}.  This
gives in particular $H_{p\to \Lambda} = - [2H^u - H^d -H^s] /\sqrt{6}$
and an analogous relation for $\tilde{H}_{p\to \Lambda}$.  We use
these relations throughout, except for $\tilde{E}$, where there are
large effects of SU(3) breaking as we shall see below.  Results
analogous to (\ref{script-H}) hold for all meson channels we consider,
see e.g.\ \cite{Goeke:2001tz,Diehl:2003ny,Diehl:2003qa}, and we have
collected the relevant combinations of GPDs in Table~\ref{tab:GPDs}.
There we have introduced $H^{\bar{q}}(x,\xi,t) = -H^q(-x,\xi,t)$ and
$\tilde{H}^{\bar{q}}(x,\xi,t) = \tilde{H}^q(-x,\xi,t)$, so that for
$x>0$ we have simple forward limits
\begin{equation}
  \label{forward-limits}
H^q(x,0,0) =q(x) , \quad
H^{\bar{q}}(x,0,0) =\bar{q}(x) , \quad
\tilde{H}^q(x,0,0) =\Delta q(x) , \quad
\tilde{H}^{\bar{q}}(x,0,0) =\Delta\bar{q}(x)
\end{equation}
in terms of the unpolarized and polarized quark and antiquark
densities in the proton.  For gluons we have $H^g(x,0,0) = xg(x)$,
which is the origin of the additional factors $x^{-1}$ in the entries
for $\rho^0$, $\omega$, $\phi$.  In addition to the replacements in
Table~\ref{tab:GPDs} one has of course to take the appropriate meson
distribution amplitude and meson decay constants in (\ref{script-H}).
For the latter we will take $f_\pi= 131 \mev$, $f_{K} = 160 \mev$, and
\begin{equation}
  \label{decay-constants}
f_\rho= 209 \mev, \quad
f_\omega = 187 \mev , \quad
f_\phi= 221 \mev , \quad
f_{K^*} = 218 \mev 
\end{equation}
from \cite{Beneke:2003zv}.  

For $\alpha_s$ in (\ref{script-H}) we will take the one-loop running
coupling at the scale $Q^2$, with three active quark flavors and
$\Lambda_{\mathrm{QCD}} =200 \mev$.  This gives $\alpha_s=0.34$ at
$Q^2=2.5 \gev^2$, where we will show most of our numerical results.
We will not attempt more refined choices of renormalization scale, as
were for instance explored in \cite{Anikin:2004jb}, since our
principal use of the leading-order calculation will be to describe the
\emph{relative} size of cross sections for different exclusive
channels.


\section{Modeling the generalized parton distributions}
\label{sec:gpd-models}

For the calculation of exclusive cross sections we use simple models
of GPDs.  They have been developed in
\cite{Musatov:1999xp,Goeke:2001tz} and been used in most
phenomenological analyses so far.  Our aim here is not to improve on
these models, but instead to see by how much predictions can vary
\emph{within} the given framework.  We take a factorizing $t$
dependence for $H$ and $\tilde{H}$,
\begin{eqnarray}
  \label{factorize-ansatz}
H^q(x,\xi,t) &=& H^q(x,\xi)\, F_1^p(t) , \qquad\qquad\qquad
H^g(x,\xi,t) \;=\; H^g(x,\xi)\, F_1^p(t) , 
\nonumber \\
\tilde{H}^q(x,\xi,t) &=& \tilde{H}^q(x,\xi)\, G_A(t)/G_A(0) ,
\end{eqnarray}
where $F_1^p(t)$ is the electromagnetic Dirac form factor of the
proton and $G_A(t)$ the isovector axial form factor of the nucleon.  A
more refined version of the model would take different combinations of
the proton and neutron form factors for $H^u$ and $H^d$, but for the
low values of $t$ dominating integrated cross sections, $F_1^n(t)$ is
much smaller than $F_1^p(t)$ and we simply neglect it.  In this sense
(\ref{factorize-ansatz}) is consistent with the sum rule for the first
moment $\int dx\, H^q(x,\xi,t)$.  The ansatz for $\tilde{H}^q$ is
consistent with the sum rule for $\int dx\, \tilde{H}^q(x,\xi,t)$ to
the extent that the (unknown) isoscalar axial form factor has the same
$t$ dependence as the isovector one.  In our numerical evaluations we
take the familiar parameterizations
\begin{equation}
  \label{dipole-ffs}
F_1^p(t) = \frac{4m_p^2-2.8 t}{4m_p^2-t}\,
    \frac{1}{[\sms 1-t/(0.71 \gev^2) \sms] \rule{0pt}{0.7em}^2} \; , 
\qquad\qquad
\frac{G_A(t)}{G_A(0)}
   = \frac{1}{[\sms 1-t/(1.05 \gev^2) \sms] \rule{0pt}{0.7em}^2}  \; .
\end{equation}
We note that for the gluon distribution $H^g$ there is no reason a
priori to take the electromagnetic form factor $F_1^p(t)$ in the
ansatz (\ref{factorize-ansatz}).  It turns out, however, that
$F_1^p(t)$ is well approximated by a dipole form $F_1^p(t) = [1 -
t/(0.98 \gev^2)]^{-2}$ for $t$ up to about $3 \gev^2$
\cite{Diehl:2004cx} and thus close to the two-gluon form factor
advocated in \cite{Frankfurt:2003td}.

It is rather certain that the ansatz (\ref{factorize-ansatz}) is too
simple and can at best reflect the correct $t$ dependence in a limited
range of $x$ and $\xi$ \cite{Goeke:2001tz,Diehl:2004cx,Hagler:2003is}.
For $x$ and $\xi$ in the valence region, say above $0.2$, the decrease
of GPDs with $t$ is most likely less steep than the one of $F_1^p(t)$
and $G_A(t)$.  Whereas there are phenomenological constraints of the
$t$ behavior of valence quark GPDs \cite{Diehl:2004cx} and for gluons
at small $x$ \cite{Frankfurt:2003td}, the behavior for sea quarks is
largely unknown, and sea quarks are important for the $\xb$ region
around $0.1$ we will be mostly concerned with.  Furthermore, the $t$
dependence of meson production at moderate $Q^2$ is strongly affected
by power corrections, as is for instance seen in the $Q^2$ dependence
of the logarithmic slope $B = (\partial /\partial t) \log (d\sigma/dt)
|_{t=0}$ for $\rho^0$ production at very high energies
\cite{Adloff:1999kg}.  We note that our ansatz
(\ref{factorize-ansatz}) gives a slope parameter $B \approx 4
\gev^{2}$, which may be quite realistic for $\xb$ around $0.1$.
Furthermore, cross section ratios should be less affected by the
insufficiency of our ansatz, since they are sensitive only to the
relative $t$ dependence of the contributions from different quark
flavors and from gluons.

For the $t$ independent functions in (\ref{factorize-ansatz}) we use
the double distribution based ansatz of \cite{Musatov:1999xp}, whose
ingredients are the usual parton densities at a given factorization
scale $\mu$ and a so-called profile parameter $b$, where $\mu$ and $b$
are to be regarded as free parameters of the model.  Explicit
expressions are given in App.~\ref{app:GPDs}.  We will not take into
account the evolution of GPDs, which should not be too problematic
since our numerical applications will stay within a rather narrow
range of $Q^2$.

The modeling of the nucleon helicity-flip distributions $E^q$ and
$E^g$ is still at an early stage of development, with the most
advanced considerations focused on the valence quark domain
\cite{Goeke:2001tz,Diehl:2004cx}.  Fortunately, contributions from $E$
enter the unpolarized meson production cross section
(\ref{H-combinations}) with prefactors that are quite small in the
kinematics we are most interested in.  Following the argumentation of
\cite{Diehl:2004wj} that $E$ is not much larger than $H$ for a given
parton species, we hence neglect $E$ altogether in our cross section
estimates.

The distributions $\tilde{E}$ cannot be neglected since they receive
contributions proportional to $\xi^{-1}$ that compensate the
kinematic prefactors in the cross section
(\ref{H-tilde-combinations}).  We model them as
\begin{eqnarray}
  \label{meson-pole}
\tilde{E}_{p\to n}(x,\xi,t) &=& 
\frac{\theta(|x| \leq \xi)}{2\xi}\,
\phi_\pi\left(\frac{x+\xi}{2\xi}\right)
\frac{2m_p f_\pi\, g_{\pi NN}}{m_\pi^2 -t}\, 
\frac{\Lambda^2-m_\pi^2}{\Lambda^2-t} ,
\nonumber \\
\tilde{E}_{p\to \Lambda}(x,\xi,t) &=& 
\frac{\theta(|x| \leq \xi)}{2\xi}\,
\phi_K\left(\frac{x+\xi}{2\xi}\right)
\frac{(m_p+m_\Lambda) f_K\, g_{KN\Lambda}}{m_K^2 -t}\,
\frac{\Lambda^2-m_K^2}{\Lambda^2-t} ,
\end{eqnarray}
where the distribution amplitudes $\phi$ are the same as those
introduced above.  For the coupling constants we take the value
$g_{\pi NN} = 2m_p\sms G_A(0) /f_\pi \approx 14.7$ from the
Goldberger-Treiman relation and $g_{KN\Lambda} \approx -13.3$ from
\cite{Frankfurt:1999xe}.  Continued to the points $t= m_\pi^2$ or $t=
m_K^2$ in the unphysical region, the expressions (\ref{meson-pole})
become the well-known results from pion or kaon exchange
\cite{Frankfurt:1999xe,Mankiewicz:1998kg,Penttinen:1999th}.  These can
only be expected to be good approximations for $t$ close to the
squared meson masses, and for $-t$ of several $0.1 \gev^2$ are to be
regarded as extrapolations.  In (\ref{meson-pole}) we have included
form factors that cut off the $1/(m_M^2 -t)$ behavior of the pure pole
terms when $-t$ becomes large.  As default value for the cut-off mass
we will take $\Lambda= 0.8 \gev$ \cite{Koepf:1995yh} and study the
sensitivity of results to the precise value of this parameter.  We
note that for $\Lambda =0.6 \gev$ and $-t \le 1 \gev^2$ the above form
of $\tilde{E}_{p\to n}$ differs by less than $15\%$ from the
corresponding term calculated in the chiral quark-soliton model as
given in Eq.~(4.39) of \cite{Penttinen:1999th}.

With this model for $\tilde{E}$ the longitudinal cross section for
$\gamma^* p\to \pi^+ n$ takes the form
\begin{eqnarray}
  \label{pion-xsec}
\frac{d\sigma_L}{dt} &=& \frac{\alpha_{em}}{Q^6}
\frac{\xb^2}{1-\xb} \Bigg\{
  (1-\xi^2) |\tilde\mathcal{H}(\xi,t) |^2 
  - 2 m_p\, \xi\, \re \tilde\mathcal{H}(\xi,t)\;
    Q^2 F_\pi(Q^2)\, \frac{g_{\pi NN}}{m_\pi^2 -t}\, 
    \frac{\Lambda^2-m_\pi^2}{\Lambda^2-t}
\nonumber \\
 && \hspace{5em} {}- \frac{t}{4}
    \Bigg[ Q^2 F_\pi(Q^2)\, \frac{g_{\pi NN}}{m_\pi^2 -t}\, 
    \frac{\Lambda^2-m_\pi^2}{\Lambda^2-t} \Bigg]^2 \,\Bigg\} \, ,
\end{eqnarray}
where 
\begin{equation}
  \label{pion-ff}
F_\pi(Q^2) = \frac{2\pi\alpha_s}{9} \frac{f_\pi^2}{Q^2} 
  \Bigg[ \int_0^1 dz \frac{1}{z(1-z)}\, \phi_{\pi}(z) \Bigg]^2
\end{equation}
is the electromagnetic pion form factor to leading order in $\alpha_s$
and $1/Q^2$.  An similar expression involving $F_{K^+}(Q^2)$ is
obtained for $\gamma^* p\to K^+ \Lambda$ according to
(\ref{H-tilde-combinations}) and (\ref{meson-pole}).  Note that the
the $|\tilde\mathcal{E}|^2$ term in $d\sigma_L/dt$ has no $\xb$
dependence other than from the explicit factor $\xb^2 /(1-\xb)$.
Within our model the $|\tilde\mathcal{H}|^2$ term reflects the
behavior of the polarized parton distributions at momentum fractions
of order $\xi$, and its contribution to $d\sigma_L/dt$ can very
roughly be represented by $[\sms \xi \Delta q(\xi) \sms]^2$.


\section{Exclusive pseudoscalar meson production}
\label{sec:pseudo-results}

In this and the next section we present numerical results for cross
sections of exclusive meson production.  Our main focus is to compare
the rates for different production channels and to investigate model
uncertainties.  Comparison with data in Sect.~\ref{sec:data-compare}
will allow us to estimate the shortcomings of the leading
approximation in $1/Q^2$ and in $\alpha_s$, on which our calculations
are based.

The factorization theorem for exclusive meson production requires $t$
to be much smaller than $Q^2$.  For definiteness we will give all
meson cross sections in this paper integrated over $-t$ from its
smallest kinematically allowed value $-t_0$ to an upper limit of $1
\gev^2$.  In generalized Bjorken kinematics we have
\begin{equation}
  \label{t-min}
- t_0 \approx \frac{2 \xi^2 (m_B^2 + m_p^2) + 2\xi (m_B^2 - m_p^2)}{
  1-\xi^2}
\end{equation}
with $\xi$ defined in (\ref{xi_definition}).  For low enough $x_B$
most of the cross section should be accumulated in this $t$ region,
whereas for large $x_B$ our cross sections decrease to the extent that
$-t_0$ approaches $1 \gev^2$.

\begin{figure}
\centering
\includegraphics[width=8.3cm]{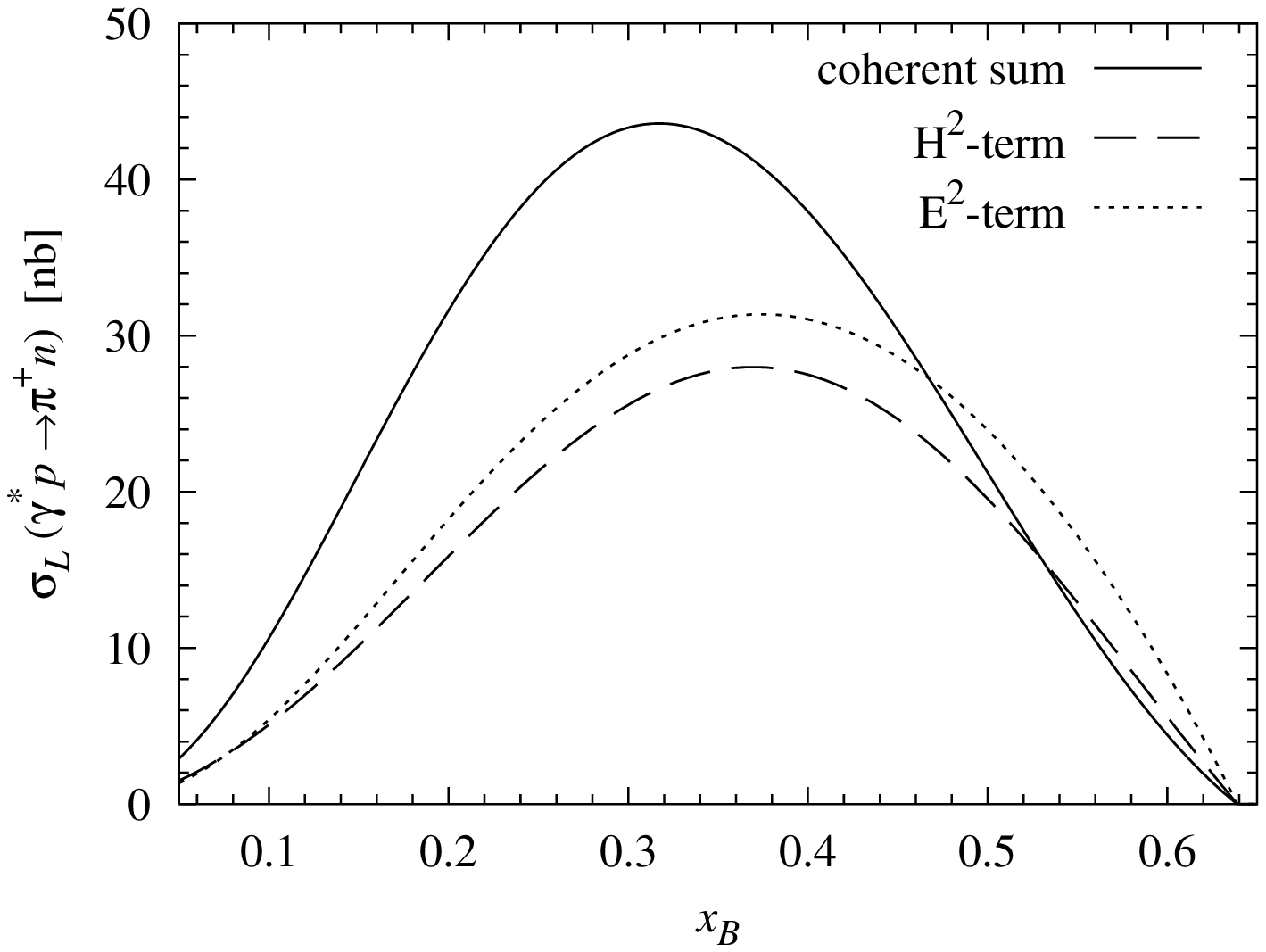}
\includegraphics[width=8.3cm]{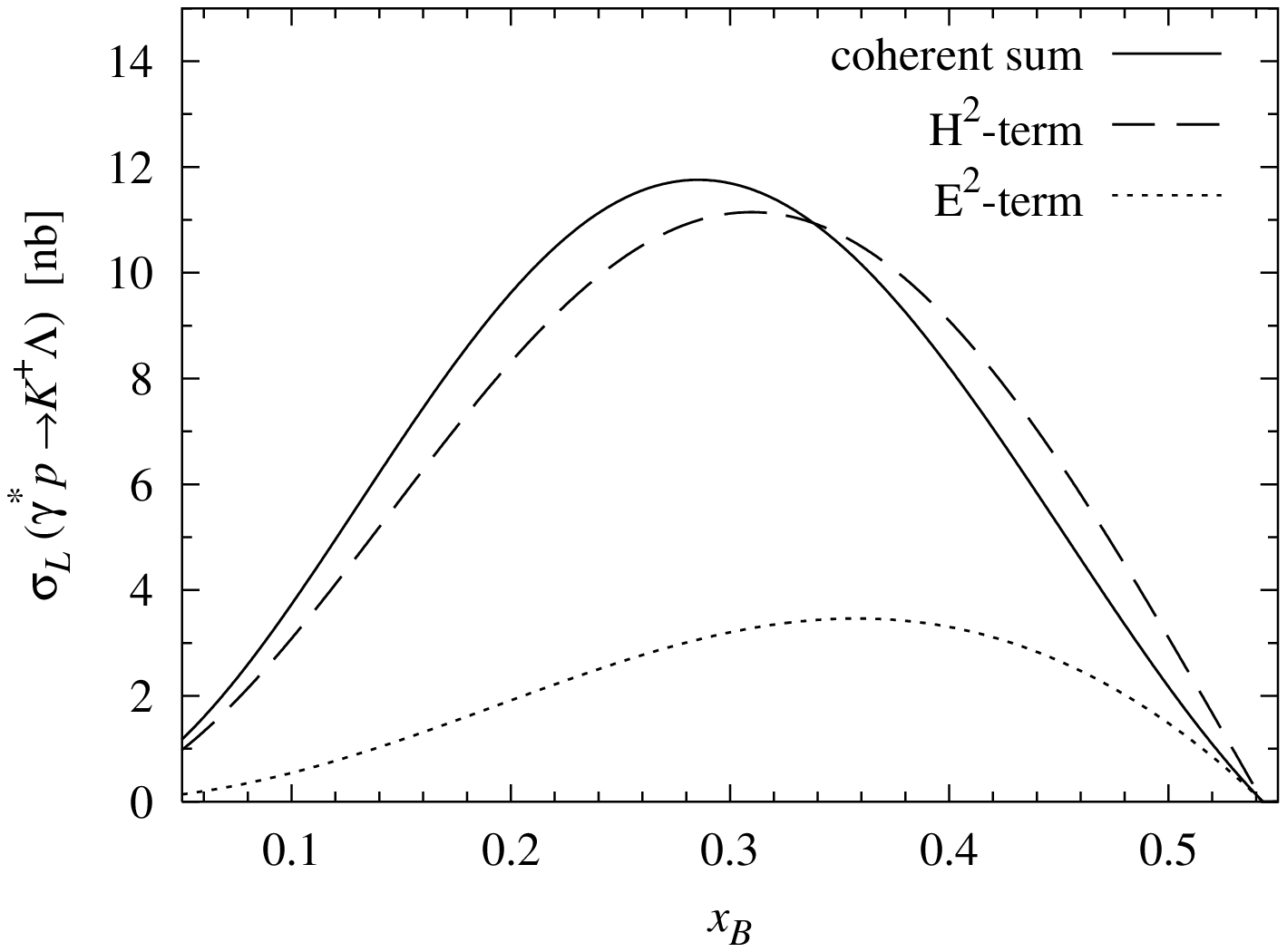}
\caption{\label{fig:EtversusHt} Leading-twist cross sections for
  $\gamma^*_L\, p\to \pi^+ n$ (left) and $\gamma^*_L\, p\to K^+
  \Lambda$ (right) for $Q^2 =2.5 \gev^2$.  An upper cut on $-t$ of
  $1\gev^2$ has been applied here and in all further plots of this
  paper.  Shown are the individual contributions from
  $|\tilde\mathcal{H}|^2$ and $|\tilde\mathcal{E}|^2$, and their
  coherent sum according to (\protect\ref{H-tilde-combinations}).}
\end{figure}

To begin with, let us investigate the relative importance of the
contributions from $\tilde\mathcal{H}$ and $\tilde\mathcal{E}$ to
$\pi^+$ and $K^+$ production with our model assumptions.  As is seen
in Fig.~\ref{fig:EtversusHt}, exclusive $\pi^+$ production receives
comparable contributions from both the $|\tilde\mathcal{H}|^2$ and the
$|\tilde\mathcal{E}|^2$ term in (\ref{H-tilde-combinations}), as well
as from the interference term proportional to $\re (
\tilde\mathcal{E}^* \tilde\mathcal{H} )$.  Note that the relative
weight of the contributions is different for $d\sigma_L /dt$, where it
strongly depends on $t$ given the characteristic $t$ dependence of the
pion pole term (\ref{meson-pole}).  In $K^+$ production the influence
of $\tilde\mathcal{E}$ is less prominent, since the pole factor
$(m_K^2 -t)^{-1}$ gives much less enhancement at small $t$ than
$(m_\pi^2 -t)^{-1}$.

\begin{figure}
\centering 
\includegraphics[width=8.3cm]{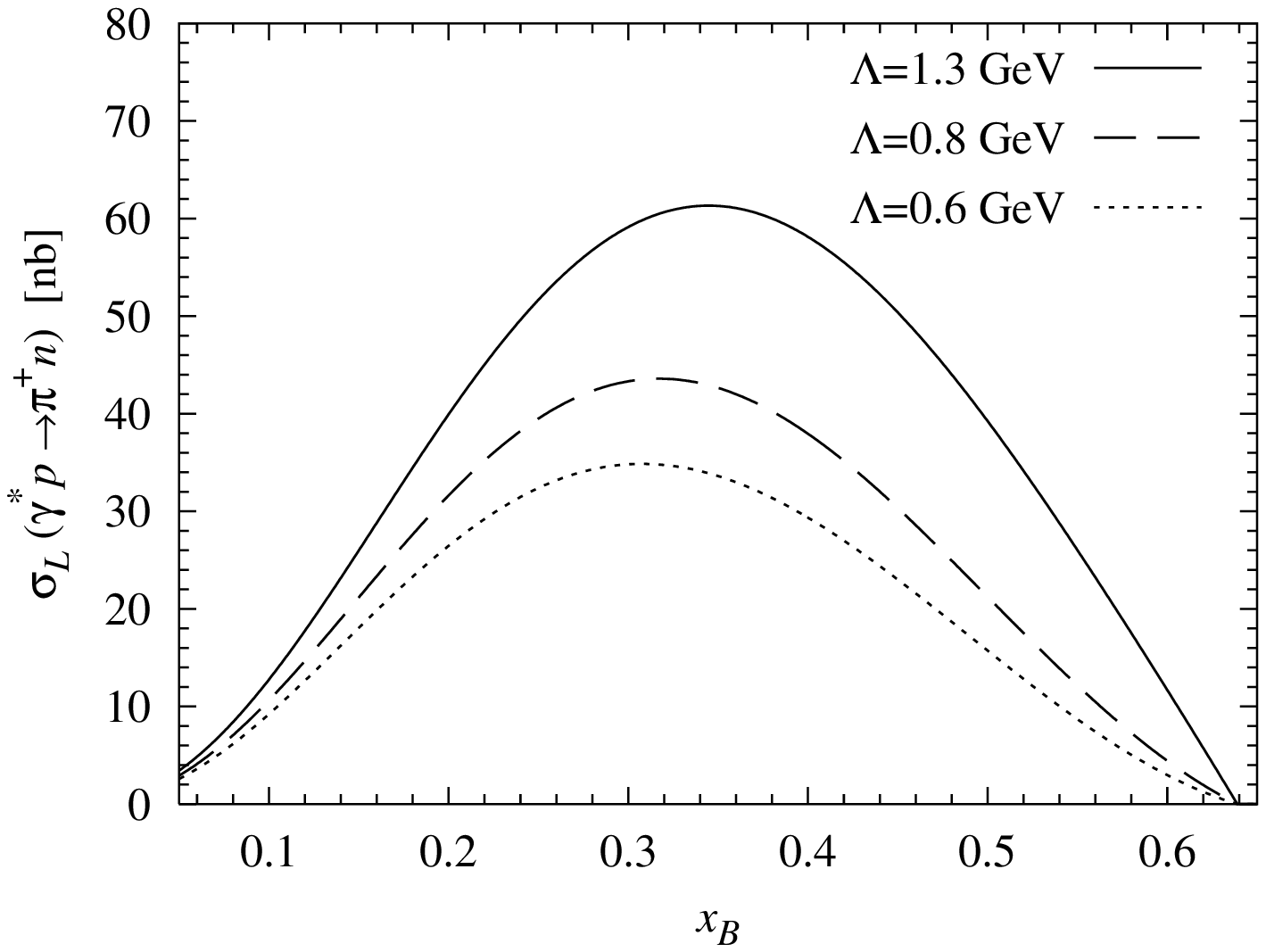}
\includegraphics[width=8.3cm]{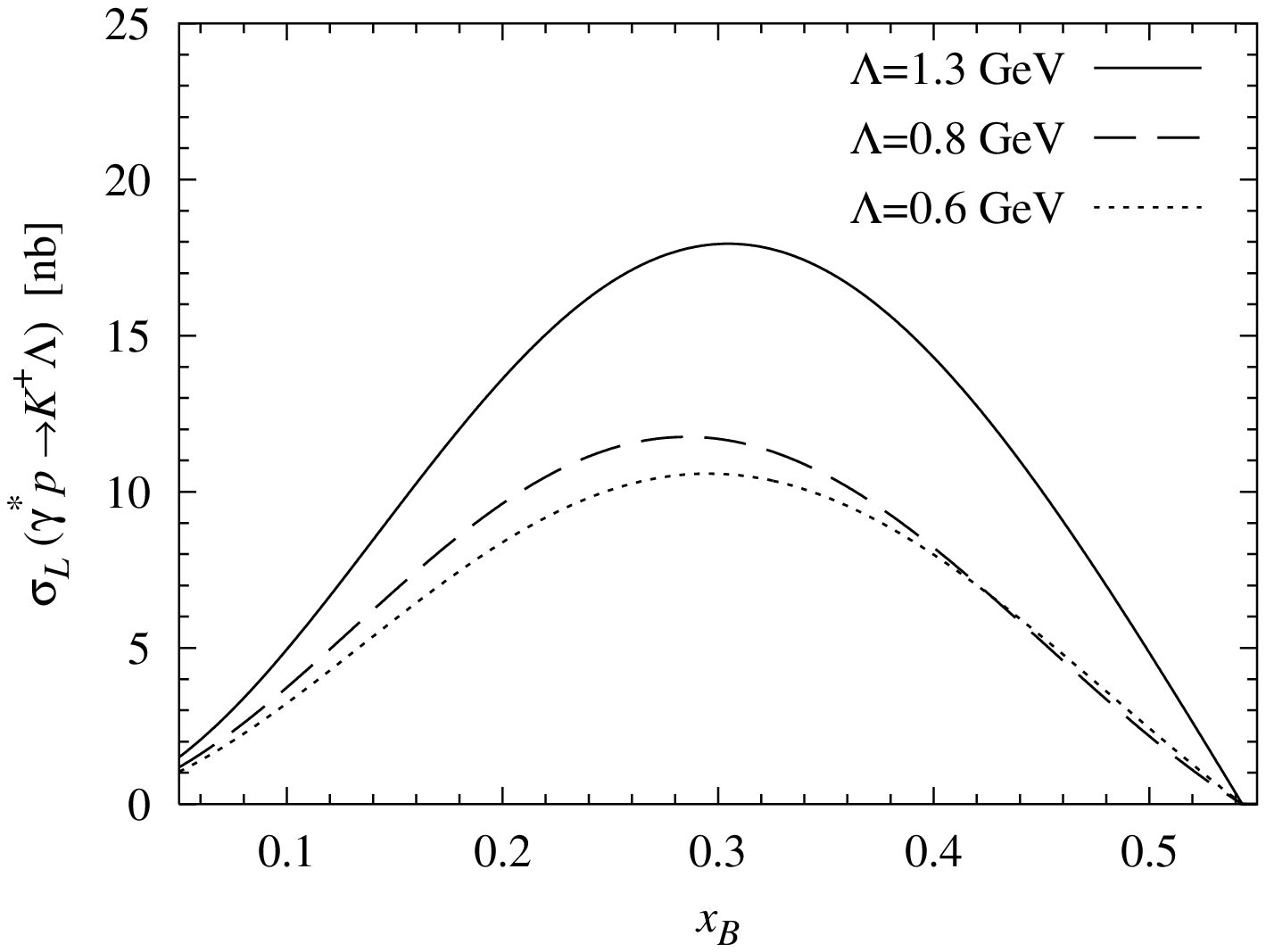}
\caption{\label{lambda-dep} Leading-twist cross section for
  $\gamma^*_L\, p\to \pi^+ n$ (left) and $\gamma^*_L\, p\to K^+
  \Lambda$ (right) at $Q^2=2.5 \gev^2$ obtained with different values
  of the parameter $\Lambda$ in the form factor multiplying the pion
  or kaon pole contribution in (\protect\ref{meson-pole}).}
\end{figure}

To obtain the curves in Fig.~\ref{fig:EtversusHt} we have made a
number of choices in the nonperturbative input to the cross section,
which we now discuss in turn.  In Fig.~\ref{lambda-dep} we show how
the cross section changes when we vary the parameter $\Lambda$ in our
model for $\tilde{E}$, where $\Lambda =1.3 \gev$ represents an upper
limit of the values discussed in the phenomenological study
\cite{Koepf:1995yh}, and $\Lambda =0.6 \gev$ approximates the form
factor dependence obtained for the pion pole contribution in
\cite{Penttinen:1999th}, as discussed at the end of
Sect.~\ref{sec:excl-mesons}.  Omitting the form factor altogether
(tantamount to setting $\Lambda\to\infty$) the $\pi^+$ cross section
would increase by more than a factor 1.4 and the $K^+$ cross section
by more than a factor 1.7 compared with our default choice
$\Lambda=0.8 \gev$.  Also, the cross sections would considerably
increase when raising the upper cutoff in the $-t$ integration above
$1 \gev^2$.  In other words, the cross section would then receive
substantial contributions from values of $t$ far away from the region
where the pion or kaon pole term can be regarded as a reasonable
approximation of $\tilde{E}$.

\begin{figure}
\centering
  \includegraphics[width=8.3cm]{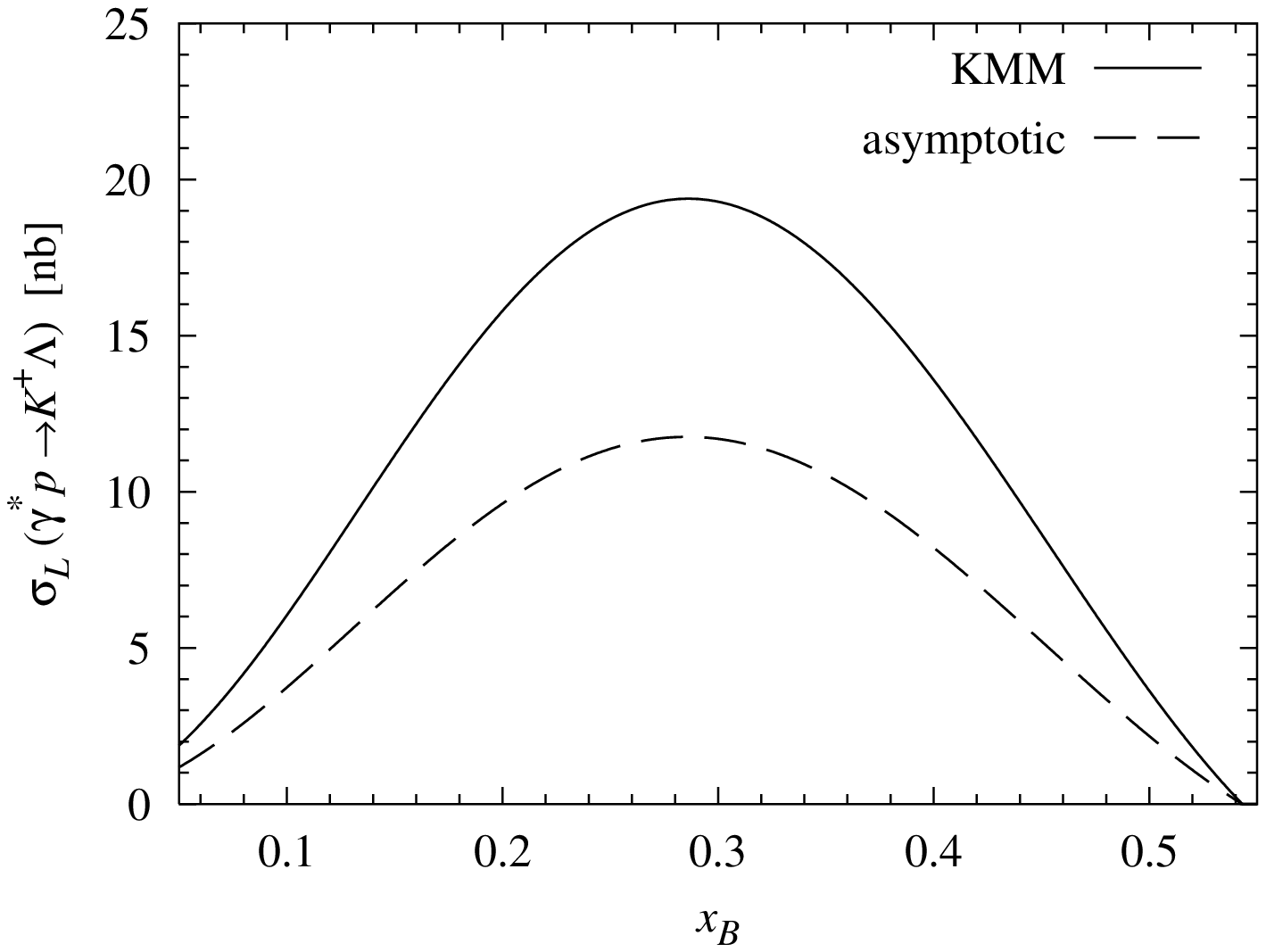}
  \includegraphics[width=8.3cm]{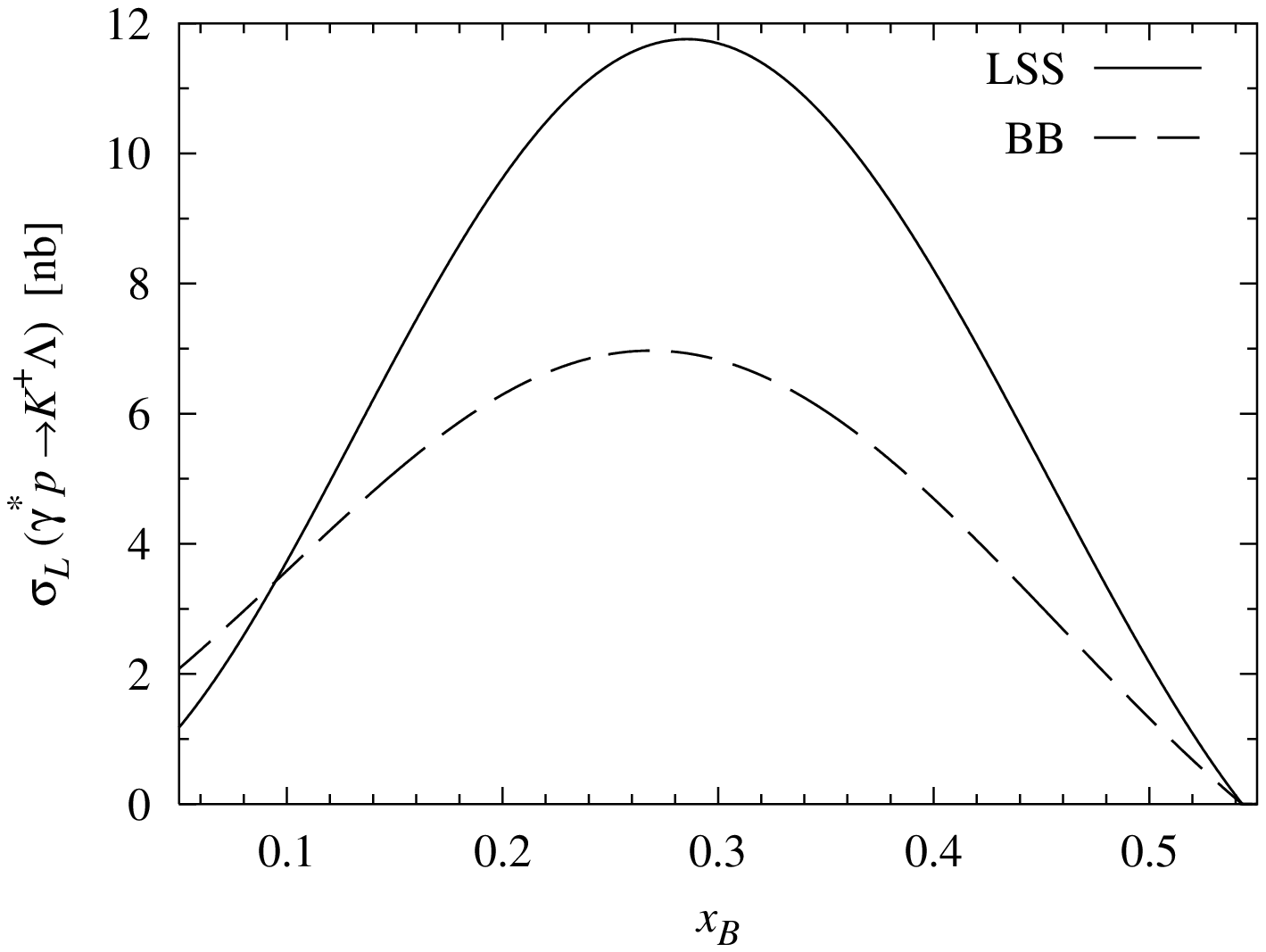}
\caption{\label{fig:kaon-studies} Left: Leading-twist cross section
  for $\gamma^*_L\, p\to K^+ \Lambda$ at $Q^2 = 2.5 \gev^2$ calculated
  with the asymptotic kaon distribution amplitude and with the one
  from Khodjamirian, Mannel and Melcher (KMM)
  \protect\cite{Khodjamirian:2004ga}.  Right: The same cross section
  calculated with different parton distributions in the model for
  $\tilde{H}$.  The distributions from Leader, Sidorov and Stamenov
  (LSS) \protect\cite{Leader:1998qv} and from Bl\"umlein and
  B\"ottcher (BB) \protect\cite{Blumlein:2002be} are taken at a scale
  $\mu = 1\gev$.}
\end{figure}

For the pion distribution amplitude we have taken the asymptotic form
$\phi_\pi(z) = 6 z(1-z)$ under scale evolution, which is close to what
can be extracted from data on the $\gamma$--$\pi$ transition form
factor, see e.g.\ \cite{Diehl:2001dg,Bakulev:2003cs}.  The study in
\cite{Bakulev:2003cs} quotes limits on $a_2 + a_4$ at scale $\mu=1
\gev$ of about $\pm 0.3$ if all other Gegenbauer coefficients are set
to zero.  This corresponds to a change of the $\gamma^* p\to \pi^+ n$
cross section by a factor $(1+a_2+a_4)^2$ between $0.5$ and $1.7$.
For the $K^+$ distribution amplitude we have taken the asymptotic form
as well.  Figure~\ref{fig:kaon-studies} shows how the $K^+$ cross
section changes if instead one takes $a_1= -0.05$ and $a_2=0.27$ at
$\mu=1 \gev$ from the QCD sum rule calculation
\cite{Khodjamirian:2004ga}.  This value of $a_1$ is compatible with
the findings of \cite{Braun:2004vf}.

For our model of $\tilde{H}$ we have taken a double distribution
ansatz with a profile parameter $b=1$ (see Sect.~\ref{sec:excl-mesons}
and App.~\ref{app:GPDs}).  Taking $b=2$ instead would decrease the
$K^+$ cross section by a factor of approximately 0.6.  The pion cross
section changes less, because of the relative weight of
$\tilde\mathcal{H}$ and $\tilde\mathcal{E}$.  A more important source
of uncertainty is however due to the polarized quark densities used as
input to model $\tilde{H}$.  As a default we have used the LO
parameterization from \cite{Leader:1998qv} at a scale $\mu = 1\gev$.
Using instead the LO parameterization in scenario~1 of
\cite{Blumlein:2002be} at the same scale, the $K^+$ cross section
changes as shown in Fig.~\ref{fig:kaon-studies}.  Note that any
uncertainty on parton distributions is amplified in the meson
production cross section, where GPDs appear squared.

Let us now comment on other pseudoscalar channels.  The cross sections
for $\gamma^* p\to K^+ \Sigma^0$ is about an order of magnitude
smaller than the one for $\gamma^* p\to K^+ \Lambda$, as is seen in
the numerical study of \cite{Frankfurt:1999xe}.  For the contribution
from $\tilde\mathcal{H}$ this can be understood from the flavor
structure in Table~\ref{tab:GPDs}, where for a rough estimate one may
concentrate on the dominant terms $\tilde{H}^u$ and $\tilde{H}^d$.
For current parameterizations of polarized parton densities the
combination $[2\tilde{H}^u - \tilde{H}^d] /\sqrt{3}$ for $\Lambda$
production is clearly larger than $\tilde{H}^d$ in the analogous
expression for the $\Sigma^0$ channel.  Concerning the contribution
from $\tilde\mathcal{E}$, the coupling $g_{KN\Sigma^0}$ is about three
times smaller than $g_{KN\Lambda}$ appearing in (\ref{meson-pole}),
see \cite{Frankfurt:1999xe}.  Along the same lines one can see that
the cross section for $\gamma^* p\to K^0 \Sigma^+$ is of similar size
as the one for $\gamma^* p\to K^+ \Sigma^0$.

The channel $\gamma^* p\to \pi^0 p$ does not receive contributions
from the pion pole term in (\ref{meson-pole}) because of charge
conjugation invariance, so that in our model it is entirely given by
the contribution from~$\tilde\mathcal{H}$.  In Table~\ref{tab:GPDs} we
see that the combination $2\tilde{H}^u + \tilde{H}^d$ for $\pi^0$
production is to be compared with $\sqrt{2} [\tilde{H}^u -
\tilde{H}^d]$ for $\gamma^* p\to \pi^+ n$, which is of comparable
size.  One thus expects the $\pi^0$ cross section to be similar to the
$|\tilde\mathcal{H}|^2$ part of the $\pi^+$ cross section.

The exclusive channels $\gamma^* p\to \eta\sms p$ and $\gamma^* p\to
\eta' p$ involve the combination $2\tilde{H}^u - \tilde{H}^d$ instead
of $2\tilde{H}^u + \tilde{H}^d$ in the $\pi^0$ case, which is somewhat
larger because the polarized distributions $\Delta u$ and $\Delta d$
have opposite sign.  The strange quark contribution to these channels
involves $\tilde{H}^s - \tilde{H}^{\bar{s}}$, which vanishes in our
model with polarized parton distributions satisfying $\Delta s(x) =
\Delta\bar{s}(x)$.  A quantitative analysis requires the appropriate
decay constants for the $\eta$ and $\eta'$, see for instance
\cite{Feldmann:1999uf}, but one can in general expect comparable cross
sections for the $\pi^0$, $\eta$ and $\eta'$ channels.


\section{Exclusive vector meson production}
\label{sec:vector}

Within our model for the GPDs, the cross section for vector meson
production is sensitive to unpolarized quark and gluon densities.  To
obtain an indication of uncertainties we have compared results with
the LO distributions from CTEQ6 \cite{Pumplin:2002vw} and the LO
distributions from MRST2001 \cite{Martin:2002dr}.  For consistency we
need LO rather than NLO parton densities, which unfortunately are not
available in several of the most recent parton fits.  We have checked
that the NLO distributions from MRST2001 are in good agreement with
those in the MRST2002 and MRST2004 analyses \cite{Martin:2002aw} for
quark and antiquark densities down to about $x\sim 10^{-2}$ and for
the gluon density down to about $x\sim 10^{-1}$.  Comparing the LO
distributions of CTEQ6 and MRST2001 at a scale $\mu^2=1.2 \gev^2$
(which is the lowest value accepted by the code for the MRST2001
parameterization) we find that the CTEQ6 gluon is larger for $x \lsim
10^{-1}$ and smaller for $x \gsim 10^{-1}$.  The $u$ quark
distribution is quite similar in the two sets for $x\gsim 10^{-2}$,
whereas the $s$ quark is significantly smaller for CTEQ6.  The
distributions for $d$, $\bar{u}$, $\bar{d}$ are quite similar for
$x\gsim 10^{-1}$ and larger for CTEQ6 at smaller $x$.  The LO
parameterization of Alekhin \cite{Alekhin:2002fv} has significantly
larger $u$ and $\bar{u}$ distributions and a smaller gluon than the
two other sets.  At $\mu^2=1.2 \gev^2$ it has however almost no
strange quarks in the proton, which we do not consider physically very
plausible and which is in clear contrast with the results of CTEQ6,
where a dedicated analysis of data constraining the strangeness
distribution was performed.  Since our study is crucially dependent on
the flavor structure of parton distributions, we have therefore not
used \cite{Alekhin:2002fv}.  Comparing the different parton sets at
the higher scale $\mu^2= 2.5 \gev^2$ we find a very similar picture.

\begin{figure}[t]
\centering
\includegraphics[width=8.3cm]
   {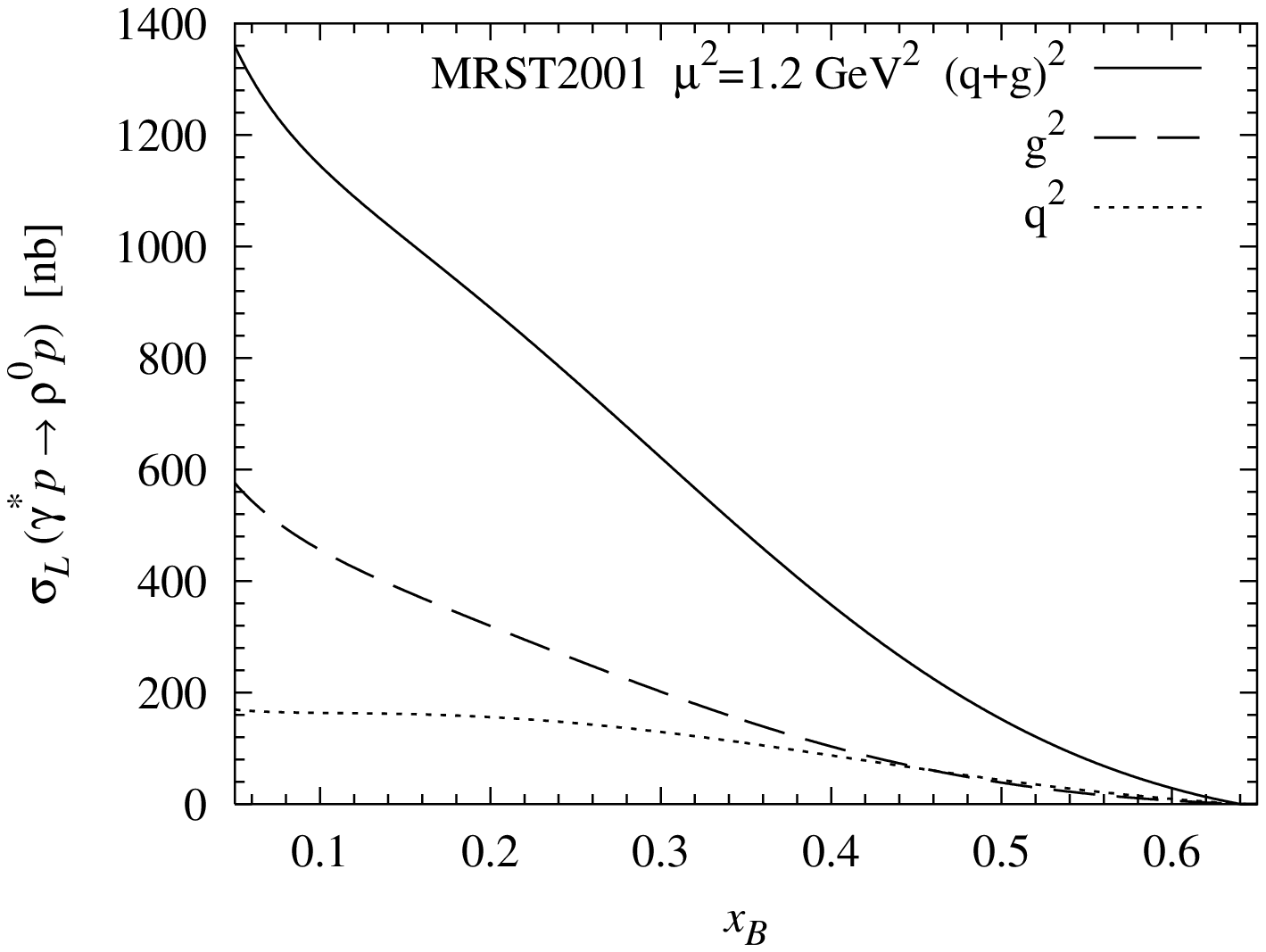}
\includegraphics[width=8.3cm]
   {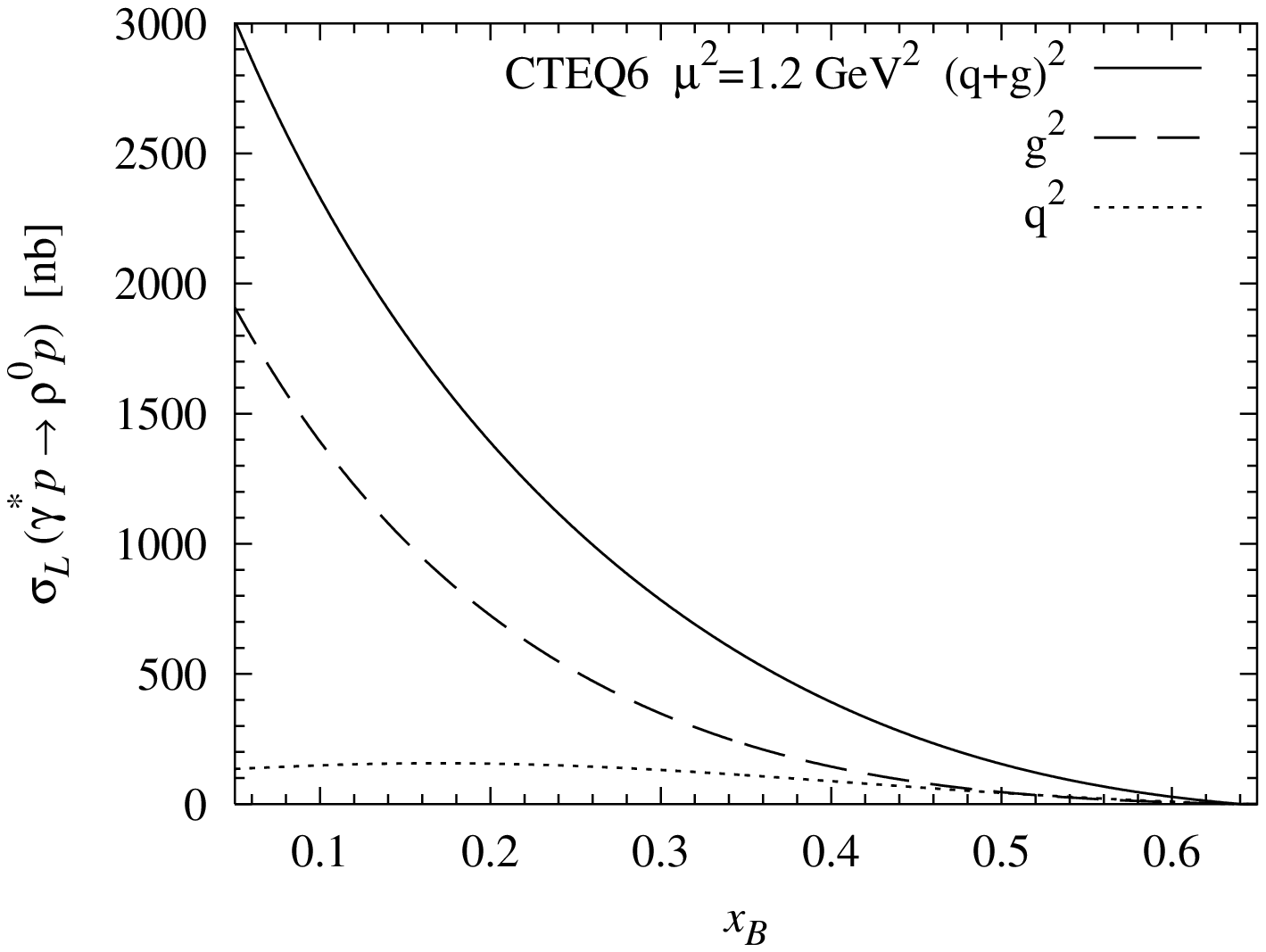}
\includegraphics[width=8.3cm]
   {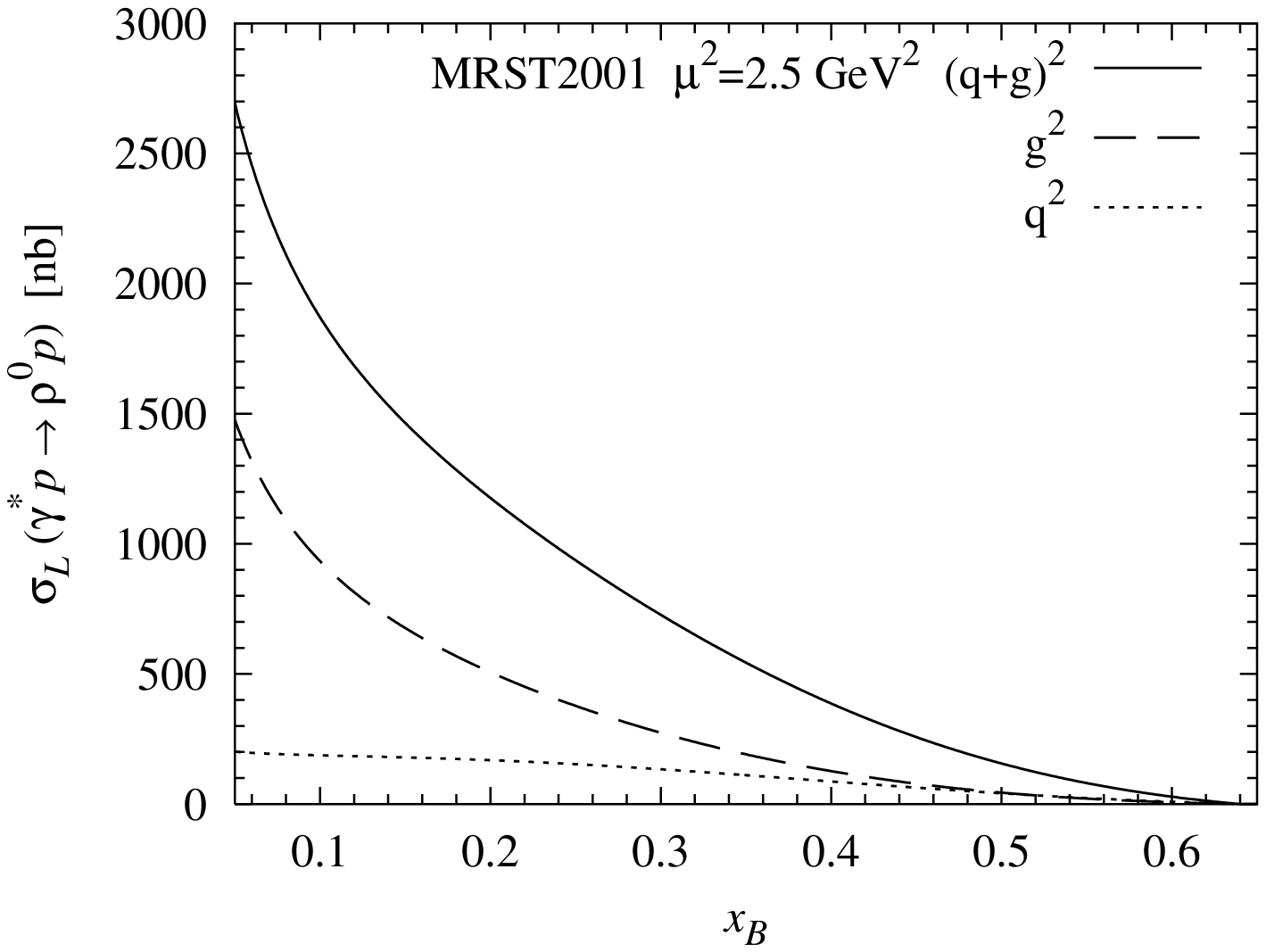}
\includegraphics[width=8.3cm]
   {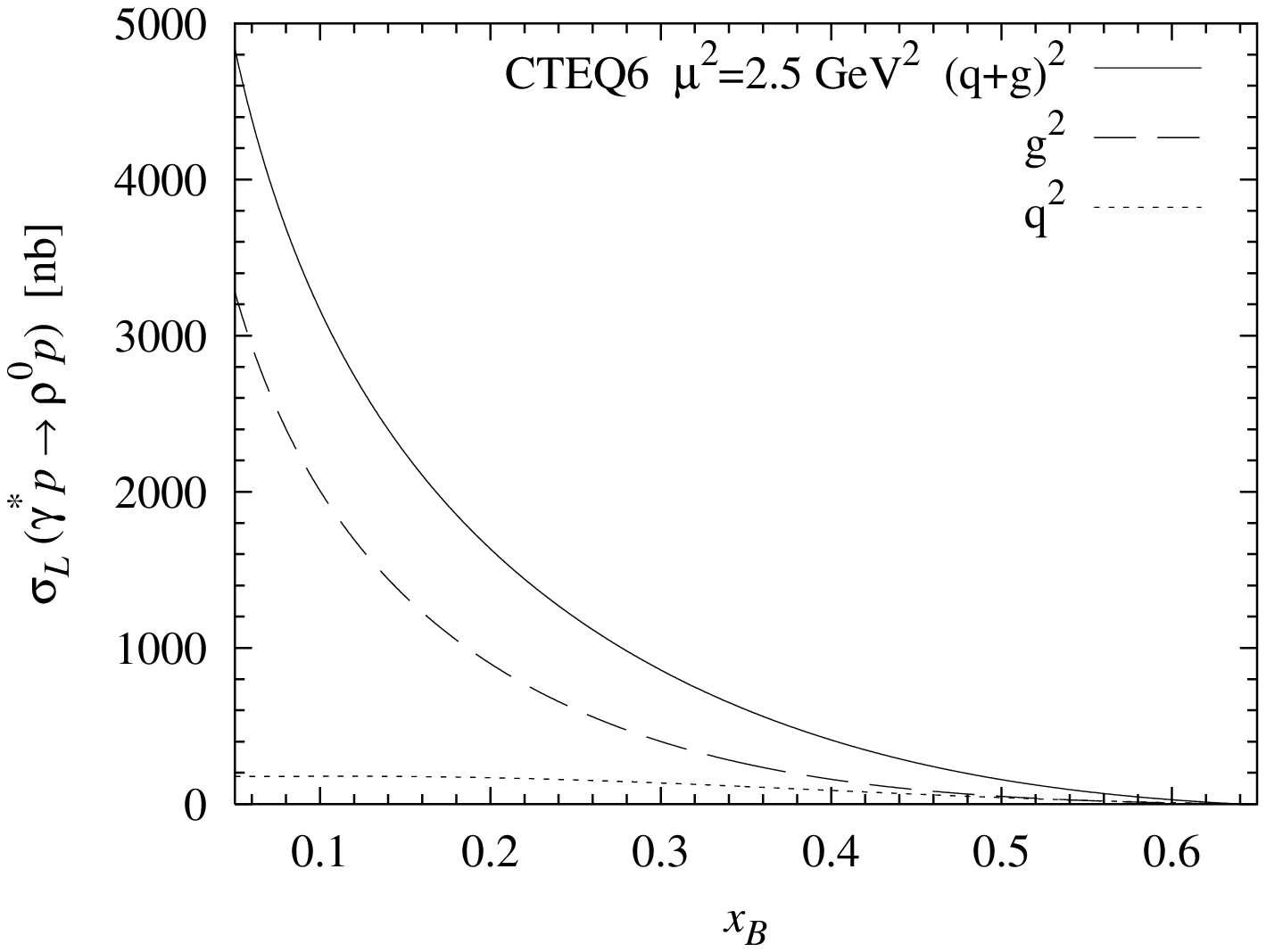}
\caption{\label{fig:rho0} Leading-twist cross section for
  $\gamma^*_L\, p\to \rho^0 p$ at $Q^2 =2.5\gev^2$.  Shown are the
  individual contributions from quark and gluon distributions and
  their coherent sum.  The parton densities in the double distribution
  model are taken at scale $\mu^2 = 1.2\gev^2$ for the upper and at
  $\mu^2 = 2.5 \gev^2$ for the lower plots.}
\end{figure}
In the double distribution model of GPDs we take the profile parameter
$b=2$ for both quark and gluon distributions.  For all mesons we take
the asymptotic shape of the distribution amplitude, given that no
direct experimental information is available for them.  Theoretical
estimates do not give stronger deviations from the asymptotic form
than those we discussed for pseudoscalar mesons, see e.g.\ the
compilation in \cite{Beneke:2003zv}.  In Fig.~\ref{fig:rho0} we show
the individual contributions from quark and gluon distributions to the
$\rho^0$ cross section as well as their coherent sum.  The clear
difference between the CTEQ6 and MRST2001 result reflects the current
uncertainty on the usual parton densities at low scales in the
relevant range of $x$.  A striking feature is the clear dominance of
the gluon distribution up to quite high values of $\xb$.  Note that
with our model of GPDs the convolutions $\mathcal{H}^q(\xi,t)$ and
$\mathcal{H}^g(\xi,t)$ are sensitive to the forward parton
distributions in a certain range of momentum fractions around $\xi$
(see App.~\ref{app:GPDs}).  The strong dominance of gluon over quark
distributions at small momentum fractions thus still shows its effect
at $\xb$ values above $10^{-1}$.  Note that we have taken the same $t$
dependence for quark and gluon GPDs in our model
(\ref{factorize-ansatz}), lacking phenomenological evidence to the
contrary.  Comparison of the $t$ dependence e.g.\ for $\rho^0$ and
$\rho^+$ production in equal kinematics could be of help here.

In our numerical calculations we have calculated the integrals
(\ref{im-part}) and (\ref{re-part}) for the meson production amplitude
with a lower cutoff at momentum fractions $x= 10^{-4}$.  The cross
section for $\xb=0.05$ changes by less than 2\% if we cut off at
$10^{-5}$ or at $10^{-3}$.  It is diminished by about 10\% with a
cutoff at $0.005$, which gives an indication of the relevance of
momentum fractions which are an order of magnitude smaller than the
$\xb$ of the process in this model.  Similar changes are observed for
the individual quark and gluon distributions.  We note that if we take
a profile parameter $b=1$ for quarks, the quark contribution to the
cross section at $\xb=0.05$ decreases by 10\% when moving the cutoff
on $x$ from $x= 10^{-4}$ to $x= 10^{-3}$ and by 35\% when moving it
from $x= 10^{-4}$ to $0.005$.  Such a strong dependence on momentum
fractions well below $\xb$ seems difficult to understand in physical
terms.  We note that in the sea quark sector there are no strong
theoretical arguments for taking $b=1$, see Sect.~4.4 of
\cite{Diehl:2003ny}.

\begin{figure}
\centering 
\includegraphics[width=8.3cm]{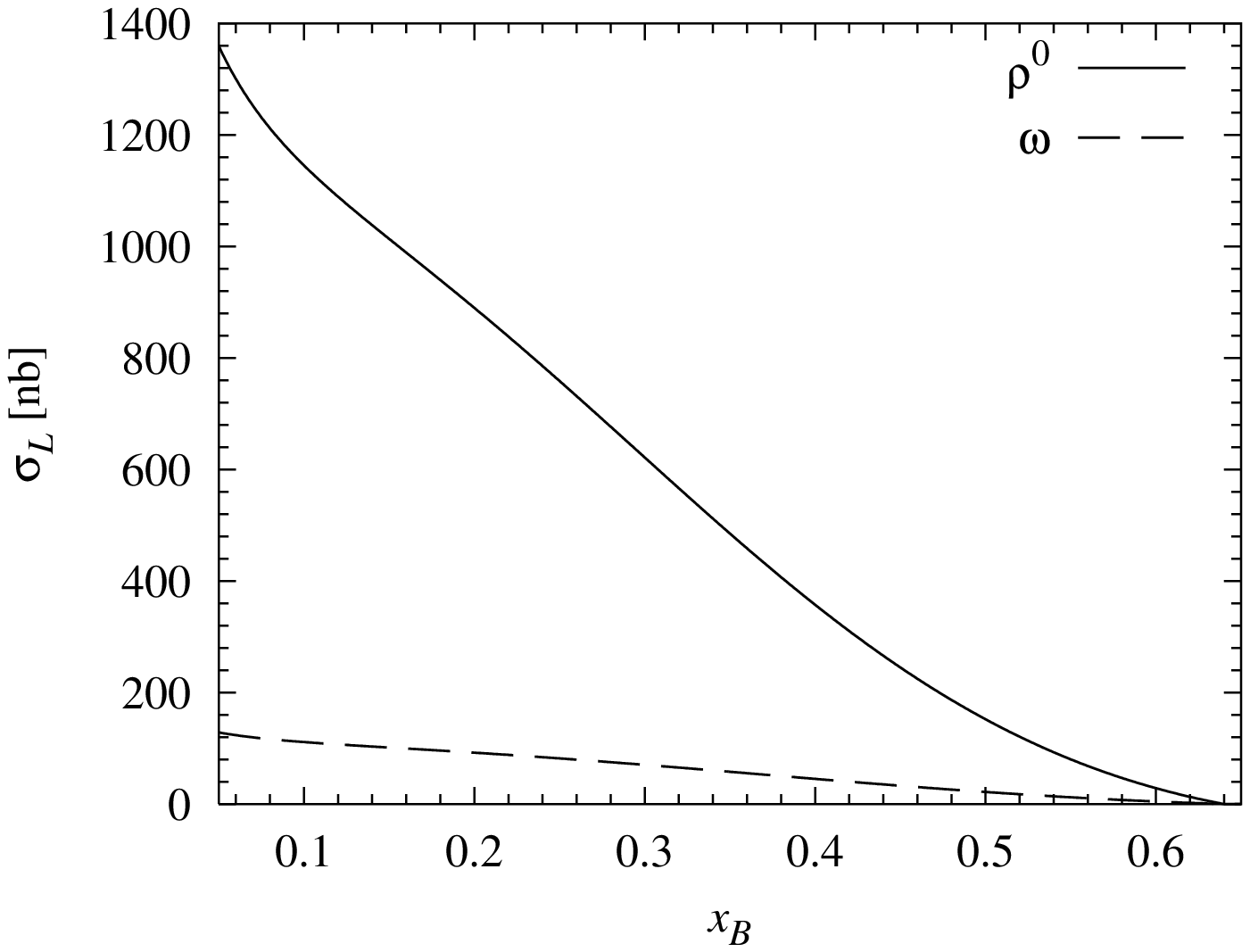}
\includegraphics[width=8.3cm]{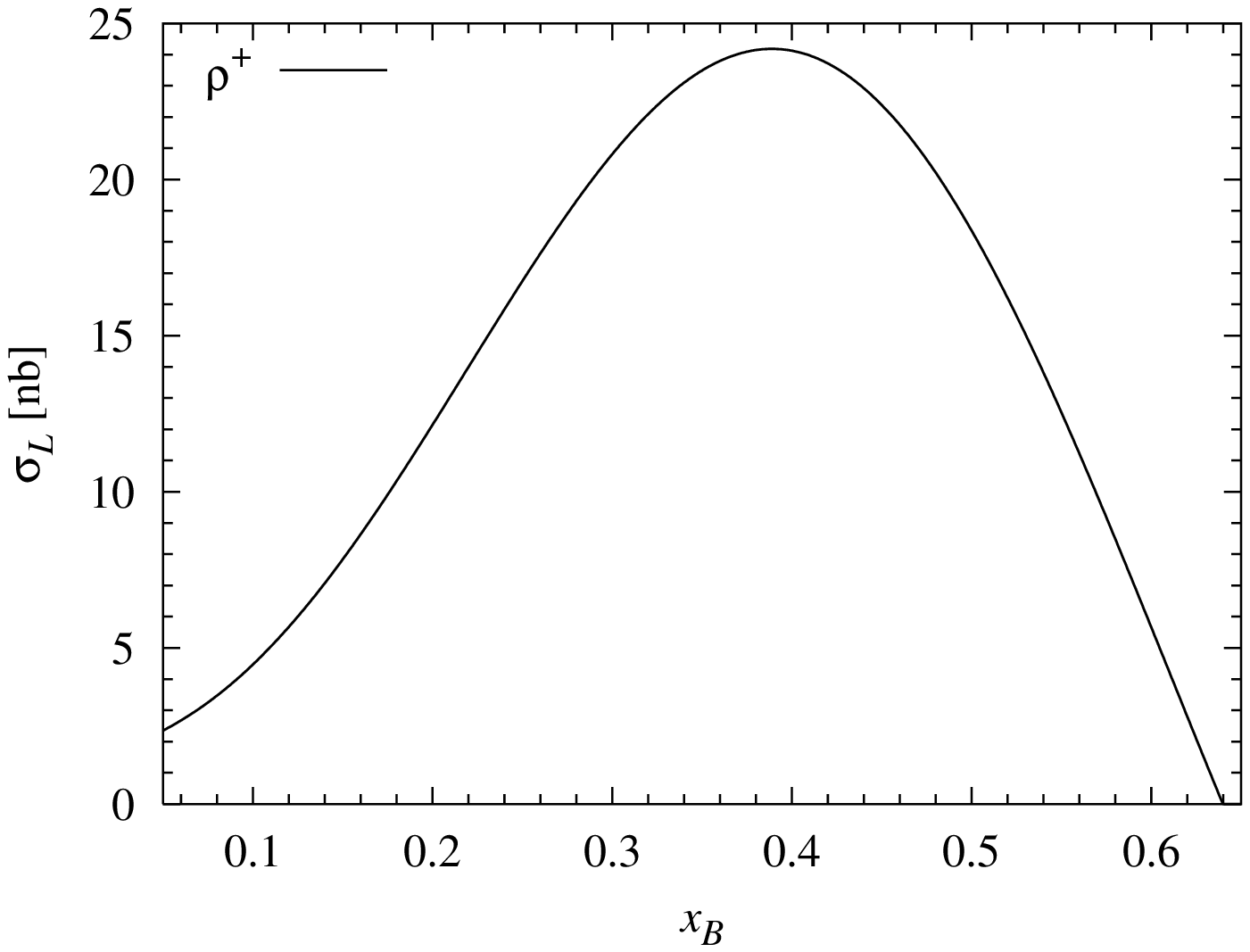}
\caption{\label{fig:rho+0} Leading-twist cross sections for
  $\gamma^*_L\, p\to \rho^0 p$ and $\gamma^*_L\, p\to \omega p$ (left)
  and for $\gamma^*_L\, p\to \rho^+ n$ (right) at $Q^2=2.5 \gev^2$,
  obtained with the MRST2001 parton densities taken at $\mu^2= 1.2
  \gev^2$.}
\end{figure}
In Fig.~\ref{fig:rho0} we also observe a significant change of the
gluon contribution to the cross section when changing the scale of the
parton distributions in the double distribution ansatz
(\ref{dd-models}).  In contrast, the quark contribution changes by at
most a factor of 1.3, reflecting the relatively weak scale evolution
of quark and antiquark distributions compared with gluons in the
relevant kinematic region.  Changing the scale of the forward
distributions in the double distribution model (\ref{dd-models}) gives
a rough indication of how the actual GPDs evolve with $\mu^2$
\cite{Musatov:1999xp}, so that the strong increase with $\mu^2$ seen
in Fig.~\ref{fig:rho0} reflects a strong scale uncertainty of the
leading-order approximation in $\alpha_s$ for channels involving gluon
exchange.  A full NLO analysis of meson production is possible using
the results of \cite{Ivanov:2004zv} but beyond the scope of this work.
We will use the smaller scale $\mu^2= 1.2 \gev^2$ in our further
studies, because the internal virtualities in the hard-scattering
graphs of Fig.~\ref{fig:meson-graphs} are typically smaller than $Q^2$
(see also the study \cite{Frankfurt:1995jw} of relevant scales in the
small-$x$ limit).  Furthermore, the MRST2001 set gives a better
description for the ratio of $\phi$ and $\rho^0$ production with our
model (see below) and will hence be our default choice in the
following.

In Fig.~\ref{fig:rho+0} we show the production cross sections for
$\rho^+$, $\rho^0$ and $\omega$.  The $\xb$ behavior of the $\rho^+$
cross section roughly follows the one of $\xi^2\sms [u(\xi) - d(\xi) +
\bar{u}(\xi) - \bar{d}(\xi)]^2$, which is a flavor nonsinglet
combination and hence does not display the strong rise of sea quarks
or gluons at small $x$.  The clear suppression of $\omega$ production
compared with the $\rho^0$ is a consequence of the relative factor in
the gluon contribution (see Table~\ref{tab:GPDs}) and at large $\xb$
of the relative size of the flavor combination $2H^u - H^d$ compared
with $2H^u + H^d$.  We remark that the exclusive channel $\gamma^* p
\to f_2\, p$ also contributes to semi-inclusive production of $\pi^+$,
$\pi^-$ and $\pi^0$.  It involves the combination $2H^u + H^d -
[2H^{\bar{u}} + H^{\bar{d}}]$, where sea quarks drop out, so that one
may expect a cross section of similar size as for $\rho^+$ production.
A numerical estimate would however require knowledge of quark and
gluon distribution amplitudes of the $f_2$, see
\cite{Lehmann-Dronke:1999aq}, and is beyond the scope of this work.

\begin{figure}
\centering
\includegraphics[width=8.3cm]
   {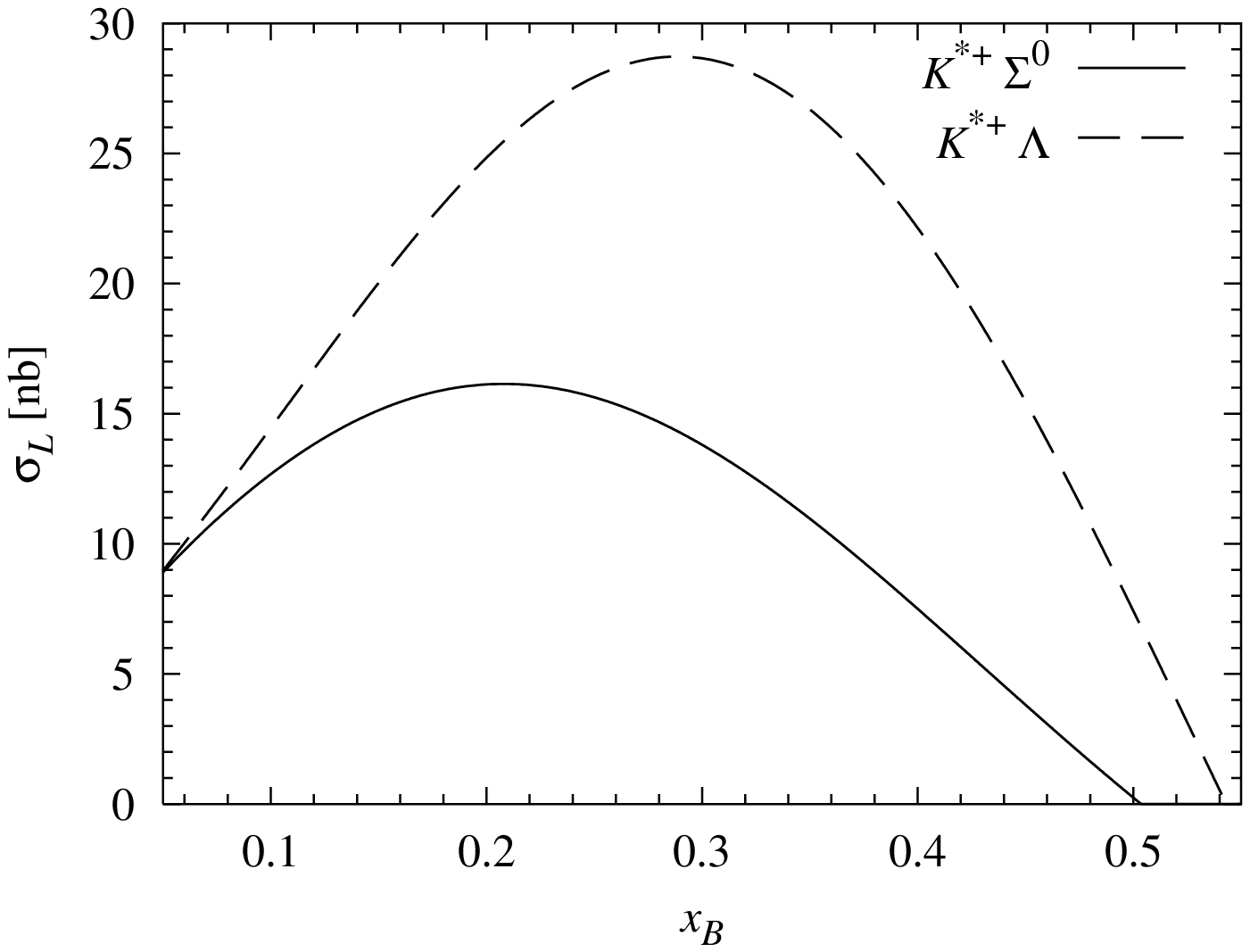}
\includegraphics[width=8.3cm]
   {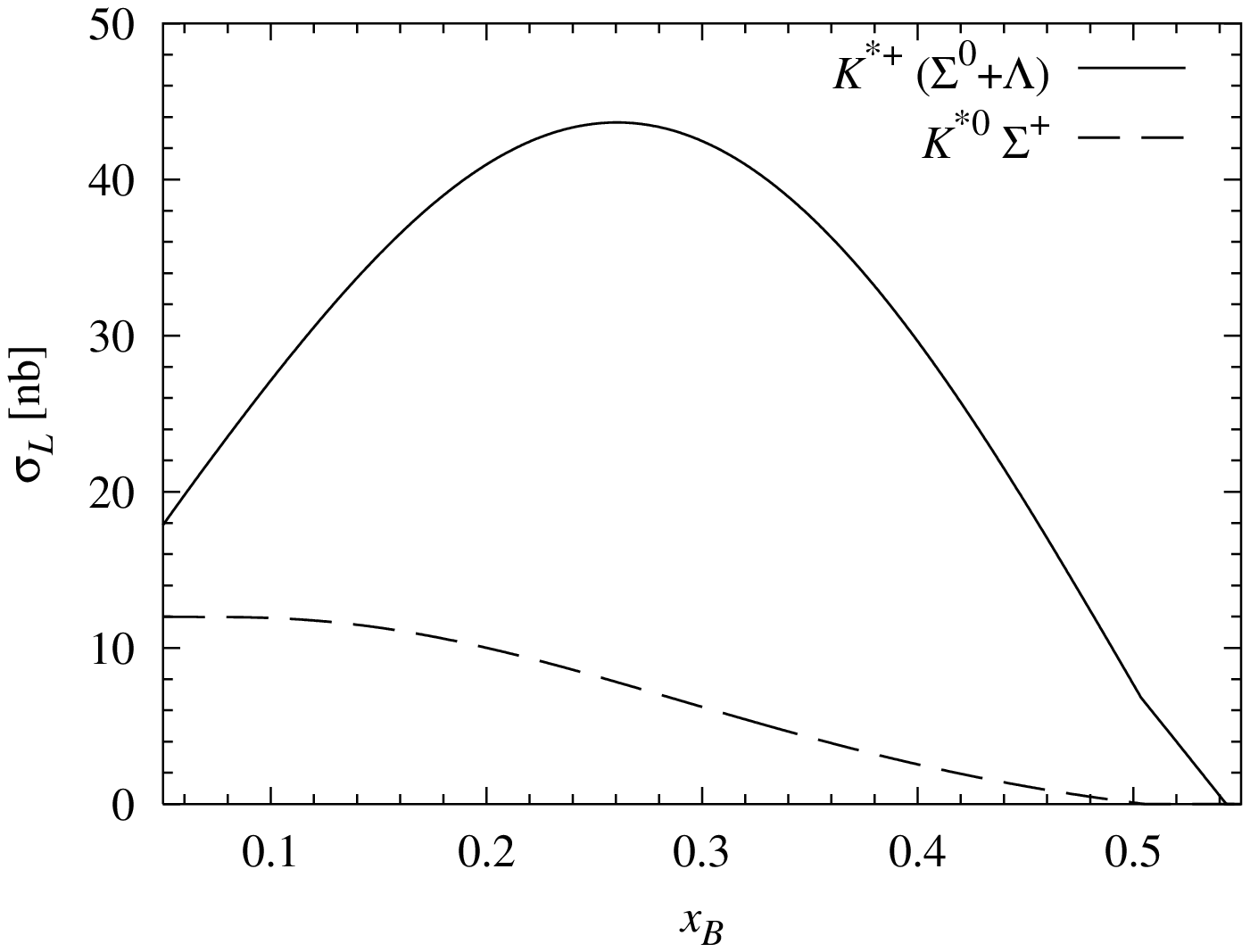}
\caption{\label{fig:Kstar} Leading-twist cross sections for
  $\gamma^*_L\, p\to K^{*+} \Lambda,\; K^{*+} \Sigma^0,\; K^{*0}
  \Sigma^+$ at $Q^2 =2.5 \gev^2$, obtained with the MRST2001 parton
  densities taken at $\mu^2= 1.2 \gev^2$.}
\end{figure}

Figure~\ref{fig:Kstar} shows our results for $K^{*+}$ and $K^{*0}$
production.  In contrast to $K^+$ production, the cross section for
the $\Lambda$ channel is not much larger than for the $\Sigma^0$
channel.  Consulting Table~\ref{tab:GPDs} we see that this is because
the contributions from $u$ and $d$ quarks partially cancel in $2 H^u -
H^d$ whereas they add in $2\tilde{H}^u - \tilde{H}^d$.  We remark that
the results for $K^{*+}$ and $K^{*0}$ production decrease by less than
25\% when instead of MRST2001 we take the CTEQ6 parameterization.  The
uncertainty due to knowledge of the parton distributions is hence much
less than for the gluon dominated channels.

\begin{figure}[t]
\centering
\includegraphics[width=8.3cm]
{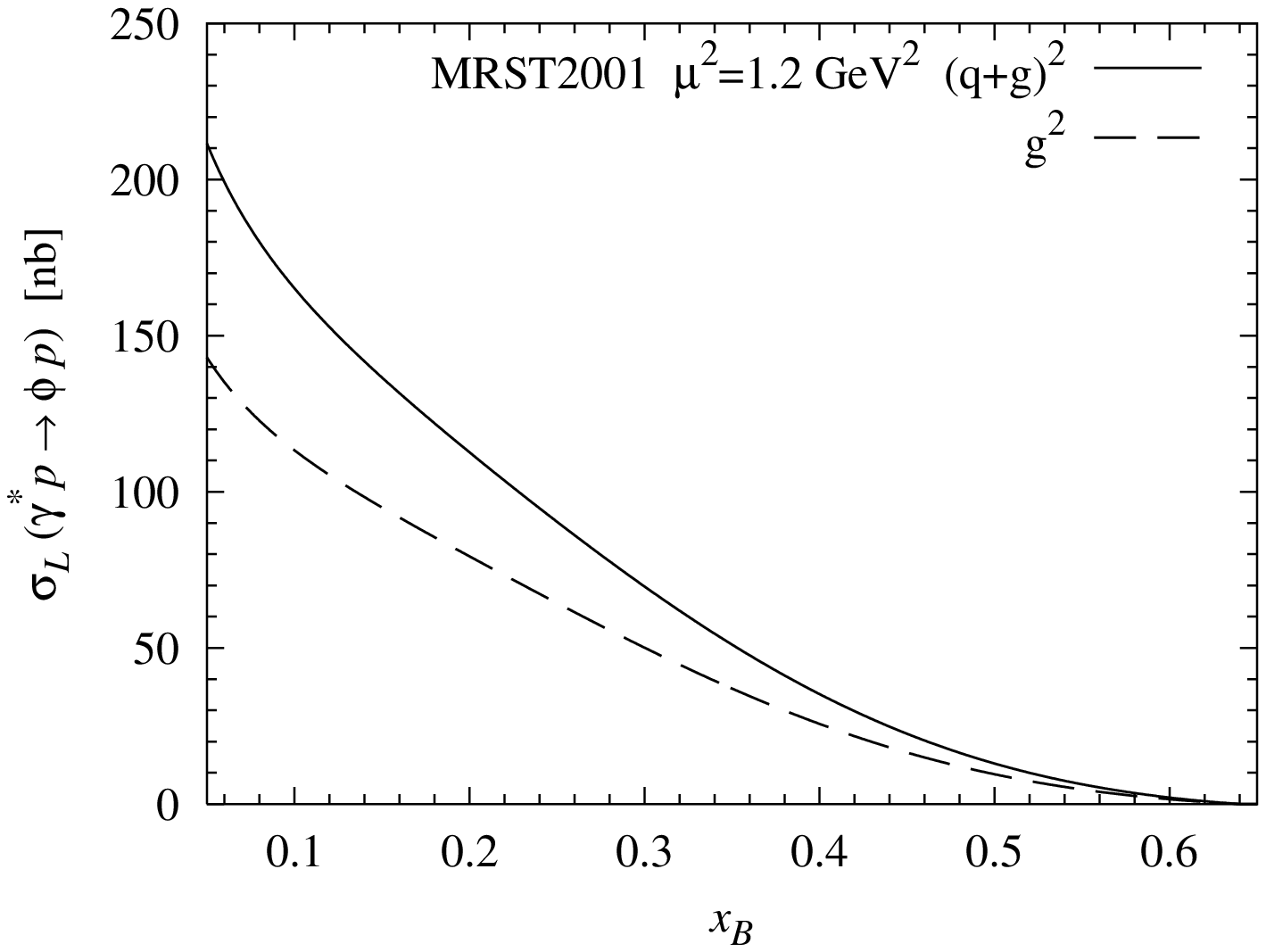}
\includegraphics[width=8.3cm]
{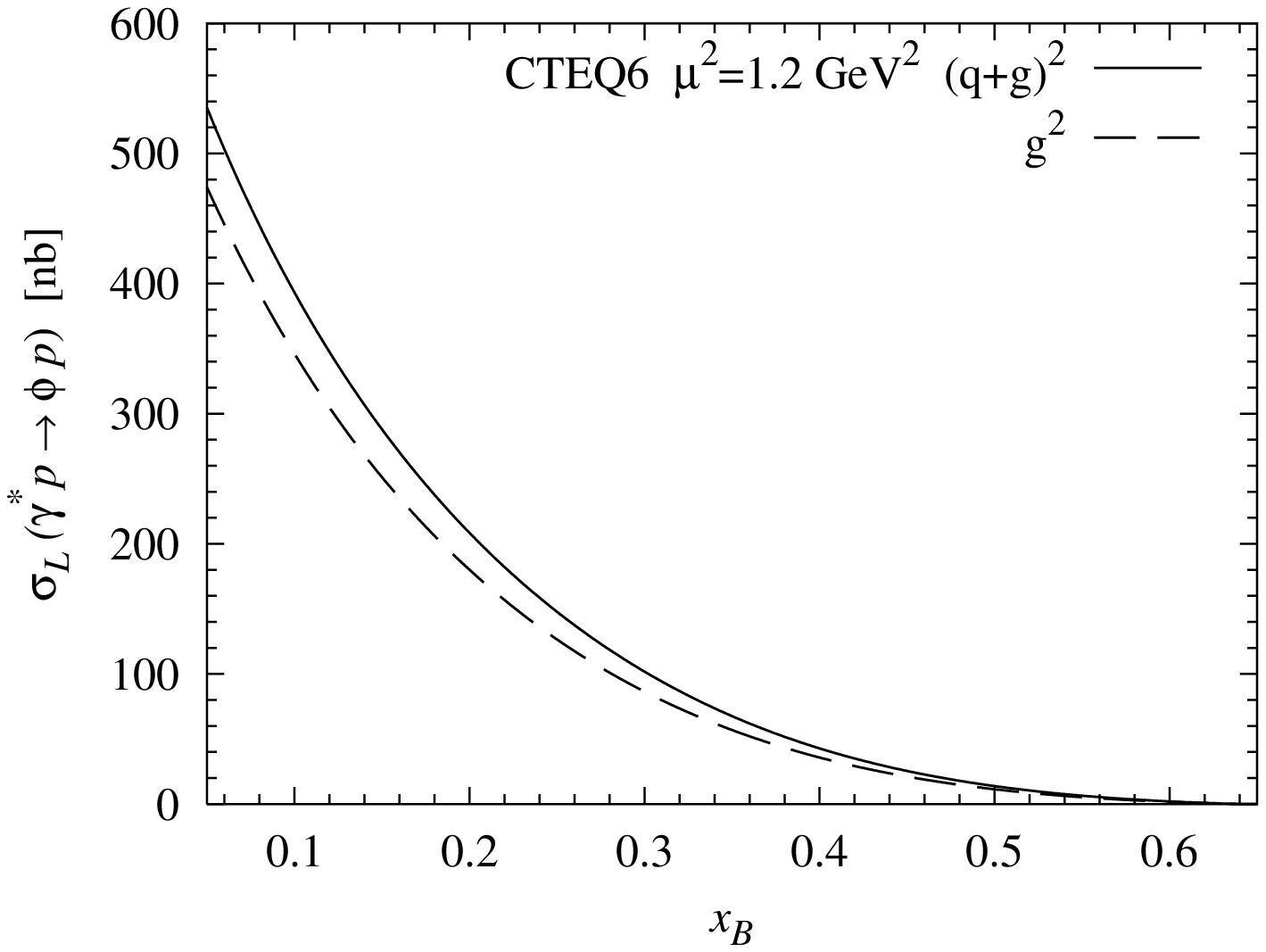}
\includegraphics[width=8.3cm]
{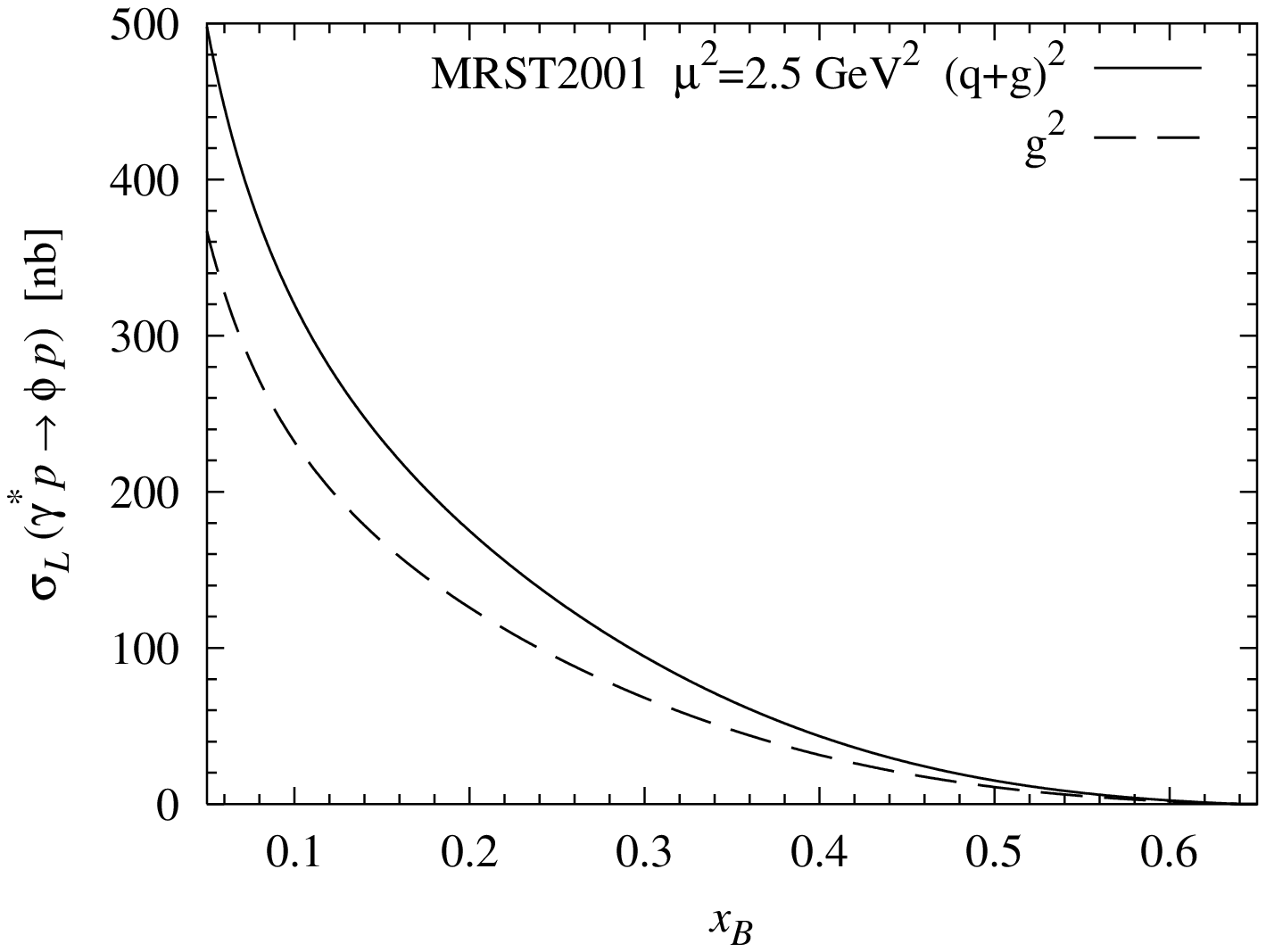}
\includegraphics[width=8.3cm]
{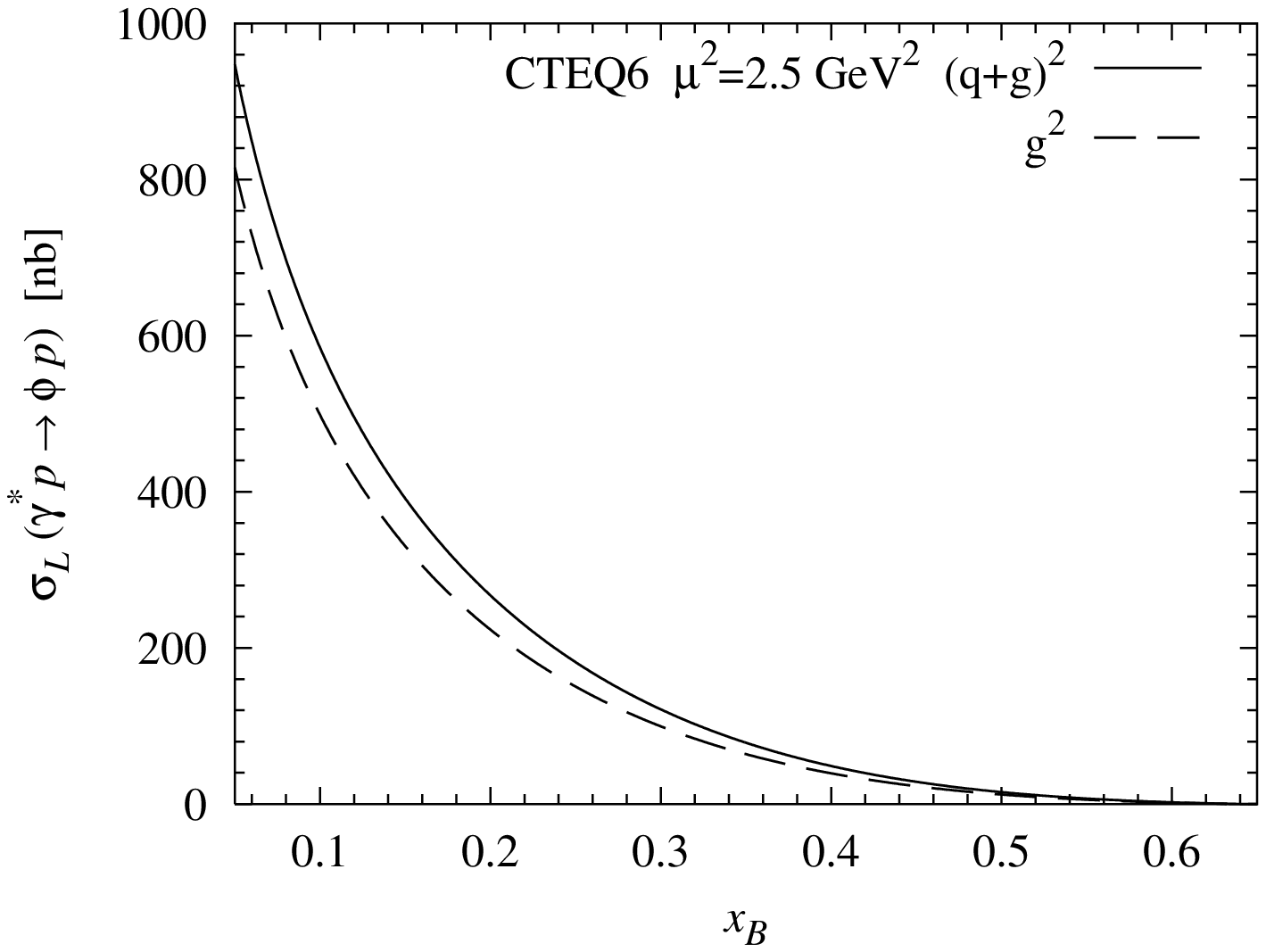}
\caption{\label{fig:phi} Leading-twist cross section for $\gamma^*_L\,
  p\to \phi p$ at $Q^2 =2.5\gev^2$.  Shown are the individual
  contributions from gluons and the coherent sum of gluons and strange
  quarks.  The upper plots are for parton densities taken at scale
  $\mu^2 = 1.2\gev^2$ in the double distribution model, and the lower
  plots for parton densities taken at $\mu^2 = 2.5 \gev^2$.}
\end{figure}

\begin{figure}
\centering
\includegraphics[width=8.3cm]
      {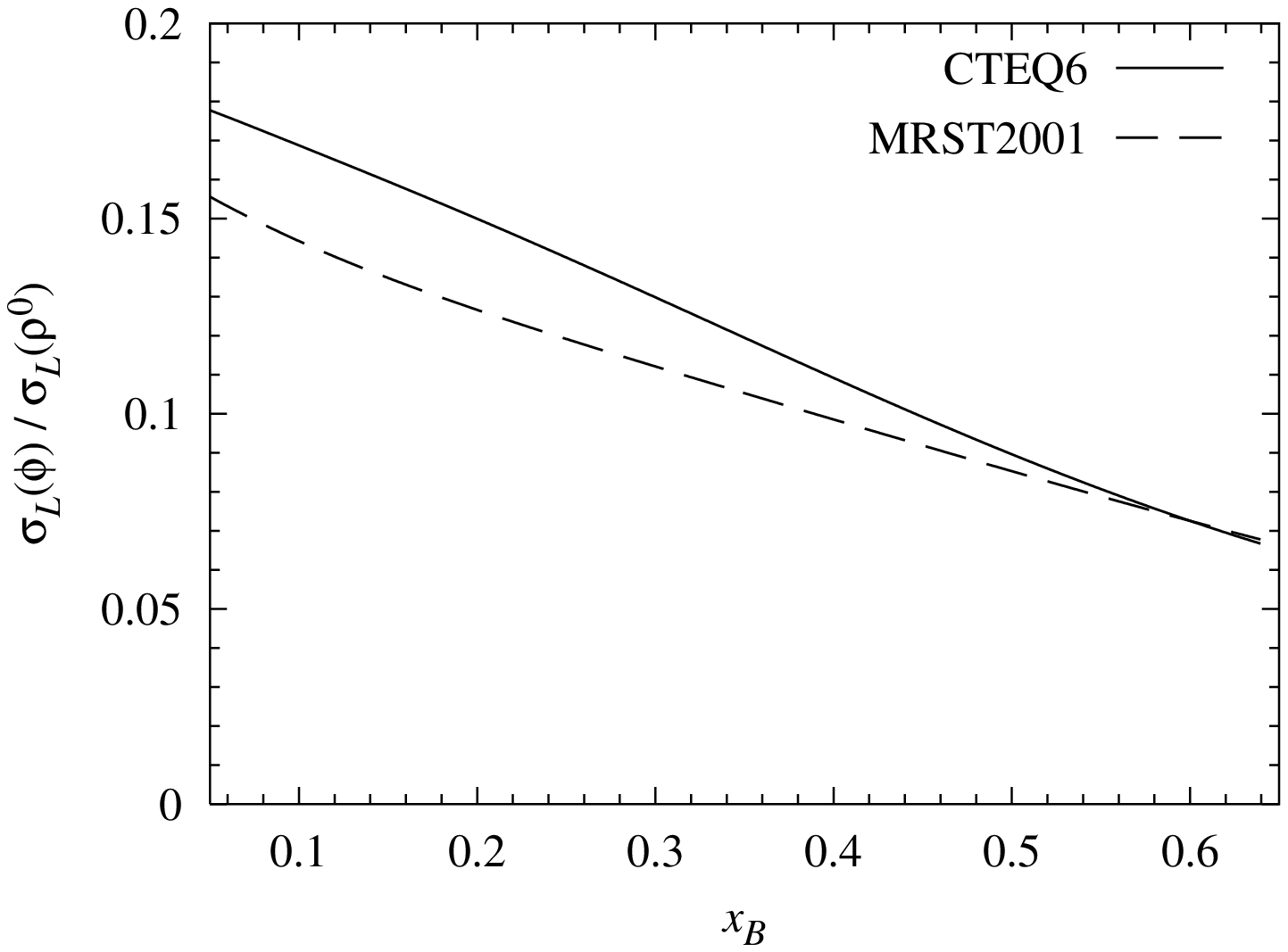}
\includegraphics[width=8.3cm]
      {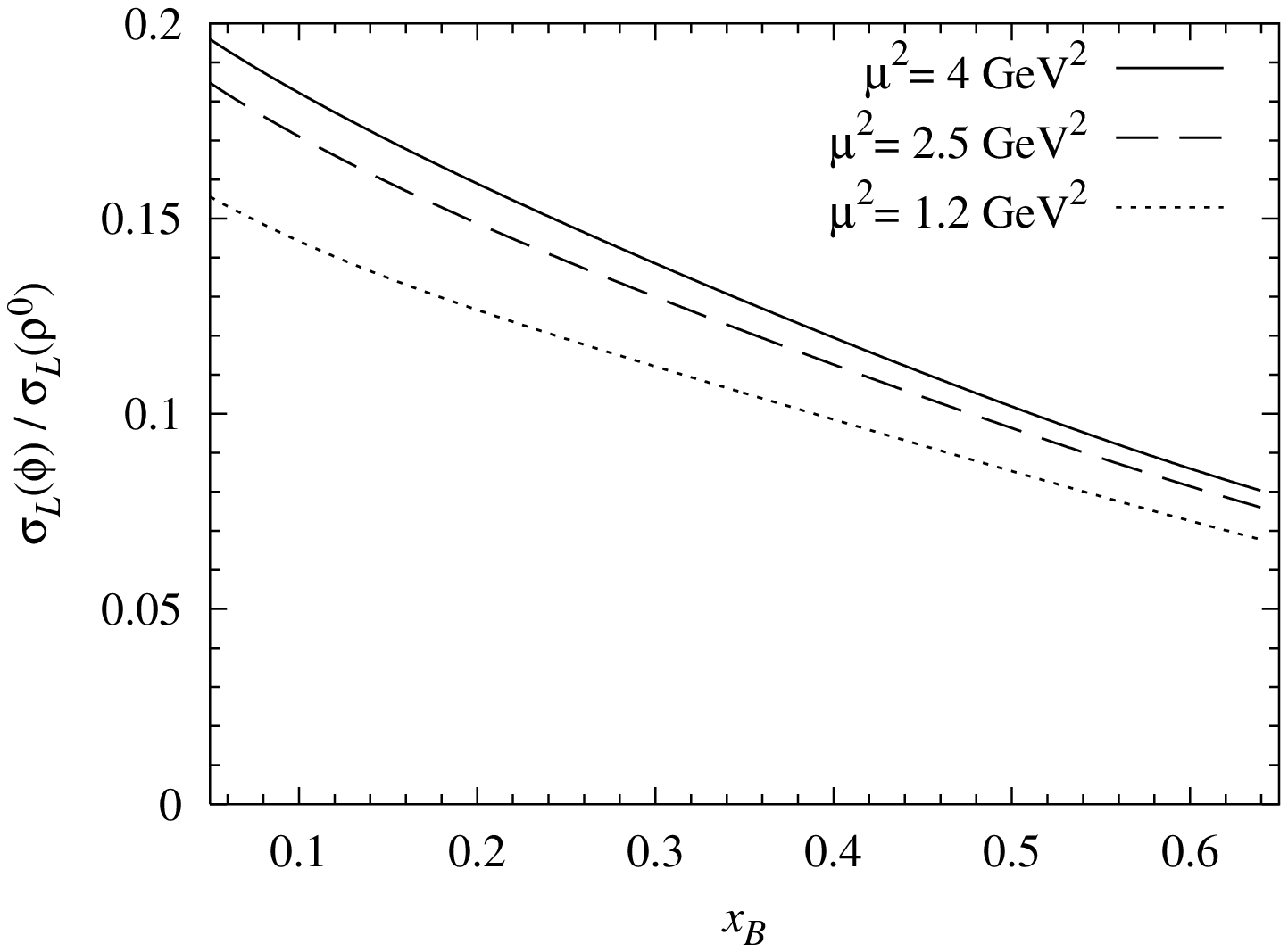}
\caption{\label{fig:phirho} Ratio of leading-twist longitudinal cross
  sections for $\phi$ and $\rho^0$ production, obtained with different
  parton distributions taken at $\mu^2 = 1.2 \gev^2$ (left) and with
  the MRST2001 distribution taken at different $\mu^2$ (right).}
\end{figure}

Results for $\phi$ production are shown in Fig.~\ref{fig:phi}.  The
dominance of the gluon over the strange quark contribution is clearly
seen, although strange quarks are not entirely negligible with the
MRST2001 parameterization.\footnote{In the study \cite{Diehl:2004wj}
strange quarks were neglected based on inspection of the CTEQ6
parameterization.}
Since gluons dominate for most $\xb$, we see the same trend concerning
differences between the parameterizations and the choice of scale as
for $\rho^0$ production.  The ratio of $\sigma_L$ for $\phi$ and
$\rho^0$ production is shown in Fig.~\ref{fig:phirho}, where the
dependence on $\mu^2$ is seen to be much milder since it partially
cancels in the ratio.  The difference between CTEQ6 and MRST2001 is
still significant and mainly due to the difference in the gluon
distributions.

Preliminary data from HERMES \cite{Borissov:2001fq,Rakness:2000th}
give a ratio of about 0.08 for the cross sections of $\phi$ and
$\rho^0$ production for $\xb=0.09$ and $Q^2=2.46 \gev^2$ and for
$\xb=0.13$ and $Q^2 =3.5 \gev^2$.  These data contain a significant
contribution from $\sigma_T$, and preliminary HERMES data
\cite{Borissov:2001fq,Tytgat:2000th,Rakness:2000th} suggest that
$R=\sigma_L /\sigma_T$ may be slightly smaller for $\phi$ than for
$\rho^0$ production at the same $Q^2$.  The $\phi$ to $\rho^0$ ratio
for $\sigma_L$ would then be somewhat larger than 0.08.  In addition,
one can expect that a narrower shape of the distribution amplitude and
power corrections due to the strange quark mass would decrease the
estimates in Fig.~\ref{fig:phirho} \cite{Diehl:2004wj}.

A complete representation of GPDs includes in addition to the double
distribution the so-called $D$-term \cite{Polyakov:1999gs}.  It
vanishes in the forward limit $\xi=0$ and does not affect the double
distribution ansatz we are using.  Its contribution to the GPDs can be
expanded in Gegenbauer polynomials as
\begin{eqnarray}
  \label{d-term-exp}
H_D^q(x,\xi,t) &=& \theta(|x| \leq \xi)\,
  \Bigg(1-\frac{x^2}{\xi^2}\Bigg) \sum_{n=0}^\infty
  d_{2n+1}^q(t)\, C_{2n+1}^{3/2}\Big(\frac{x}{\xi}\Big) , 
\nonumber \\[0.7ex]
H_D^g(x,\xi,t)
  &=& \theta(|x| \leq \xi)\, \frac{3\xi}{2}
  \Bigg(1-\frac{x^2}{\xi^2}\Bigg)^2\, \sum_{n=0}^\infty
  d^g_{2n+1}(t)\, C_{2n}^{5/3}\Big(\frac{x}{\xi}\Big)
\end{eqnarray}
for $\xi>0$.  Such terms contribute to the real part of the
convolution integrals needed in the meson production amplitudes as
\begin{eqnarray}
  \label{d-term-amp}
I^q_D &=& \int_{-1}^1 dx\, \frac{H_D^q(x,\xi,t)}{\xi-x-i\epsilon}
 \:=\: \int_{-1}^1 dx\, 
   \frac{H_D^{\bar{q}}(x,\xi,t)}{\xi-x-i\epsilon}
 \:=\: 2 \sum_{n=0}^\infty d_{2n+1}^q(t) \,,
\nonumber \\
I^g_D &=& \int_{-1}^1 dx\, \frac{H_D^g(x,\xi,t)}{x}\, 
          \frac{1}{\xi-x-i\epsilon}
 \:=\: 2 \sum_{n=0}^\infty d_{2n+1}^g(t) \,.
\end{eqnarray}
These terms give a $\xi$ independent contribution to
$\mathcal{H}(\xi,t)$, in contrast to the contributions from the double
distribution part, which very roughly follow the behavior of $\xi
q(\xi)$, $\xi \bar{q}(\xi)$ or $\xi g(\xi)$ and hence grow as $\xi$
becomes smaller.  In \cite{Kivel:2000fg} the first three coefficients
in the quark $D$-term at $t=0$ have been extracted from a calculation
in the chiral quark-soliton model of the nucleon, giving $d_1^u(0)
\approx -2.0$, $d_3^u(0) \approx -0.6$, $d_5^u(0) \approx -0.2$ and
equal values for $d$ quarks, referring to a scale $\mu$ of a few GeV
\cite{Schweitzer:2002nm}.  The gluon $D$-term is parametrically
subleading at the low scale intrinsic to the model, but evolution to
$\mu$ of a few GeV can give values similar in size to those for
quarks.  The values $I^u_D = I^d_D =-5.6$ turn out to be similar in
size and opposite in sign to the real parts of the corresponding
integrals from the double distribution part in our model.  The effect
of such a $D$-term is however much weaker on the square
$|\mathcal{H}|^2$ appearing the cross section, which is dominated by
the imaginary parts of the integrals in a large region of $\xb$.
Taking the above values for the quark $D$-term and $I^g_D =-11.2$ as
an order-of-magnitude estimate, we find that the change of the various
vector meson cross sections is at the 10\% level for $\xb=0.1$ and not
more than a factor 1.5 in either direction for $\xb$ below 0.3, where
for definiteness we have taken the MRST2001 distributions at
$\mu^2=1.2 \gev^2$.


\section{Comparison with data and discussion of power corrections}
\label{sec:data-compare}

Our calculation of meson production cross sections is based on the
leading-twist approximation.  It is known that corrections in $1/Q^2$
to leading-twist meson cross sections can be substantial for $Q^2$ of
a few $\gev^2$.  A systematic treatment of such power corrections
remains an unsolved problem.  There is however a number of approaches
that allow one to model particular sources of power corrections, see
e.g.\ \cite{Diehl:2003ny} for a discussion and references.  For vector
meson production, a considerable suppression of the leading-twist
result at moderate $Q^2$ is found when including in the
hard-scattering kernel the transverse momentum of the quarks in the
meson \cite{Frankfurt:1995jw,Vanderhaeghen:1999xj,Goloskokov:2005sd}.
This means that the transverse resolution power of the virtual photon
cannot be neglected compared with the transverse size of the meson.
Similarly, the finite transverse momentum of the partons coming from
the proton gives rise to power corrections, when it is included in the
hard-scattering kernel.  Estimating both effects in a calculation
considering only quark GPDs \cite{Vanderhaeghen:1999xj}, a suppression
of the leading-twist cross section for $\rho^0$ production by factors
of $3.3$, $4.9$ and $9.2$ was found at $Q^2= 5 \gev^2$ for $\xb=0.3$,
$0.45$ and $0.6$, respectively.  The recent study
\cite{Goloskokov:2005sd} for small $\xb$, where
only gluon GPDs were retained and only the transverse quark momentum
in the meson was taken into account, found corresponding suppression
factors of $4.6$ and $6.6$ for respective values of $\xb=2.95 \times
10^{-3}$ and $10^{-4}$ at $Q^2= 4.8 \gev^2$.  For $Q^2= 10.9 \gev^2$
and $\xb=4.3 \times 10^{-3}$ this factor decreases to $1.9$.  The
discrepancy of the calculation including power suppression and
experimental data is less than 35\% in all three cases.\footnote{The
leading-order formula (90) in \protect\cite{Goloskokov:2005sd} with
which we obtained the suppression factors just quoted contains in
addition an approximation for small $\xb$, which should however not be
the dominant effect comparing to the full calculation.}

In Table~\ref{tab:vm-data} we compare our leading-oder results for
$\rho^0$ and $\phi$ production with data from HERMES
\cite{Airapetian:2000ni,Borissov:2001fq,Rakness:2000th,Borissov:2005pc}.
The discrepancy between our calculation and the $\rho^0$ data is well
in the range of what can be explained by suppression from quark
transverse momentum (considering in addition the uncertainties of our
model for the GPDs).  The stronger discrepancy with the $\phi$ data
corresponds to our overestimating the $\phi$ to $\rho^0$ production
ratio, discussed in the previous subsection.

\begin{table}
\caption{\label{tab:vm-data} Experimental values
  $\sigma_L^{\mathrm{exp}}$ of the longitudinal cross section for the
  production of $\rho^0$ \protect\cite{Airapetian:2000ni} and $\phi$
  \protect\cite{Borissov:2001fq,Rakness:2000th,Borissov:2005pc} from
  HERMES, and the ratio between our leading-twist results
  $\sigma_L^{\mathrm{thy}}$ (obtained with the MRST2001 distributions
  taken at $\mu^2=1.2 \gev^2$) and the data.  The data for $\phi$
  production is preliminary.}
\renewcommand{\arraystretch}{1.3}
$$
\begin{array}[t]{cccc} \hline
 \multicolumn{4}{c}{\gamma^* p\to \rho^0 p} \\
 Q^2 [\gev^2] & \xb & \sigma_L^{\mathrm{exp}} [\mu \mathrm{b}] & 
       \sigma_L^{\mathrm{thy}} /\sigma_L^{\mathrm{exp}} \\ \hline
2.3 & 0.1~~ & 0.21 \pm 0.04 & 7.1 \\
2.3 & 0.075 & 0.21 \pm 0.04 & 7.6 \\
4.0 & 0.16~ & 0.09 \pm 0.02 & 2.1 \\
4.0 & 0.12~ & 0.06 \pm 0.01 & 3.5 \\ \hline
\end{array}
\hspace{3em}
\begin{array}[t]{cccc} \hline
 \multicolumn{4}{c}{\gamma^* p\to \phi\sms p 
   \mbox{~~(preliminary data)}} \\
 Q^2 [\gev^2] & \xb & \sigma_L^{\mathrm{exp}} [\mathrm{nb}] & 
       \sigma_L^{\mathrm{thy}} /\sigma_L^{\mathrm{exp}} \\ \hline
2.3 & 0.087 & 15.6 \pm 3.1  & 14.9 \\
3.8 & 0.136 &  6.2 \pm 1.24 & 5.5 \\ \hline
\end{array}
$$
\end{table}

The CLAS collaboration has published results for $\rho^0$ production
at $\xb= 0.31$ and at $\xb= 0.38$, with $Q^2$ values between $1.5$ and
$2.3 \gev^2$ \cite{Hadjidakis:2004zm}, and for $\phi$ production at
$\xb= 0.29$ and $Q^2=1.7 \gev^2$ \cite{Lukashin:2001sh}.  We consider
that this kinematics, where the hadronic invariant mass $W$ is below
$2.3 \gev$, is too close to threshold for comparison with a
leading-twist calculation.  A recent CLAS measurement
\cite{Morand:2005ex} of $\omega$ production at higher energies, with
$Q^2$ up to $5.1 \gev^2$ and $W$ up to $2.8 \gev$, found that helicity
conservation between the $\gamma^*$ and the $\omega$ is strongly
violated, in contrast with the predicted behavior in the large-$Q^2$
limit.  This prevented the extraction of $\sigma_L$ and was ascribed
to a strong contribution from $\pi^0$ exchange (which is absent in the
$\rho^0$ and $\phi$ channels).

Let us now turn to $\pi^+$ production, where the situation is quite
different.  For the contribution from $\tilde\mathcal{H}$ one expects
a similar suppression from quark transverse momentum as in the case of
$\mathcal{H}$ in vector meson production, which was indeed found in
the numerical study \cite{Vanderhaeghen:1999xj}.  The pion exchange
contribution from $\tilde\mathcal{E}$ is described in terms of the
pion form factor according to (\ref{pion-xsec}), and this relation
persists beyond the leading approximation in $1/Q^2$ to the extent
that the pion emitted from the nucleon is not too far from off-shell.
The power corrections for $\tilde\mathcal{E}$ are then the same as
those for the pion form factor.  The leading-twist expression
(\ref{pion-ff}) for $F_\pi(Q^2)$, with our choice of $\alpha_s$
specified after (\ref{decay-constants}), undershoots the data of
\cite{Volmer:2000ek} by a factor 0.53 at $Q^2= 1 \gev^2$ and a factor
$0.41$ at $Q^2= 1.6 \gev^2$.  For $Q^2= 3.3 \gev^2$ we find this
factor to be between 0.34 and 0.77 within the large error bars in
\cite{Bebek:1977pe}.  We will not attempt here to summarize the
detailed theoretical and phenomenological work on the pion form
factor, but remark that in addition to the leading-twist perturbative
contribution there is a contribution from the Feynman mechanism, where
the photon hits a quark carrying almost all of the pion momentum.
This leads to a considerable enhancement over the leading-twist
approximation.  The calculations of $F_\pi(Q^2)$ in
\cite{Jakob:1994hd}, which take this effect into account using
different methods, give for instance results larger than our
leading-twist value by factors between 2 and 4, even at $Q^2=
10\gev^2$.  Note that these factors are to be squared in the
contribution of $|\tilde\mathcal{E}|^2$ to the $\pi^+$ production
cross section.  For the production of the neutral pseudoscalars
$\pi^0$, $\eta$, $\eta'$, where there is no pion exchange
contribution, one expects that power corrections will decrease the
cross section, similarly to the case of vector meson production.

We have further compared our leading-twist calculation of
$\epsilon\sigma_L$ for $\gamma^* p\to \pi^+ n$ with preliminary data
on $\sigma_T + \epsilon\sigma_L$ from HERMES \cite{Hadjidakis:2004jc}.
The HERMES data are presented for three different bins in $\xb$, with
the average values of $Q^2$ and $\xb$ for individual data points
ranging from $1.5 \gev^2$ and $0.1$ to $4.2 \gev^2$ and $0.17$ in the
first bin, from $2.5 \gev^2$ and $0.21$ to $6.3 \gev^2$ and $0.25$ in
the second bin, and from $4.5 \gev^2$ and $0.34$ to $10.5 \gev^2$ and
$0.45$ in the third bin \cite{Cynthia}.  Averaging the ratio between
theoretical and experimental cross sections for the data points in
each bin, we find ratios of $0.42$, $0.19$ and $0.12$ in the first,
second and third bin, respectively.  The large discrepancy at large
$Q^2$ and large $\xb$ (which are strongly correlated in the data) is
striking, but not too surprising given the size of corrections just
estimated for the pion form factor.  The much better agreement at
smaller $Q^2$ and $\xb$ might be accidental, given that we expect
comparable contributions from $\tilde\mathcal{H}$ and
$\tilde\mathcal{E}$, for with the power corrections go in different
directions.  Help in clarifying this issue could come from the spin
asymmetry for transverse target polarization, which gives access to
the relative size of $\tilde\mathcal{H}$ and $\tilde\mathcal{E}$
\cite{Goeke:2001tz}.

The case of $\pi^+$ production (and also the findings in the $\omega$
channel mentioned above) show that there are specific power
corrections which will not cancel in cross section ratios for
different processes.  The situation is however better for channels
that are sufficiently similar, as the example of $\rho^0$ and $\phi$
production shows.  Corrections due to quark transverse momentum (as
well as the overall normalization uncertainty from the scale of
$\alpha_s$ in our leading-order calculation) will tend to cancel in
that case.  We hence expect that the overall pattern of differences
between various meson cross sections we estimated at leading order
will not be overturned in a more realistic treatment, given that these
differences are largely controlled by the relevant combinations of
quark and gluon distributions.


\section{Exclusive channels in semi-inclusive pion and kaon
production}
\label{sec:exclu-inclu}

In semi-inclusive hadron production one considers processes of the
type
\begin{equation}
\label{sidis-reac}
e(k) \; + \; p(p) \;\; \to \;\; e(k') \; + \; h(q_h) \; + \; X ,
\end{equation}
where $h$ is a specified hadron and $X$ an unspecified inclusive final
state. A basic observable is the distribution of the produced hadron
over the variable
\begin{equation}
  \label{z-def}
z = \frac{q_h\sms p}{q\sms p} ,
\end{equation}
which measures the fraction of the virtual photon energy carried by
the produced hadron in the target rest frame.  In the Bjorken limit of
large $Q^2$ at fixed $\xb$ and fixed $z$, semi-inclusive hadron
production can be treated within a QCD factorization approach.  The
differential cross section factorizes into the distribution of partons
of type $i$ in the target, the cross section for the virtual photon
scattering off this parton, and the fragmentation function $D^{i\to
h}(z)$ describing the fragmentation of the struck parton into the
hadron $h$, which carries a fraction $z$ of its longitudinal momentum.
To leading order in $\alpha_s$ one has
\begin{equation}
  \label{sidis-xsec}
\frac{d\sigma}{dQ^2\, d\xb\, dz}
 = 2\pi\alpha_{em}\, \frac{y^2}{1-\epsilon}\, 
   \frac{1}{\xb\sms Q^4}\,
   \sum_{q} e_q^2\, 
   \Big[ q(\xb) D^{q\to h}(z) 
       + \bar{q}(\xb) D^{\bar{q}\to h}(z) \Big] ,
\end{equation}
where the sum is over quark flavors.  Note that (\ref{sidis-xsec}) has
the same $Q^2$-dependence as the inclusive DIS cross section in the
Bjorken regime.  The fragmentation functions are process-independent
and describe not only semi-inclusive DIS but also $e^+ e^-$
annihilation into hadrons and the distribution of leading hadrons in
high-$p_T$ jets.  Their scale evolution is governed by evolution
equations analogous to the DGLAP equations for the parton distribution
functions.  Various parameterizations of the fragmentation functions
have been presented in the literature, which fit $e^+ e^-$
annihilation and semi-inclusive DIS data at higher scales.

\begin{figure}
\centering
\includegraphics[width=15cm]{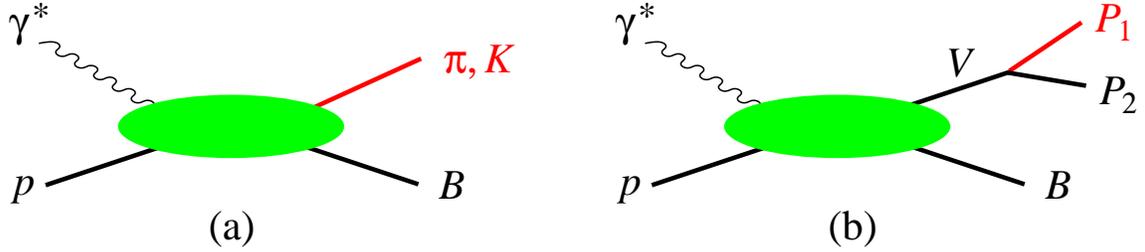}
\caption{\label{fig:exclusive_channels} Contributions of exclusive
  channels to semi-inclusive pion and kaon production, calculated at
  leading order.  (a)~Direct exclusive production of pseudoscalar
  mesons.  (b)~Exclusive production of a vector meson with subsequent
  decay into pseudoscalars.}
\end{figure}

The derivation of semi-inclusive factorization relies on the fact that
in the Bjorken limit the inclusive final state $X$ has a large
invariant mass
\begin{equation}
  \label{mx}
m_X^2 = Q^2\, \frac{1-z}{\xb} + m_p^2 - (q-q_h)^2 ,
\end{equation}
and thus a large average multiplicity (note that the squared momentum
transfer $(q-q_h)^2$ is always negative).  The semi-inclusive cross
section is thus obtained by summing over many individual channels.  In
practice $m_X^2$ is not very large for $Q^2$ values of a few $\gev^2$
typical of fixed-target experiments, for instance at Jefferson Lab or
HERMES, especially at high $z$.  At the same time, for moderate $Q^2$
the suppression of the cross sections for exclusive channels relative
to the semi-inclusive cross section (see below) may not yet be
effective.  One thus can reach a situation in which the cross sections
of individual exclusive channels becomes comparable to the
semi-inclusive one.  It is interesting to compare the semi-inclusive
cross section (\ref{sidis-xsec}) with the cross sections of exclusive
channels contributing to semi-inclusive production.  In the following
we investigate the role of exclusive channels in semi-inclusive $\pi$
and $K$ production on a proton target.  We study two types of
exclusive channels (see Fig.~\ref{fig:exclusive_channels}) at a
quantitative level:
\begin{enumerate}
\item[\textit{i)}] direct exclusive production of pseudoscalar mesons,
  $\gamma^\ast p \rightarrow \pi^+ n$ and $\gamma^\ast p \rightarrow
  K^+ \Lambda$,
\item[\textit{ii)}] exclusive production of neutral or charged vector
  mesons $\rho, \phi, K^{*}$ with subsequent decay into pseudoscalars.
\end{enumerate}

%
%

In the direct production of pseudoscalar mesons
(Fig.~\ref{fig:exclusive_channels}a), the energy fraction  $z$
carried by the produced meson is related to the invariant momentum
transfer to the nucleon $t$ by
\begin{equation}
1-z = \xb\, \frac{m_B^2-m_p^2-t}{Q^2}
\label{z_from_t}
\end{equation}
according to (\ref{mx}).  Exclusive production in the limit of large
$Q^2$ at fixed $\xb$ and fixed $t$ thus corresponds to values of $z$
very close to $1$.  For example, $\pi^+$ and $K^+$ production
corresponds to $z > 0.94$ in typical HERMES kinematics of $\xb = 0.1$
and $Q^2 = 2.5 \gev^2$ with a maximum momentum transfer $|t| = 1\,
\gev^2$ (since the cross section drops rapidly with $t$, most events
have $z$ values yet closer to unity).  Such exclusive contributions
can usually be separated experimentally from the semi-inclusive events
at smaller values of $z$.

The situation is different for the contribution to semi-inclusive
production resulting from the decay of exclusively produced vector
mesons (Fig.~\ref{fig:exclusive_channels}b).  Since the decay products
share the energy of the vector meson, such contributions result in an
extended $z$ distribution for the pion or kaon, even in the Bjorken
limit.  With the approximations described in Appendix~\ref{app:decay},
the $z$ spectrum of the pseudoscalar meson $P_1$ from the decay $V\to
P_1 P_2$ can be written as
\begin{equation}
  \label{dsig-dz-sim}
\frac{d\sigma(ep\to P_1 + P_2\sms B)}{dQ^2\, d\xb\, dz} =
\frac{\alpha_{em}}{2\pi}\,
\frac{y^2}{1-\epsilon}\,
\frac{1-\xb}{\xb\sms Q^2}\,
\Big[\,
   \epsilon\sigma_L(\gamma^* p\to V B)\, D_L(z)
 + \sigma_T(\gamma^* p\to V B)\, D_T(z)
\,\Big]
\end{equation}
with
\begin{equation}
D_L(z) = \frac{3}{2 \zeta^3}\, (z - z_0)^2 , \qquad\qquad
D_T(z) = \frac{3}{4 \zeta^3}\, (z - z_1) (z_2 - z) .
\end{equation}
Here $z_1 \le z \le z_2$ with
\begin{equation}
z_0 = \frac{E_{P1}}{m_{V}} \,,
\qquad\qquad
z_1 = z_0 - \zeta \,,
\qquad\qquad
z_2 = z_0 + \zeta \,,
\qquad\qquad
\zeta = \frac{|\sms\mbox{\boldmath{$q$}}_{P1}|}{m_{V}}
\end{equation}
up to corrections of order $\xb\sms m_p^2 /Q^2$.  For brevity we have
not explicitly indicated the dependence of $D_L$ and $D_T$ on $\xb$
and $Q^2$ due to these corrections.  The energy and three-momentum of
$P_1$ in the rest frame of the vector meson
\begin{eqnarray}
  \label{meson-masses}
E_{P1} &=& \frac{m_{V}^2 +m_{P1}^2 -m_{P2}^2}{2m_{V}} ,
\nonumber \\
|\sms\mbox{\boldmath{$q$}}_{P1}| &=& 
  \frac{\sqrt{ m_{V}^4  +m_{P1}^4 +m_{P2}^4
  -2\, (m_{V}^2 m_{P1}^2 
  +m_{V}^2 m_{P2}^2 +m_{P1}^2 m_{P2}^2) }}{2m_{V}}
\end{eqnarray}
depend only on the meson masses, and so do $z_0$, $z_1$ and $z_2$ in
the limit of large $Q^2$.  In particular, the smallest and largest
possible values of $z$ for pions from the decay $\rho\to \pi\pi$ are
$z_1 = 0.04$ and $z_2 = 0.96$ in that limit.  The corresponding values
for kaons from $\phi\to K K$ are $z_1 = 0.37$ and $z_2 = 0.63$.  For the
kaon from $K^*$ decay one has $z_1 = 0.32$ and $z_2 = 0.96$, and for
the pion from $K^*$ decay one has $z_1 = 0.04$ and $z_2 = 0.68$.

According to (\ref{H-combinations}), (\ref{H-tilde-combinations}) and
(\ref{dsig-dz-sim}), the contribution of exclusive vector meson
production to the cross section $d\sigma/(dQ^2\, d\xb\, dz)$
asymptotically scales as $1/Q^8$ at fixed $x_B$ and $z$ and is thus
suppressed by $1/Q^4$ compared with the leading behavior
(\ref{sidis-xsec}) of the semi-inclusive cross section.  Notice that
(\ref{sidis-xsec}) corresponds to transverse photon polarization, with
contributions to the longitudinal cross arising at the level of
$\alpha_s$ and of $1/Q^2$ corrections, just as in the familiar case of
inclusive DIS.  The situation is opposite for hard exclusive meson
production, where $\sigma_L$ dominates over $\sigma_T$ in the
large-$Q^2$ limit.  Measurements show however that at $Q^2$ in the few
$\gev^2$ region the ratio $R =\sigma_L/\sigma_T$ is still of order 1
in $\rho^0$ and $\phi$ production
\cite{Borissov:2001fq,Tytgat:2000th,Rakness:2000th}.

We first consider the semi-inclusive production of pions.  Depending
on the pion charge, exclusive channels contributing here are direct
production $ep \to e \pi^+ n$ and the production and decay of $\rho$
and $K^*(892)$.  The $\rho$ decays to almost 100\% as $\rho^0 \to
\pi^+ \pi^-$ and $\rho^+ \to \pi^+ \pi^0$, and the $K^*(892)$ decays
to almost 100\% into $K\pi$, with branching fractions
\begin{eqnarray}
  \label{kstar-decay} 
B(K^{*+}\to K^+ \pi^0) &=& \textstyle\frac{1}{3} , 
\qquad\qquad
B(K^{*+}\to K^0 \pi^+) \;=\; \textstyle\frac{2}{3} , 
\nonumber \\
B(K^{*0}\to K^+ \pi^-) &=& \textstyle\frac{2}{3} , 
\qquad\qquad
B(K^{*0}\,\to K^0 \pi^0)\, \;=\; \textstyle\frac{1}{3}
\end{eqnarray}
following from isospin symmetry.  Note that in quark fragmentation one
has $\sigma(\pi^0) = \half [\sigma(\pi^+) + \sigma(\pi^-)]$, which
follows directly from the isospin relations between the pion
fragmentation functions.  This relation also holds for the
contributions from $K^*$ decay, but it is strongly violated for $\rho$
decay.  For the $\rho^0$ this effect was investigated in
Ref.~\cite{Szczurek:2000mj} in connection with the separation of
$\bar{u}$ and $\bar{d}$ distributions in the proton using
semi-inclusive DIS.

\begin{figure}[t]
  \centering \includegraphics[width=8.3cm]{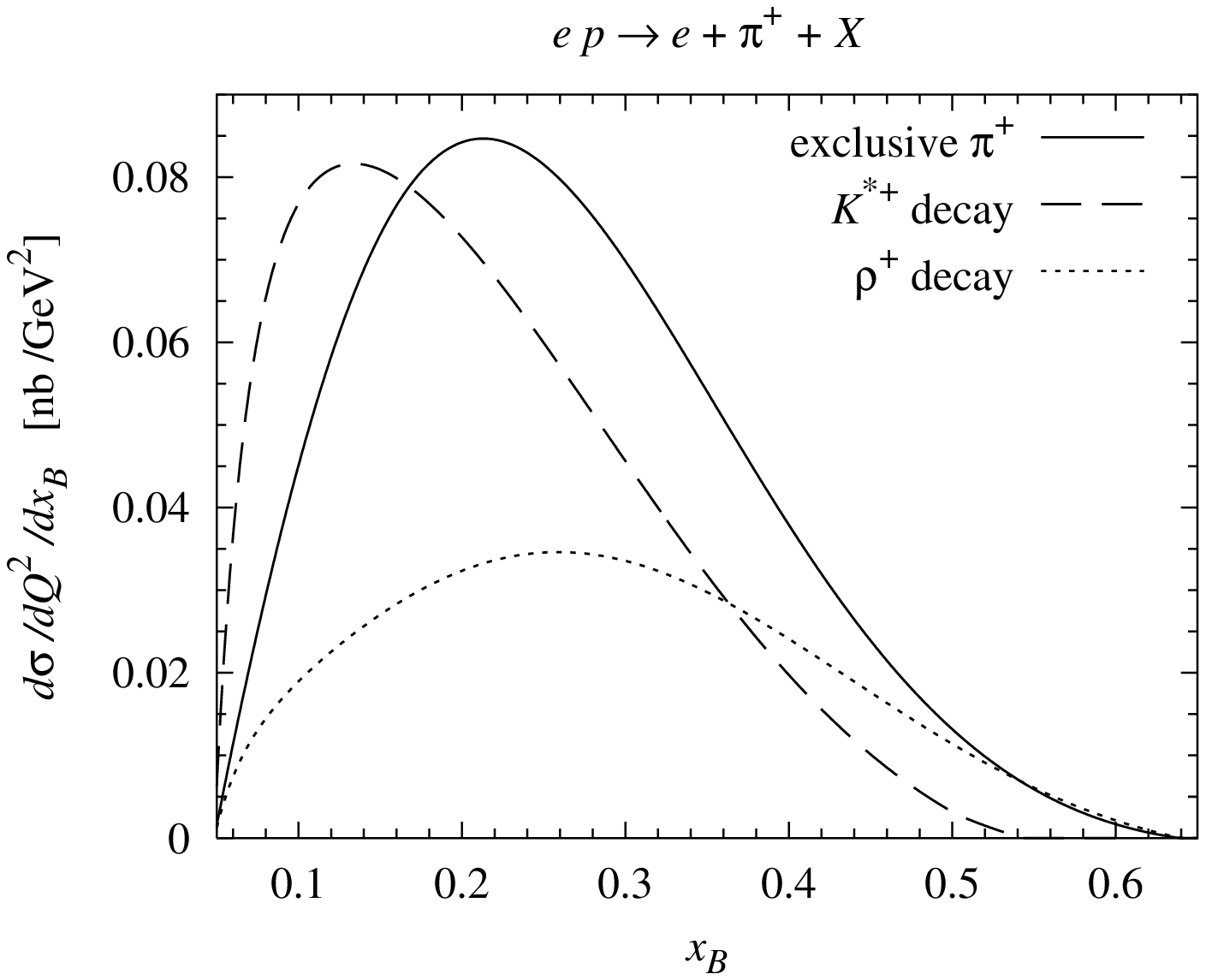}
  \includegraphics[width=8.3cm]{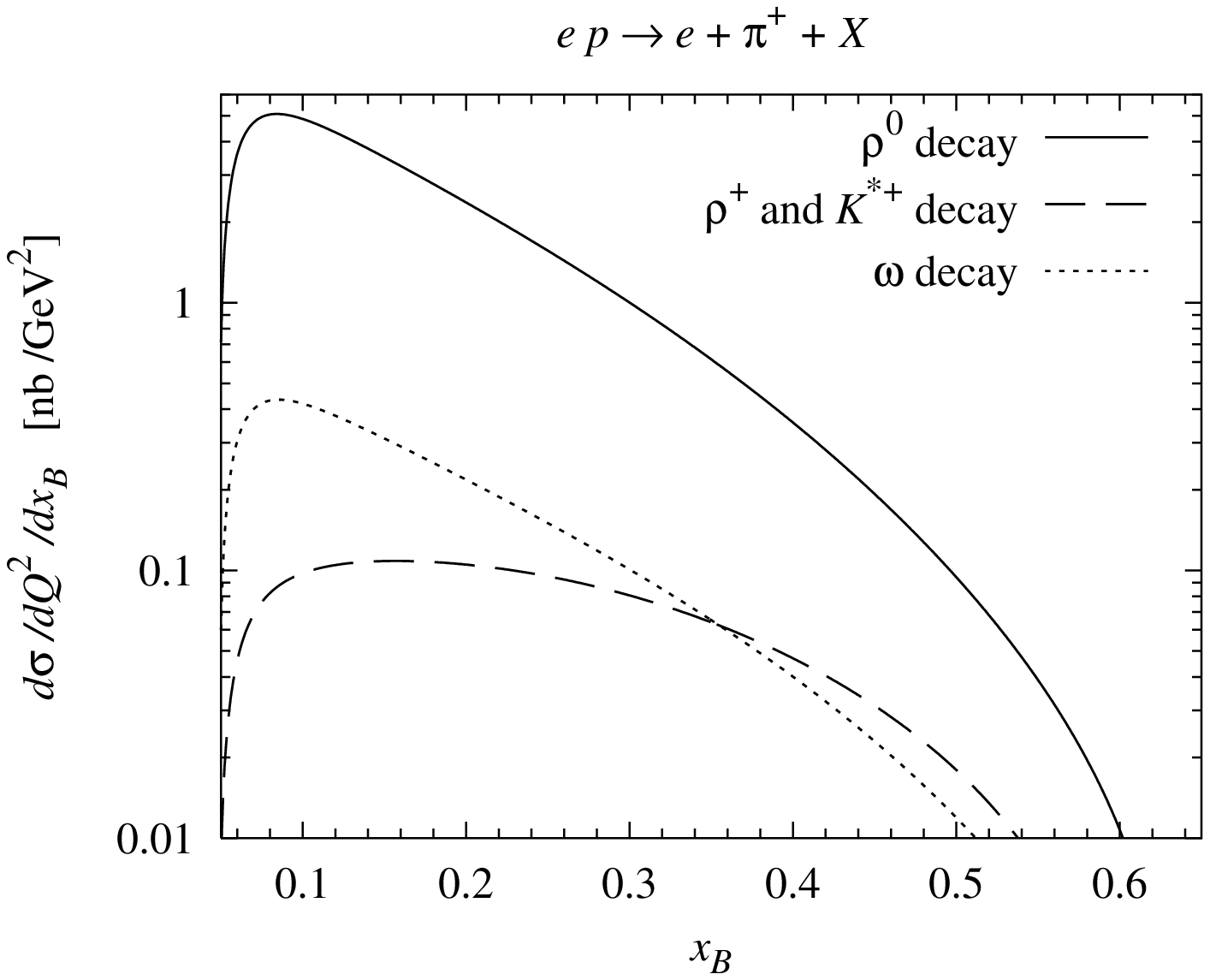}
\caption{\label{fig:pion-ep} Exclusive contributions to the $\pi^+$
  electroproduction cross section at $Q^2=2.5 \gev^2$, obtained from
  our leading-twist calculation of $\sigma_L$.  This and the following
  plots are for a lepton beam energy of $27.5 \gev$ in the target rest
  frame.  Left: direct exclusive production, and contributions from
  $K^{*+}$ and $\rho^+$ decay.  Right: sum of contributions from
  $K^{*+}$ and $\rho^+$ decay compared with contributions from
  $\omega$ and $\rho^0$ decay.}
\end{figure}
\begin{figure}
\centering
\includegraphics[width=8.3cm]{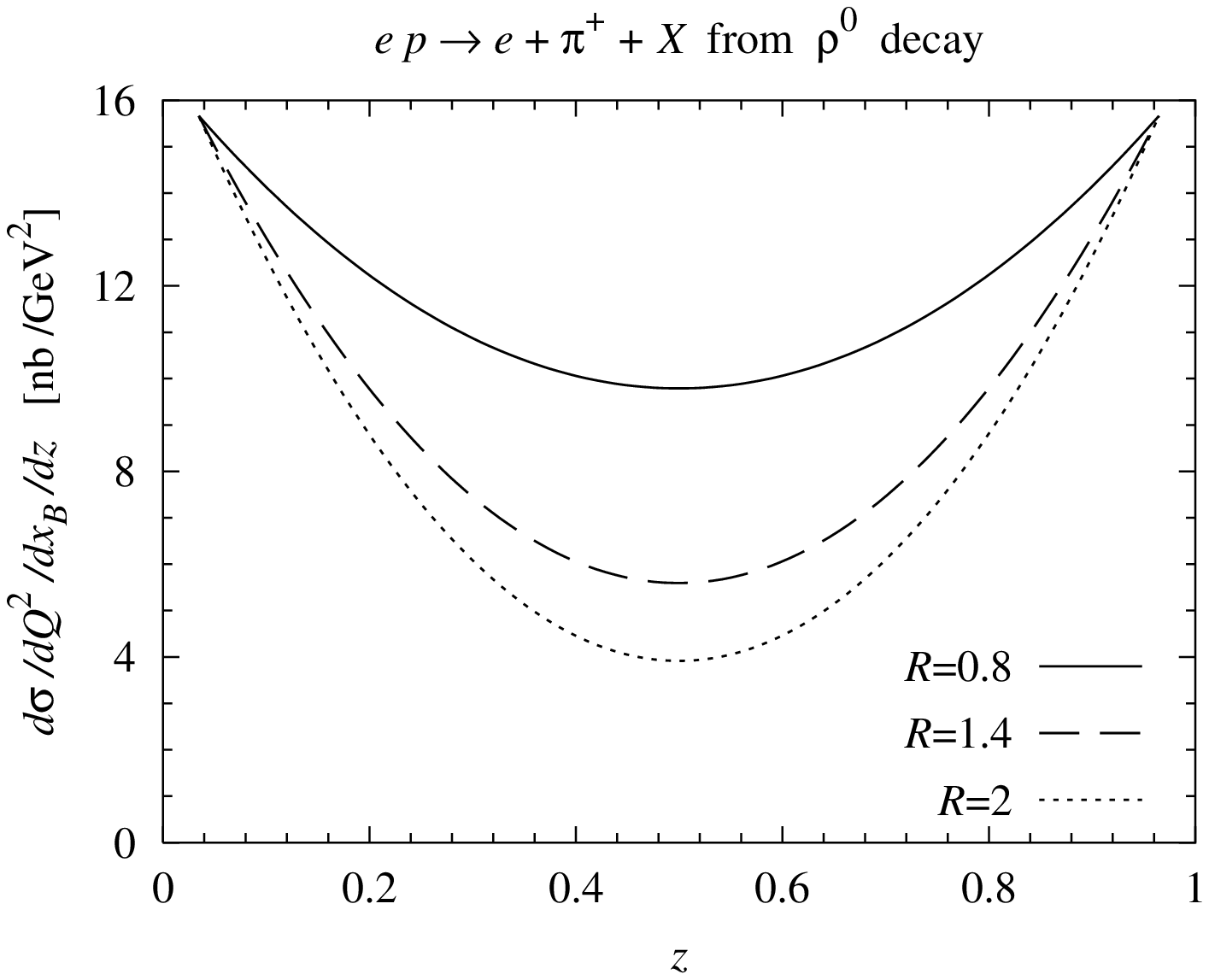}
\includegraphics[width=8.3cm]{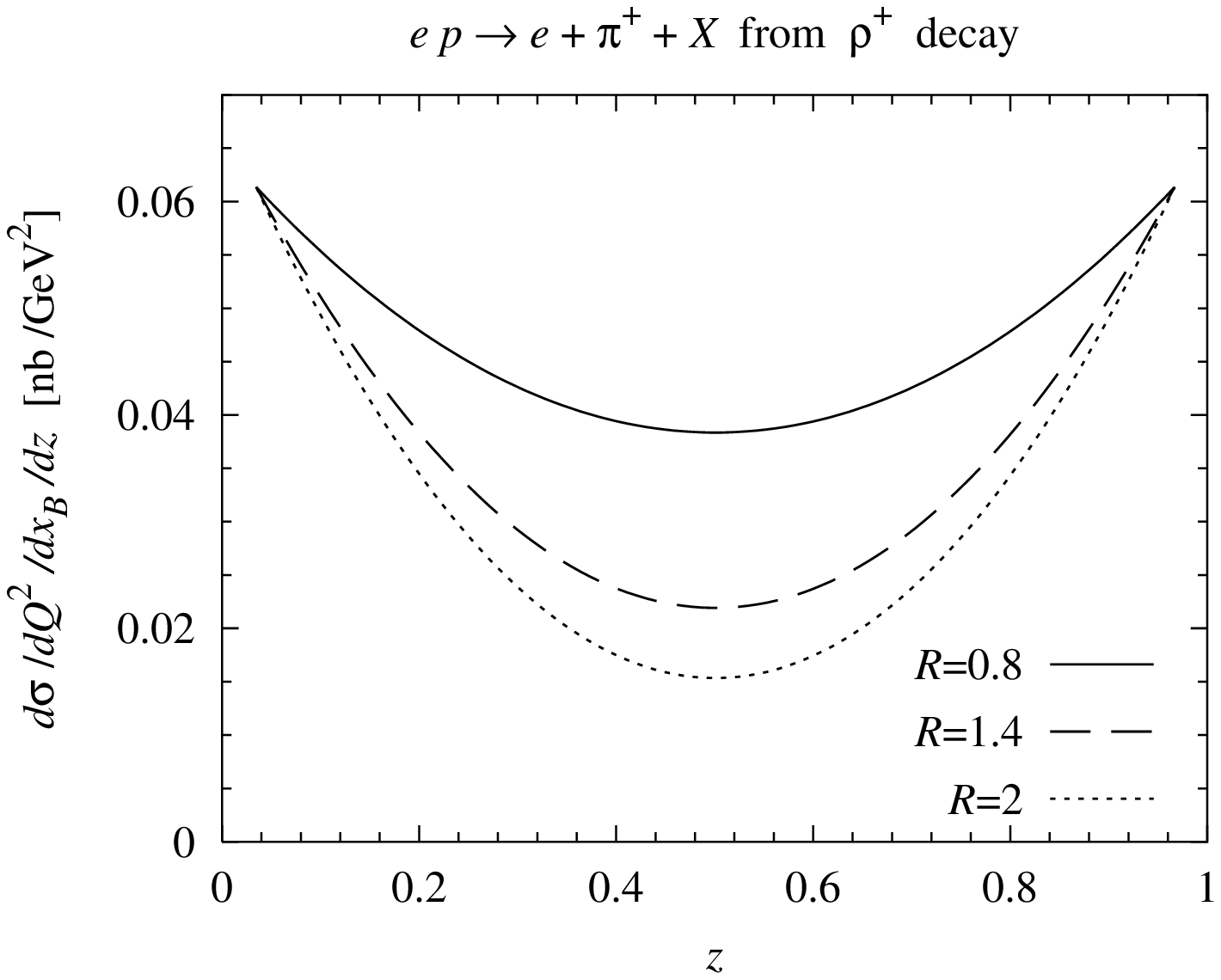}
\includegraphics[width=8.3cm]{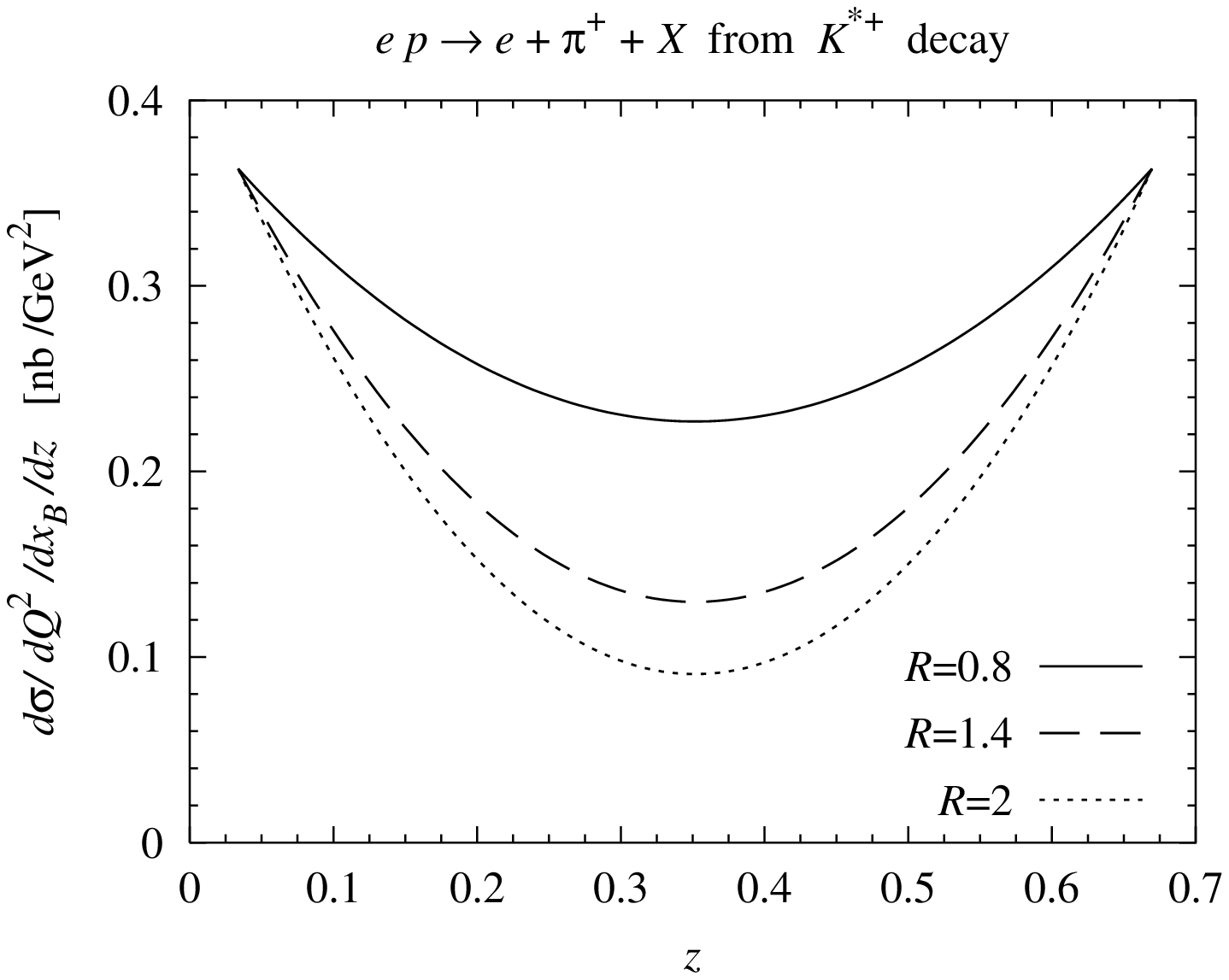}
\includegraphics[width=8.3cm]{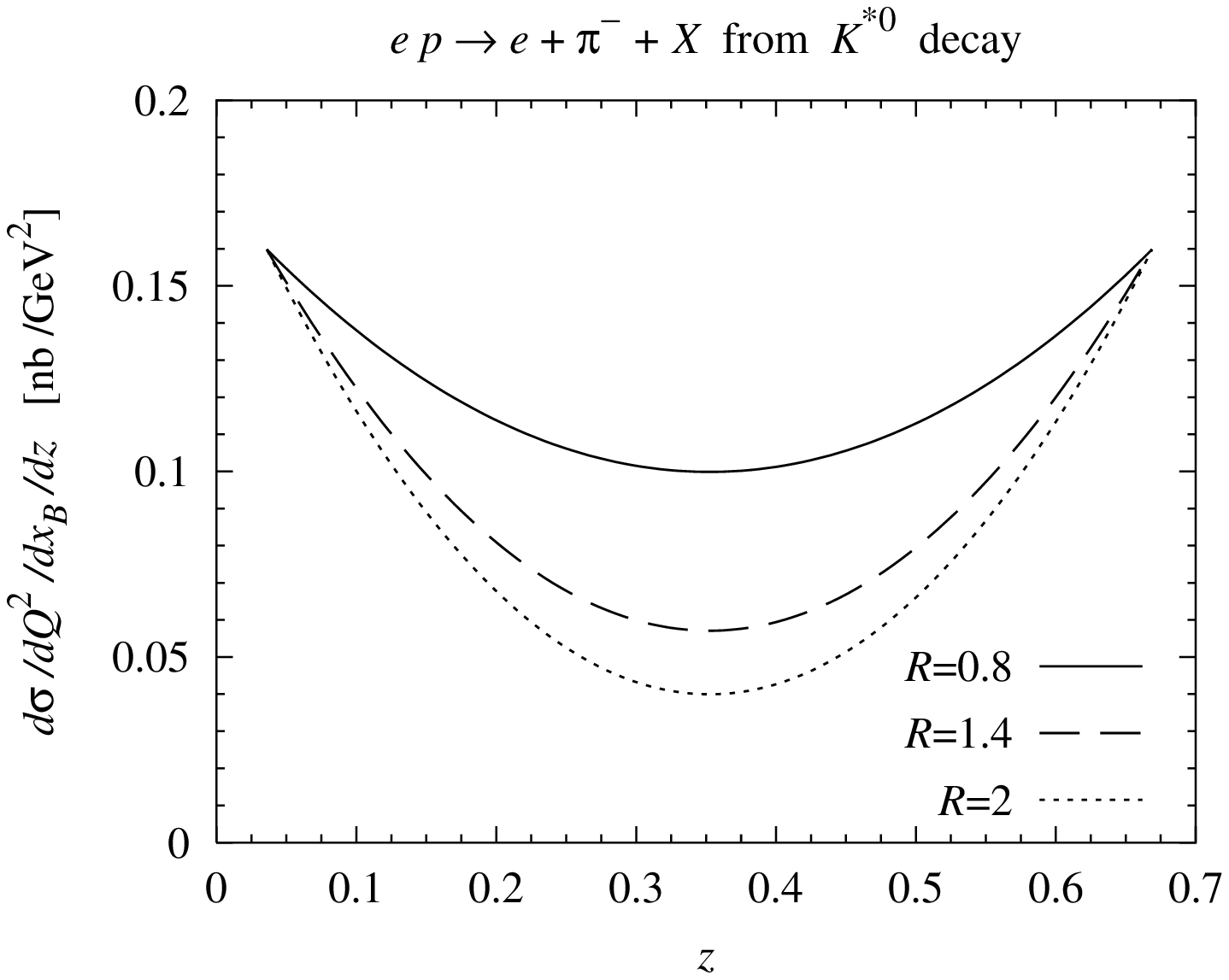}
\caption{\label{fig:pi-vec-z} Contributions to the semi-inclusive 
pion electroproduction cross section from decays of exclusively
produced vector mesons, for $Q^2=2.5 \gev^2$ and $\xb=0.1$. Shown
are the results corresponding to our leading-twist calculation of 
$\sigma_L$ for vector meson production, and a value of 
$R= \sigma_L /\sigma_T$ chosen as indicated in the figure.
Top left: $\pi^+$ from $\rho^0$ decay. 
Top right: $\pi^+$ from $\rho^+$ decay. 
Bottom left: $\pi^+$ from $K^{*+}$ decay. 
Bottom right: $\pi^-$ from $K^{*0}$ decay.
The contribution to $\pi^0$ production from $K^{*+}$ and $K^{*0}$ decays
is given by the average of the corresponding curves in the 
two lower plots, according to (\protect\ref{kstar-decay}).
}
\end{figure}

In Fig.~\ref{fig:pion-ep} we show the result of our leading-twist
calculation from Sects.~\ref{sec:pseudo-results} and \ref{sec:vector}
for the $ep$ cross section of $\pi^+$ production.  We give all $ep$
cross section for a lepton beam energy of $27.5 \gev$ in the target
rest frame, corresponding to the HERMES experiment, and recall that
all our exclusive cross sections are calculated with an upper cutoff
of $1\gev^2$ on $|t|$.  We see that the $\rho^0$ channel is clearly
dominating.  The $\omega$, which decays to almost 100\% into
$\pi^+\pi^-\pi^0$, is much less prominent.  According to our
discussion in Sects.~\ref{sec:pseudo-results} and
\ref{sec:data-compare} one expects that the contribution from $ep \to
e \pi^0 p$ to $\pi^0$ production is smaller than in the case of
direct $\pi^+$ production.  The same holds for the production and
decay of $\eta$ and $\eta'$, which have several three-body decays
contributing to all three pion channels.  As we argued in
Sect.~\ref{sec:vector} the production of $f_2(1270)$, which
predominantly decays into $\pi^+\pi^-$ and $\pi^0\pi^0$, may
contribute at a similar or lower level as $\rho^+$ decays.  In
Fig.~\ref{fig:pi-vec-z} we show the $z$
spectrum arising from different vector meson decays.  Whereas $\rho$
decays contribute in almost the entire $z$ range, pions from $K^*$
decays are limited to $z$ values below $0.7$.  Note that due to charge
conjugation invariance the $z$ spectrum from $\rho^0$ decays is
identical for the $\pi^+$ and $\pi^-$, and by isospin invariance the
same holds for the $\pi^+$ and $\pi^0$ spectra from the decay of
$\rho^+$.  To illustrate the dependence of the $z$ spectrum on the
ratio $R$ of longitudinal and transverse meson production cross
sections we have taken values which correspond to the range measured
in $\rho^0$ and $\phi$ production at $Q^2= 2.5 \gev^2$
\cite{Borissov:2001fq,Tytgat:2000th,Rakness:2000th,Hadjidakis:2004zm}.

We now compare the contribution from exclusive channels with the
semi-inclusive cross section obtained from the leading-order
expression~(\ref{sidis-xsec}) for quark fragmentation.  We use the LO
parton densities from MRST2001 and the LO fragmentation functions of
Kretzer \cite{Kretzer:2000yf}, both at a scale $Q^2= 2.5 \gev^2$.  Let
us first take a look at the high-$z$ tail of the spectrum, where
direct exclusive production contributes.  Integrating the
semi-inclusive cross section for $\pi^+$ production for $z>0.9$ at
$Q^2=2.5 \gev^2$ and $\xb=0.1$, we obtain $d\sigma/(dQ^2\, d\xb) = 0.19
\nb \gev^{-2}$ from (\ref{sidis-xsec}).  This number should be
understood as a naive extrapolation: the fragmentation functions are
not well known for $z$ close to 1, and in the above kinematics $z>0.9$
corresponds to an invariant mass $m_X <1.84 \gev$ according to
(\ref{mx}), where leading-twist fragmentation can be just marginally
valid.  Our leading-twist result for $ep\to e \pi^+ n$ gives
$d\sigma/(dQ^2\, d\xb) = 0.045 \nb\gev^{-2}$, which according to our
comparison with preliminary HERMES data (Sect.~\ref{sec:data-compare})
undershoots the actual cross section by a factor of about 0.4.  We
thus find that for $z>0.9$ direct exclusive pion production may be a
substantial part of the semi-inclusive cross section, but cannot be
more quantitative given the uncertainties just discussed.

For a realistic estimate of exclusive vector meson production we
divide our leading-twist results for $\sigma_L$ by a factor 7, except
for $\phi$ production.  According to Table~\ref{tab:vm-data} this
brings us close to the HERMES measurement for $\rho^0$ production at
$Q^2=2.3 \gev^2$ and $\xb=0.1$, and according to our arguments in
Sect.~\ref{sec:data-compare} it should give a reasonable estimate for
the other channels.  In other words, we assume that the ratio of
vector meson channels is sufficiently well described by our
leading-twist calculation.  Only for $\phi$ production do we divide
our leading-twist results by a different factor, namely by 15,
following our comparison in Table~\ref{tab:vm-data} with preliminary
HERMES data in this channel.  One might argue that for the production
of a $K^*$, which has one light and one strange quark, power
corrections are between those for the $\rho^0$ and for the $\phi$, but
we refrain from such refinements here.  Possible changes by a factor
of 2 would in fact not change our conclusions regarding the role of
$K^*$ decays.  For a prediction of $\sigma_T$ we divide $\sigma_L$
obtained as just described by the value $R=\sigma_L/\sigma_T = 1.2$
from preliminary HERMES data for $\rho^0$ production in the relevant
kinematics \cite{Borissov:2001fq,Tytgat:2000th}, except for the $\phi$
channel, where instead we take $R= 0.8$ from the parameterization of
preliminary HERMES data in \cite{Borissov:2001fq,Rakness:2000th}.
Variation of $R$ as shown in Fig.~\ref{fig:pi-vec-z} 
would not affect the conclusions we shall draw.

%
%
\begin{figure}
\centering
\includegraphics[width=8.3cm]{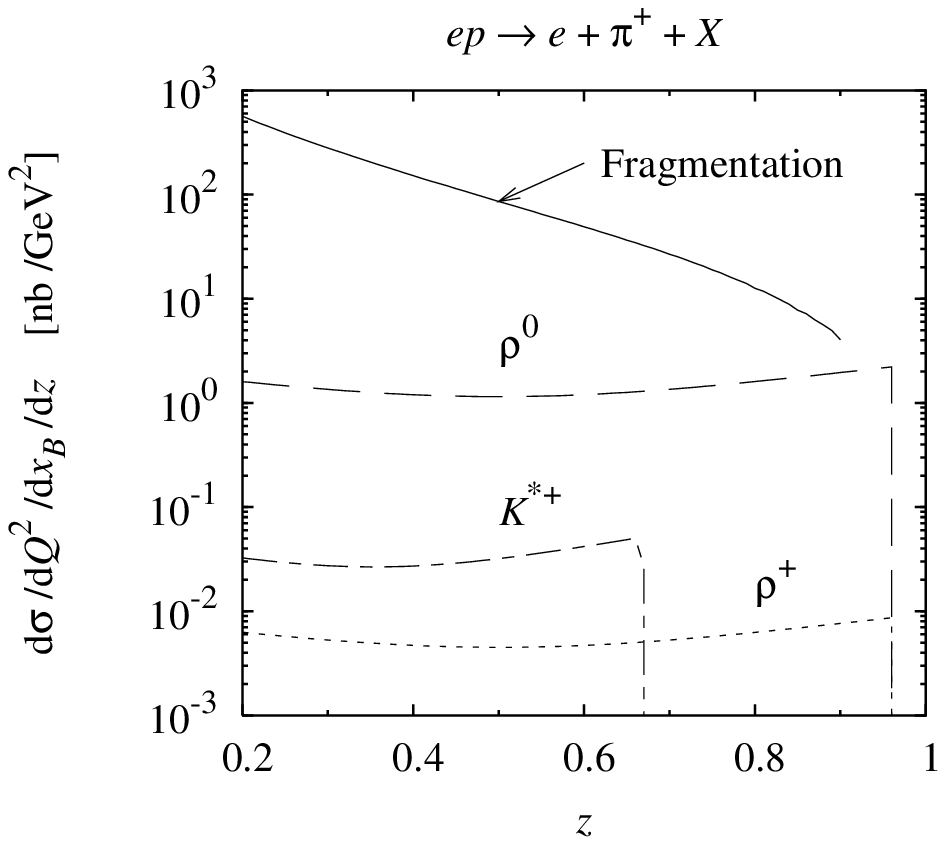}
\includegraphics[width=8.3cm]{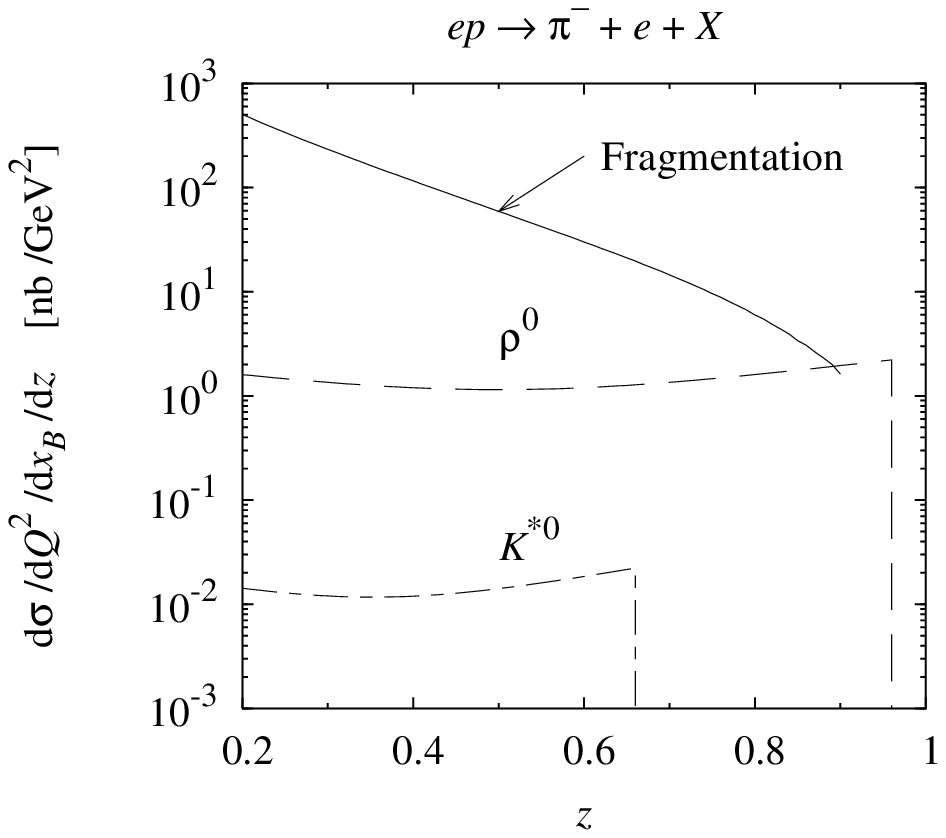}
\caption{\label{fig:pi_compare} The cross section for semi-inclusive
  electroproduction of $\pi^+$ (left) and $\pi^-$ (right), as a
  function of $z$, at $\xb = 0.1$ and $Q^2 = 2.5 \gev^2$.  The cross
  sections from quark fragmentation were calculated using the LO
  fragmentation functions of Kretzer \cite{Kretzer:2000yf} 
  and the MRST 2001 LO parton distributions. The contributions from 
  vector meson decays were obtained by adjusting our leading--twist results
  for the vector meson production cross sections, as explained in the text.  
  The value taken for the $\rho^0$ cross section matches the HERMES 
  measurement \protect\cite{Airapetian:2000ni}.} 
\end{figure}

Integrating the $\rho^0$ decay contribution to $\pi^+$ production for
$z>0.9$ at $Q^2=2.5 \gev^2$ and $\xb=0.1$, we find $d\sigma/(dQ^2\,
d\xb) = 0.13 \nb\gev^{-2}$, which is surprisingly close to the
extrapolation of the fragmentation result given above.  The
fragmentation formula for $\pi^-$ production gives $d\sigma/(dQ^2\,
d\xb) = 0.07 \nb\gev^{-2}$ when integrated over $z>0.9$, so that
in this case the $\rho^0$ contribution slightly overshoots the naive
fragmentation result.  In Fig.~\ref{fig:pi_compare} we show the $z$
spectrum of semi-inclusive $\pi^+$ and $\pi^-$ production, comparing
the fragmentation result with the contributions from vector meson
decays.  Following our above discussion we do not show the cross
section from fragmentation for $z$ above 0.9.  We see that $\rho^0$
production gives a sizable contribution to semi-inclusive production
for $z$ greater than about 0.8.  According to our estimate, $\rho^+$
production is suppressed relative to $\rho^0$ by two orders of
magnitude and cannot compete with the cross section from quark
fragmentation even at large $z$.  The $K^*$ decay contribution is
somewhat larger in size but limited to $z$ below 0.7.  The
fragmentation result for $\pi^0$ production is just the average of
$\pi^+$ and $\pi^-$ because of isospin invariance.  With $\rho^0$
decay being absent and the contributions from other vector mesons
being comparatively small, we find no exclusive channel that is
prominent in semi-inclusive $\pi^0$ production for the kinematics
discussed here.  We expect direct exclusive production $ep\to e \pi^0
p$ to be much less important at high $z$ than in the case of $\pi^+$
production, to the extent that power corrections enhance the $\pi^+$
but suppress the $\pi^0$ compared with the leading approximation in
$1/Q^2$.  Given the relative size of cross sections in
Fig.~\ref{fig:pion-ep} it is clear that pions from $\omega$ production
are significantly smaller than the fragmentation result for all $z$,
and we shall not analyze the kinematics of the corresponding
three-body decay here.

\begin{figure}
\centering
\includegraphics[width=8.3cm]{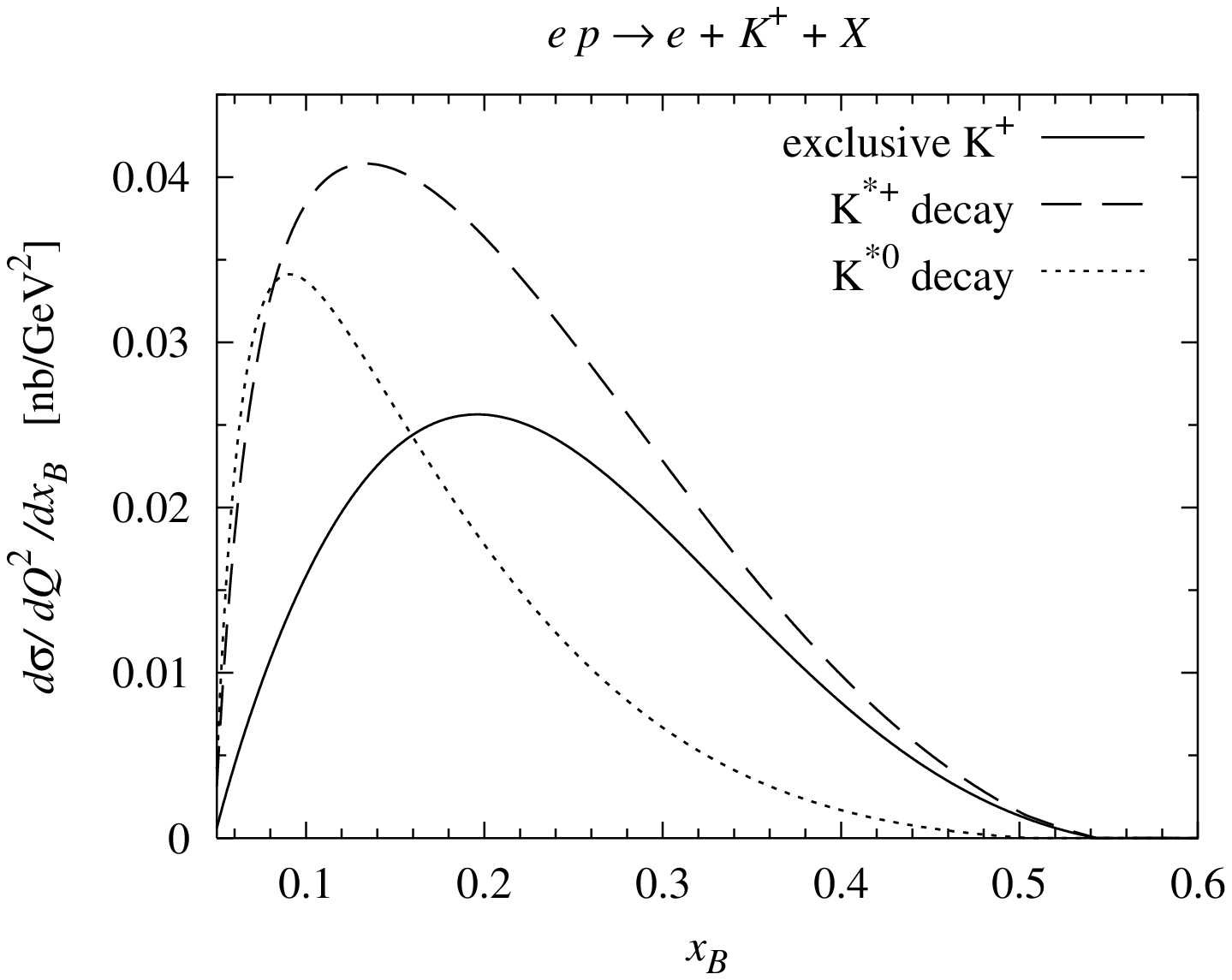}
\includegraphics[width=8.3cm]{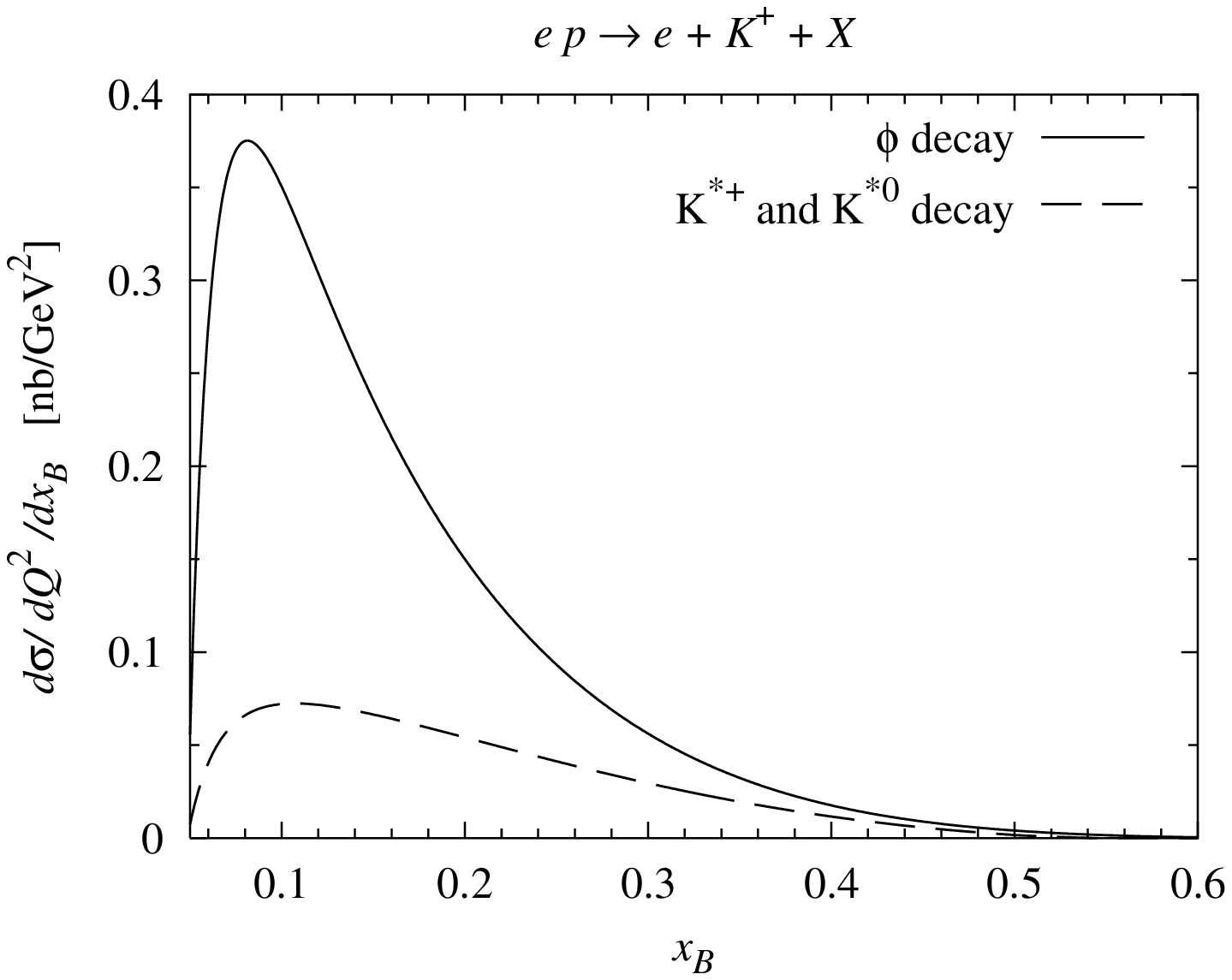}
\caption{\label{fig:kaon-ep} Same as Fig.~\protect\ref{fig:pion-ep}, but for
  electroproduction of $K^+$.}
\end{figure}

\begin{figure}
\centering
\includegraphics[width=8.3cm]{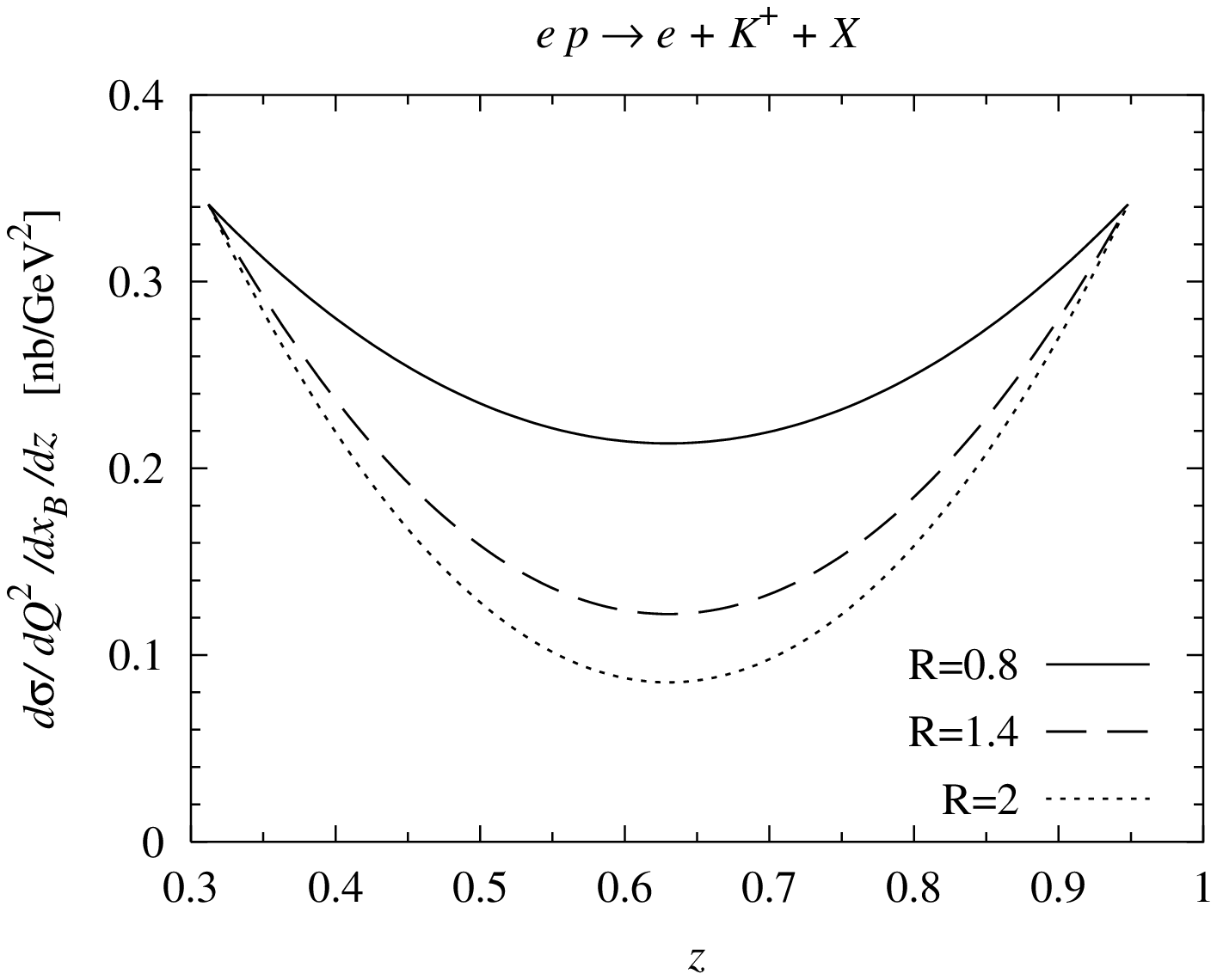}
\includegraphics[width=8.3cm]{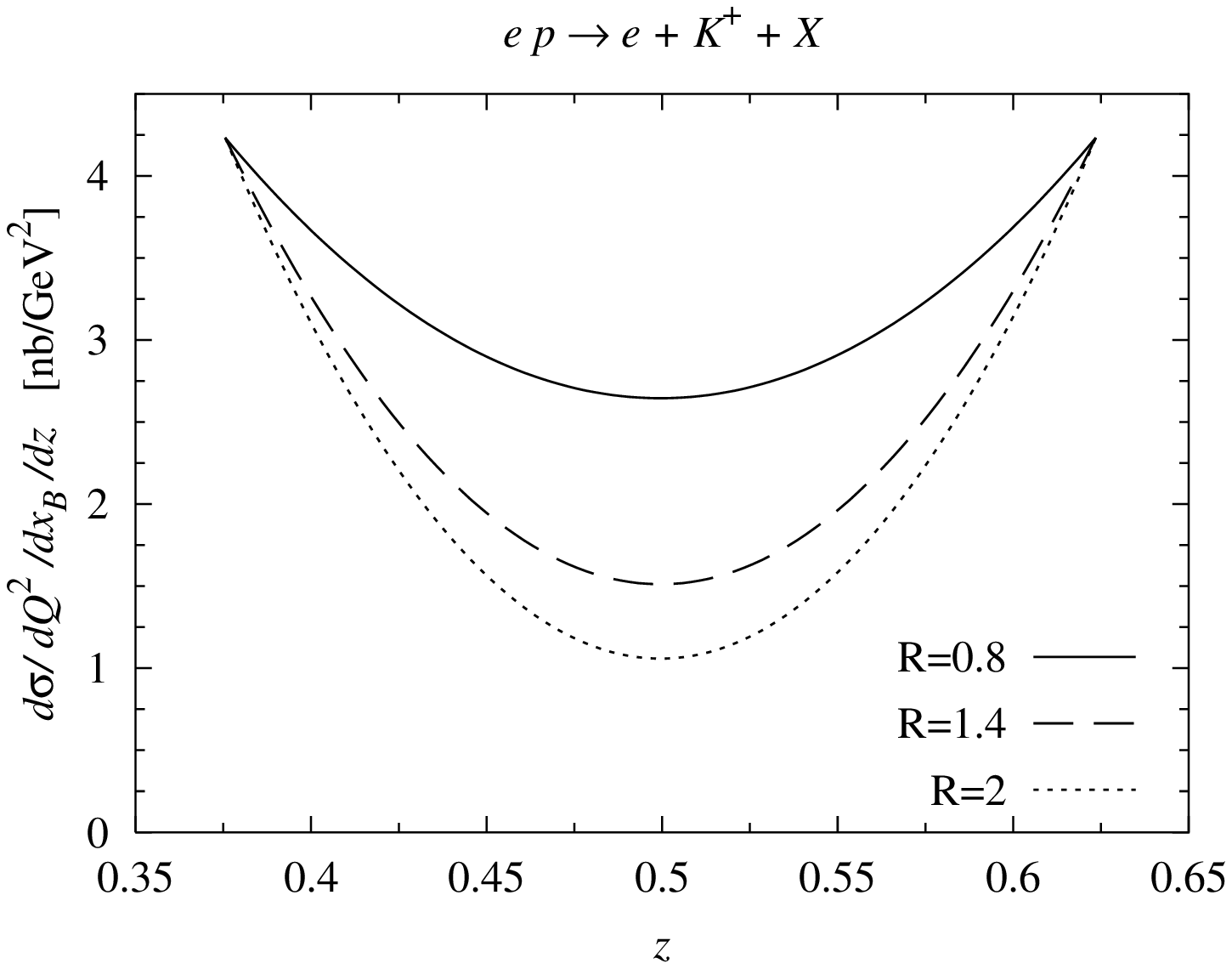}
\caption{\label{fig:kaon-z} Same as Fig.~\protect\ref{fig:pi-vec-z}, but for
  the contribution from $K^{*+}$ and $K^{*0}$ decays to $K^+$
  production (left) and for the contribution of $\phi$ decay to $K^+$
  or $K^-$ production (right).}
\end{figure}

Turning to semi-inclusive $K^+$ and $K^-$ production, we show in
Fig.~\ref{fig:kaon-ep} the contributions of the relevant exclusive
channels to the $ep$ cross section, obtained from our leading-twist
calculation in Sect.~\ref{sec:vector}.  The production of $\phi$,
which decays to approximately 50\% into $K^+ K^-$, is clearly dominant
for $K^+$ production, and it is the only channel contributing to $K^-$
production.  As is seen in the $z$-spectra of Fig.~\ref{fig:kaon-z},
it is however only the $K^+$ from $K^{*}$ decays that extends to $z$
values above 0.65.

Integrating the leading-order fragmentation formula (\ref{sidis-xsec})
for $K^+$ production for $z>0.9$ at $Q^2=2.5 \gev^2$ and $\xb=0.1$, 
we obtain $d\sigma/(dQ^2\, d\xb) = 0.048 \nb\gev^{-2}$, which is to be
understood as an extrapolation as in the pion case discussed above.
Our leading-twist estimate for direct exclusive $K^+$ production in
Sect.~\ref{sec:pseudo-results} gives $d\sigma/(dQ^2\, d\xb) = 0.016
\nb /\gev^2$.  Following our discussion in
Sect.~\ref{sec:data-compare} one expects that power corrections will
lead to weaker enhancement than in the case of $ep\to e\pi^+ n$ (or
possibly even to a suppression), because $\tilde\mathcal{H}$ is more
important in the leading-twist cross section for $K^+$ production than
$\tilde\mathcal{E}$.  Nevertheless, the above numbers suggest that
direct exclusive $K^+$ production may be of some significance at the
high-$z$ end of the spectrum.

\begin{figure}[t]
\centering
\includegraphics[width=8.3cm]{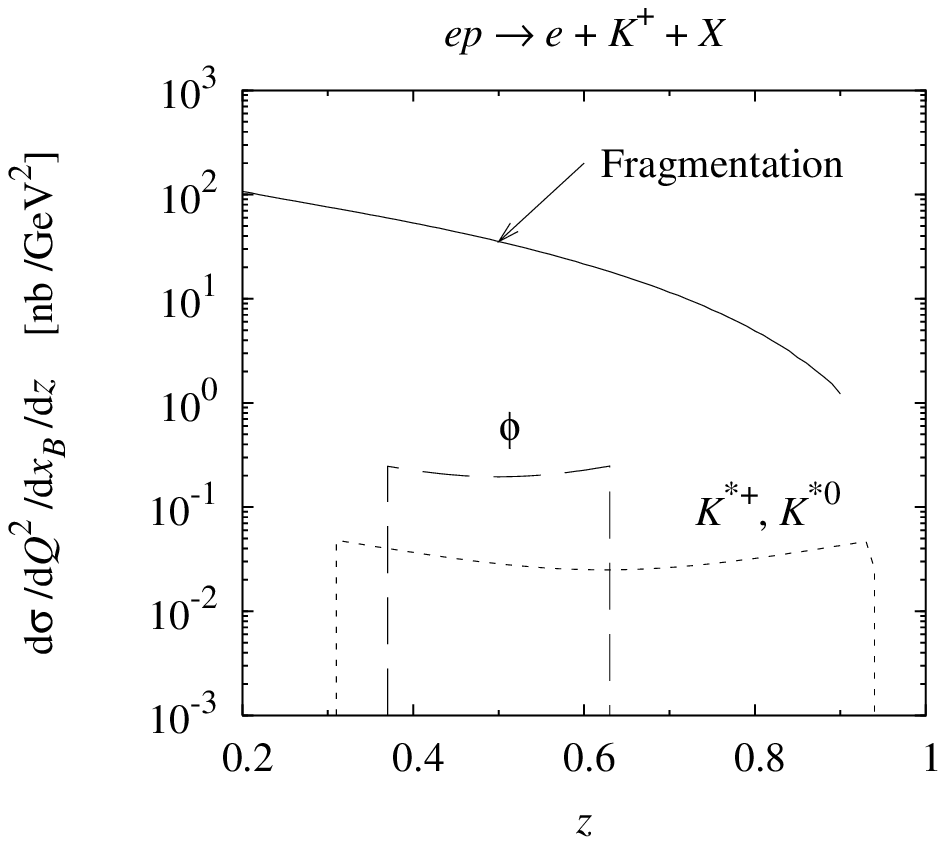} 
\includegraphics[width=8.3cm]{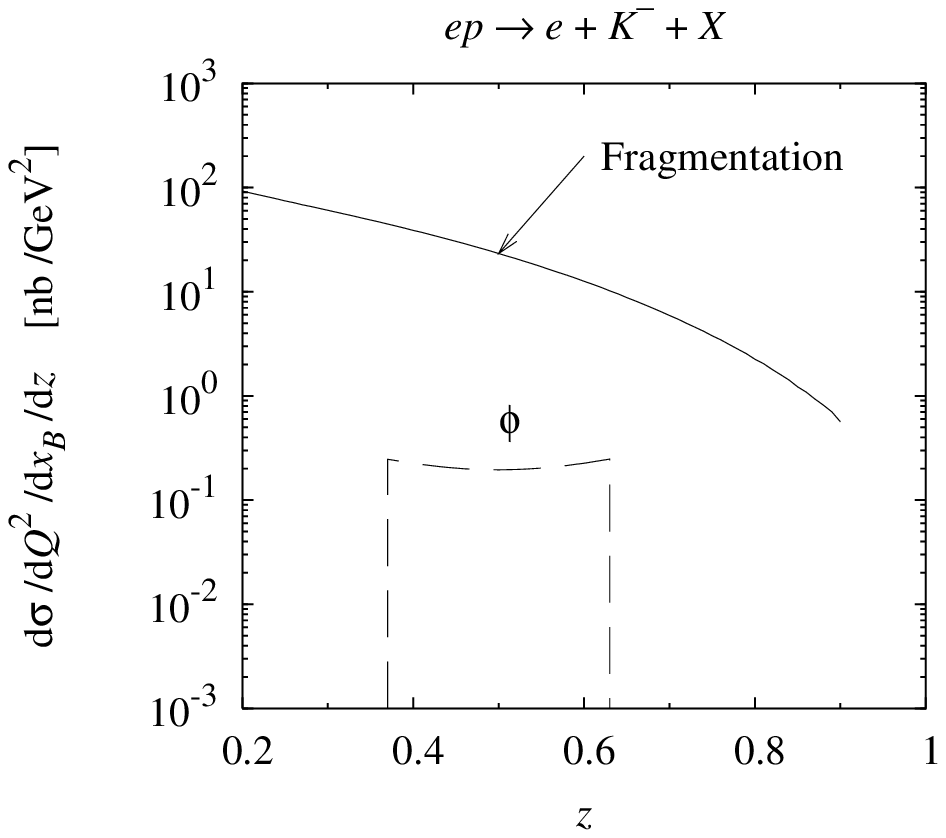}
\caption{\label{fig:K_compare} Same as Fig.~\protect\ref{fig:pi_compare},
  but for semi-inclusive production of $K^+$ (left) and $K^-$ (right).
  As discussed in the text, the $\phi$ production cross section used
  in this plot matches the preliminary results of a HERMES measurement 
  \protect\cite{Borissov:2001fq,Rakness:2000th}.}
\end{figure}

If we integrate the $K^+$ spectrum from $K^{*+}$ and $K^{*0}$ decays
for $z>0.9$, dividing our leading-twist result by 7 and accounting for
the transverse cross section as described above, we obtain
$d\sigma/(dQ^2\, d\xb) = 0.0021 \nb\gev^{-2}$, which is well below our
extrapolation from leading-twist fragmentation.  In
Fig.~\ref{fig:K_compare} we compare the fragmentation result for
semi-inclusive $K^+$ and $K^-$ production with the individual
contributions from $K^*$ and $\phi$ decays.  We conclude that, even
within the uncertainties of our estimates, contributions from $K^*$
production are only a fraction of the fragmentation result at any $z$,
and that $\phi$ production, despite its larger cross section, is
always well below the semi-inclusive cross section.  Our finding
concerning the $\phi$ contribution agrees with a recent study of
measured kaon multiplicities in \cite{Bino}.  On one hand, kaon
production by quark fragmentation is less suppressed compared with
pion production than is exclusive $\phi$ production compared with
production of $\rho^0$, and on the other hand $\phi$ decays only
contribute in a $z$-range where the fragmentation functions are still
large.  Apart possibly from direct $K^+$ production at $z$ close to 1,
we thus find no exclusive channel dominating $K^+$ or $K^-$ production
in typical HERMES kinematics.
%
%

So far we have compared exclusive channels with quark fragmentation at
$\xb=0.1$ and $Q^2= 2.5 \gev^2$.  We have also performed the
comparison of Figs.~\ref{fig:pi_compare} and \ref{fig:K_compare} for
$\xb=0.3$ at the same $Q^2$.  For the vector meson cross sections we
used the same values of $R$ and the same correction factors as for
$\xb=0.1$, dividing our leading-twist cross sections by 7 for all
vector mesons except the $\phi$, where we divide by 15.  In doing
this, we assume that the leading-twist approximation gives a realistic
description of the $\xb$ dependence in this region.  We find that our
qualitative conclusions do not change when going to the larger value 
of $\xb$.

A comment is in order concerning the treatment of exclusive channels
in the analysis of semi-inclusive DIS data when extracting quark
fragmentation or distribution functions.  It is by no means clear that
by subtracting contributions of individual exclusive channels from the
total yield one obtains an observable more suitable for comparison
with the quark fragmentation formulae. In fact, the derivation of
factorization theorems relies on the sum over all channels $X$ in
(\ref{sidis-reac}) to be complete.  At sufficiently large $Q^2$, each
individual exclusive channel by itself is a power correction which may
or may not be included in the leading-twist analysis.  This situation
is similar to the one with the contribution from individual nucleon
resonances to the cross section of inclusive DIS.  A way to think
about the relation of exclusive channels to the leading-twist cross
section is quark-hadron duality.  It remains a challenge to formulate
the problem of quark-hadron duality for semi-inclusive DIS in a
quantitative fashion.

Symmetry properties like $\sigma(\pi^0) = \half [\sigma(\pi^+) +
\sigma(\pi^-)]$ can emerge from summing over many channels which do
not fulfill this relation individually.  If however a single channel
like $\rho^0$ production dominates the semi-inclusive cross section,
it clearly becomes more and more difficult for the remaining channels
to compensate the missing symmetry between charged and neutral pion
production.  In such a situation, parton-hadron duality must cease to
work.  The outcome of our study is that indeed the only channel whose
contribution can become dangerously large in HERMES kinematics, is the
$\rho^0$ contribution to $\pi^+$ and $\pi^-$ production.


\section{Summary}
\label{sec:summary}

We have evaluated the cross section for a variety of exclusive meson
production channels for moderate to large $\xb$ at leading order in
$1/Q^2$ and in $\alpha_s$.  Cross sections change significantly when
varying the nonperturbative input, generalized parton distributions
and meson distribution amplitudes, within plausible limits of current
model building.  On one hand this implies an uncertainty in predicting
these cross sections, but on the other hand it implies that their
measurement can ultimately help to constrain the nonperturbative
functions, provided theoretical control over corrections to the
leading-order formulae.  We find the largest cross section
uncertainties for $\rho^0$, $\omega$ and $\phi$ production, which is
sensitive to the generalized gluon distribution over a large range of
$\xb$, reflecting the current uncertainty of the unpolarized gluon
density at low scales.  A strong dependence on the factorization scale
in these channels underlines the need for analysis at next-to-leading
order in $\alpha_s$.  Comparing our leading-twist cross section with
experimental data, we confirm that for $Q^2$ of a few $\gev^2$ power
corrections are substantial.  In particular, the suppression of vector
meson cross sections we find is consistent with what has been
estimated in the literature from the effects of parton transverse
momentum in the hard-scattering subprocess.  A consistent description
of such effects together with next-to-leading order corrections in
$\alpha_s$ remains a challenge for theory.  As is seen for the ratio
of $\phi$ and $\rho^0$ production, the most serious theoretical
uncertainties cancel however in cross section ratios for sufficiently
similar channels (the main distinction being between channels with and
without $t$-channel pion exchange).

Rescaling our leading-twist cross sections such as to be consistent
with experimental data for $\rho^0$ and $\phi$ production, we have
compared their contribution to semi-inclusive pion or kaon production
with the result of leading-twist quark fragmentation, focusing on the
typical kinematics of HERMES measurements, where $Q^2\sim 2.5 \gev^2$
and $\xb\sim 0.1$.  Within large uncertainties, direct exclusive
production of $\pi^+$ and possibly $K^+$ appears to be comparable with
the fragmentation result extrapolated to the bin $0.9 <z< 1$.  Through
their decays, exclusively produced $\rho$, $\phi$ and $K^*$ contribute
in a wide range of $z$.  Pions from $K^*$ decay and kaons from $\phi$
decay are however limited to $z$ below 0.7.  With this and the
relative size of cross sections, our estimates indicate that in
typical HERMES kinematics the only exclusive channel whose cross
section can compete with quark fragmentation is the $\rho^0$. The
$\rho^0$ saturates the quark fragmentation result for semi-inclusive
$\pi^+$ and $\pi^-$ production at large $z$. Since the $\rho^0$ does
not contribute to $\pi^0$ and to kaon production, there is no
corresponding ``dangerous'' vector channel in these cases.


\section*{Acknowledgments} 

We are indebted to E.-C.~Aschenauer, H.~Avakian, A.~Borissov,
C.~Hadjidakis, D.~Hasch, A.~Hillenbrand, M.~Strikman, M.~Vanderhaeghen
and A.~Vinnikov for valuable discussions and information.  This work is 
supported by the Helmholtz Association, contract number VH-NG-004. 
This work is supported by U.S.\ Department of Energy Contract 
DE-AC05-84ER40150, under which the Southeastern Universities 
Research Association (SURA) operates the Thomas Jefferson 
National Accelerator Facility.


\begin{appendix}

\section{Integrals over GPDs within the double distribution model}
\label{app:GPDs}

The $t$ independent functions in the ansatz (\ref{factorize-ansatz})
for quark and gluon GPDs are modeled as
\begin{eqnarray}
  \label{dd-models}
H^q(x,\xi) &=& \int_{-1}^1 d\beta \int_{-1+|\beta|}^{1-|\beta|} d\alpha\;
  \delta(x-\beta-\xi\alpha)\, h(\beta,\alpha)
  \Big[ \theta(\beta)\, q(\beta)\, 
      - \theta(-\beta)\, \bar{q}(-\beta) \Big] ,
\nonumber \\
H^g(x,\xi) &=& \int_{-1}^1 d\beta \int_{-1+|\beta|}^{1-|\beta|} d\alpha\;
  \delta(x-\beta-\xi\alpha)\, h(\beta,\alpha)\, \beta 
  \Big[ \theta(\beta)\, g(\beta)\, 
      - \theta(-\beta)\, g(-\beta) \Big] ,
\nonumber \\
\tilde{H}^q(x,\xi) &=& 
  \int_{-1}^1 d\beta \int_
{-1+|\beta|}^{1-|\beta|} d\alpha\;
  \delta(x-\beta-\xi\alpha)\, h(\beta,\alpha)
  \Big[ \theta(\beta)\, \Delta q(\beta) 
      + \theta(-\beta)\, \Delta\bar{q}(-\beta) \Big] ,
\end{eqnarray}
where $\theta$ denotes the usual step function, $q$, $\bar{q}$,
$\Delta q$, $\Delta\bar{q}$ the unpolarized and polarized quark and
antiquark distributions, and $g$ the unpolarized gluon distribution.
The profile function
\begin{equation}
  \label{profile}
h(\beta,\alpha) = \frac{\Gamma(2b+2)}{2^{2b+1}\Gamma^2(b+1)}\,
\frac{[ (1-|\beta|)^2- \alpha^2 ]^b}{(1-|\beta|)^{2b+1}}
\end{equation}
depends on a parameter $b$, which we chose to be either $b=1$ or $b=2$
in this work.

For meson production amplitudes we need the integrals
\begin{eqnarray}
  \label{amp-integrals}
I^q(\xi) &=& \int_{-1}^1dx\;
            \frac{H^q(x,\xi)}{\xi-x-i\epsilon} \, ,
\qquad\qquad
I^{\bar{q}}(\xi) \;=\; \int_{-1}^1dx\;
            \frac{H^{\bar{q}}(x,\xi)}{\xi-x-i\epsilon} 
            \;=\; [\sms I^q(-\xi) \sms]^{*} ,
\nonumber \\
\tilde{I}^q(\xi) &=& \int_{-1}^1dx\;
            \frac{\tilde{H}^q(x,\xi)}{\xi-x-i\epsilon} \, ,
\qquad\qquad
\tilde{I}^{\bar{q}}(\xi) \;=\; \int_{-1}^1dx\;
            \frac{\tilde{H}^{\bar{q}}(x,\xi)}{\xi-x-i\epsilon} 
            \;=\; - [\sms \tilde{I}^q(-\xi) \sms]^{*} ,
\end{eqnarray}
where we used the definitions $H^{\bar{q}}(x,\xi)= - H^q(-x,\xi)$ and
$\tilde{H}^{\bar{q}}(x,\xi)= \tilde{H}^q(-x,\xi)$ together with the
fact that these functions are even in $\xi$.  The required integral
for gluons can be brought into the same form as $I^q(\xi)$ by
rewriting
\begin{equation}
  \label{amp-g-integrals}
I^g(\xi) 
 = \int_{-1}^1 dx\, \frac{H^g(x,\xi)}{x} \, \frac{1}{\xi-x-i\epsilon}
 = \frac{1}{\xi} \int_{-1}^1 dx\, \frac{H^g(x,\xi)}{\xi-x-i\epsilon}
 \, ,
\end{equation}
where we used that $H^g(x,\xi)$ is even in $x$.  The imaginary parts
of these integrals are readily converted into integrals over $\beta$,
with
\begin{eqnarray}
  \label{im-part}
\im I^q(\xi) &=& \int_{0}^{\frac{2\xi}{1+\xi}} d\beta\, 
  I(\beta,\xi)\, q(\beta) ,
\nonumber \\
I(\beta,\xi) &=& \frac{\pi \Gamma(2b+2)}{2^{2b+1}\Gamma^2(b+1)}\,
   \frac{(1-\xi^2)^b}{\xi^{2b+1}}\,
   \frac{1}{(1-\beta)^{2b+1}}\,
   \Big( \frac{2\xi}{1+\xi} - \beta \Big)^b\, \beta^b
\end{eqnarray}
for $\xi>0$.  The function $I(\beta,\xi)$ vanishes at the endpoints of
the integration region, which in particular ensures the convergence of
the integral at $\beta=0$ for common parameterizations of quark
densities.  To obtain the analogous expressions for $I^{\bar{q}}$,
$\tilde{I}^q$, $\tilde{I}^{\bar{q}}$ and $I^g$ one has to replace
$q(\beta)$ with $\bar{q}(\beta)$, $\Delta q(\beta)$,
$\Delta\bar{q}(\beta)$ and $\beta g(\beta)$, respectively.

\begin{figure}[t]
\centering
\includegraphics[width=8.3cm]{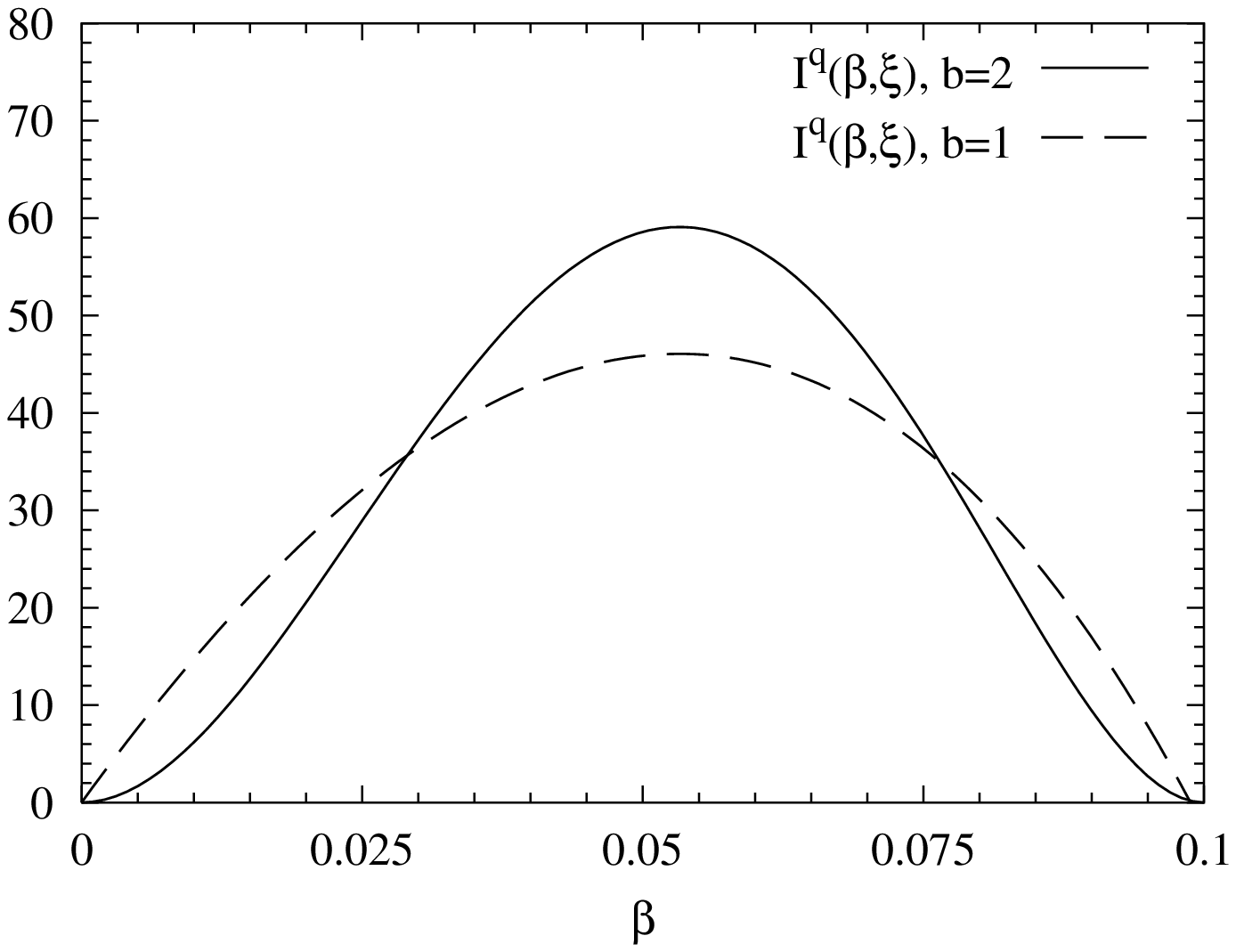}
\includegraphics[width=8.3cm]{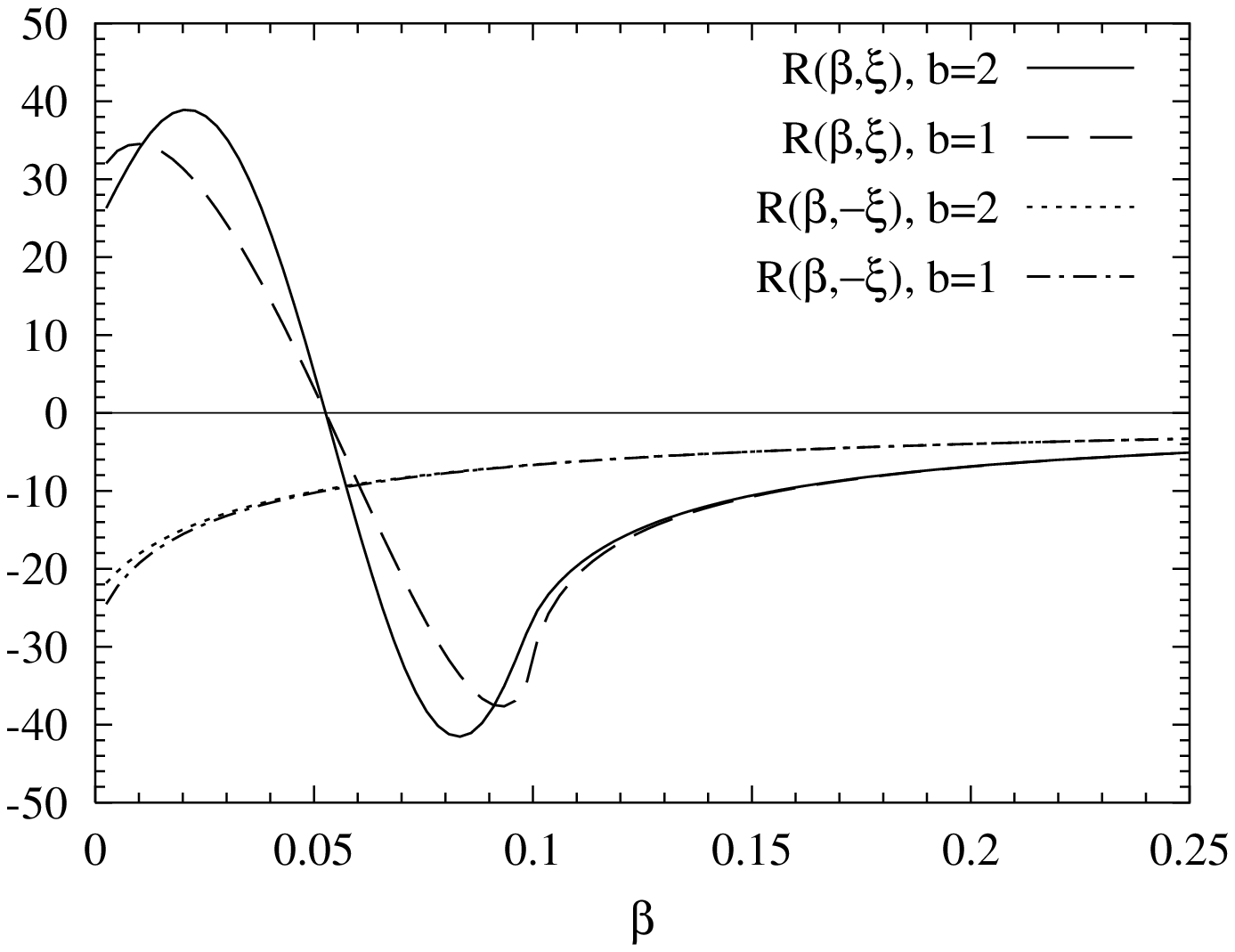}
\includegraphics[width=8.3cm]{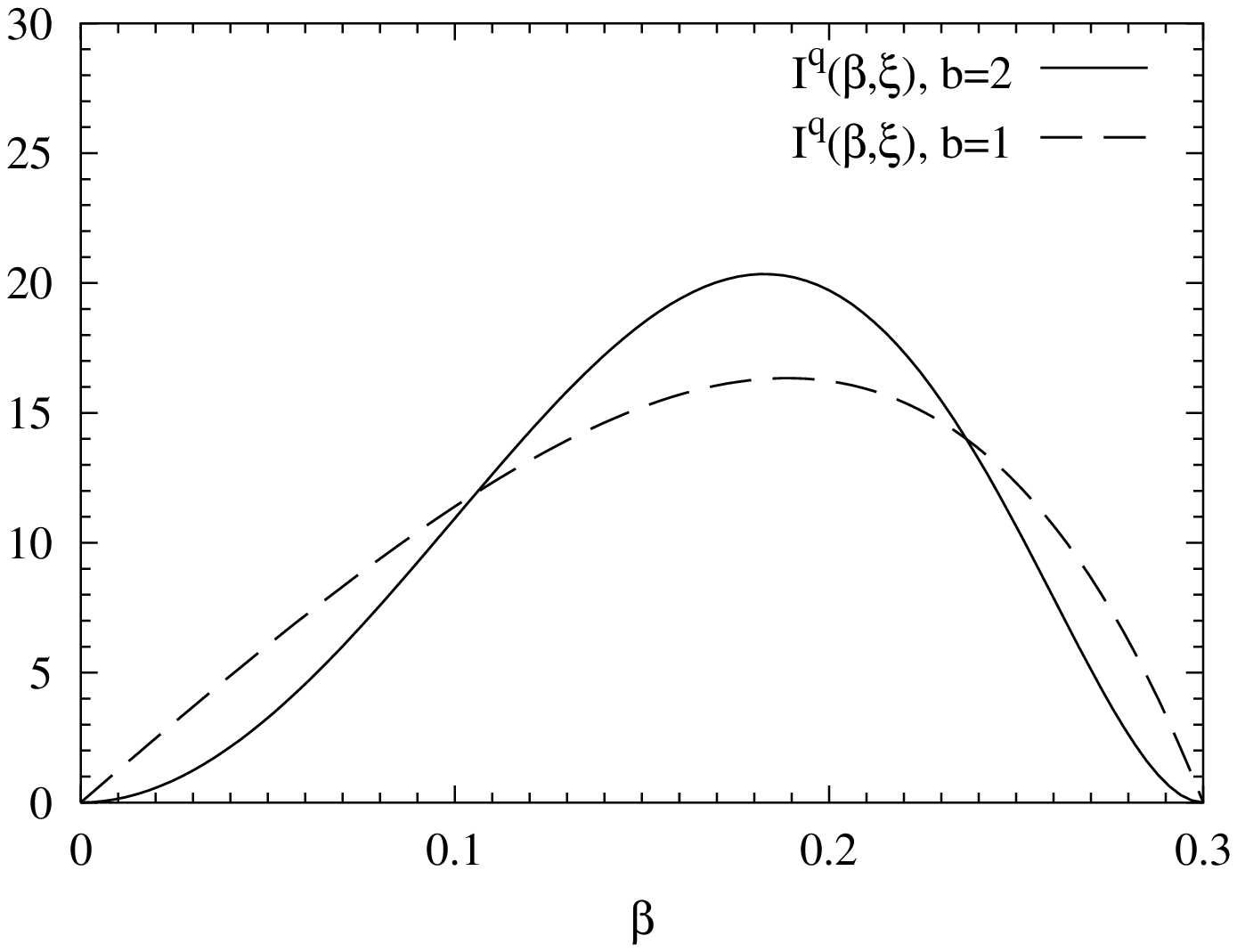}
\includegraphics[width=8.3cm]{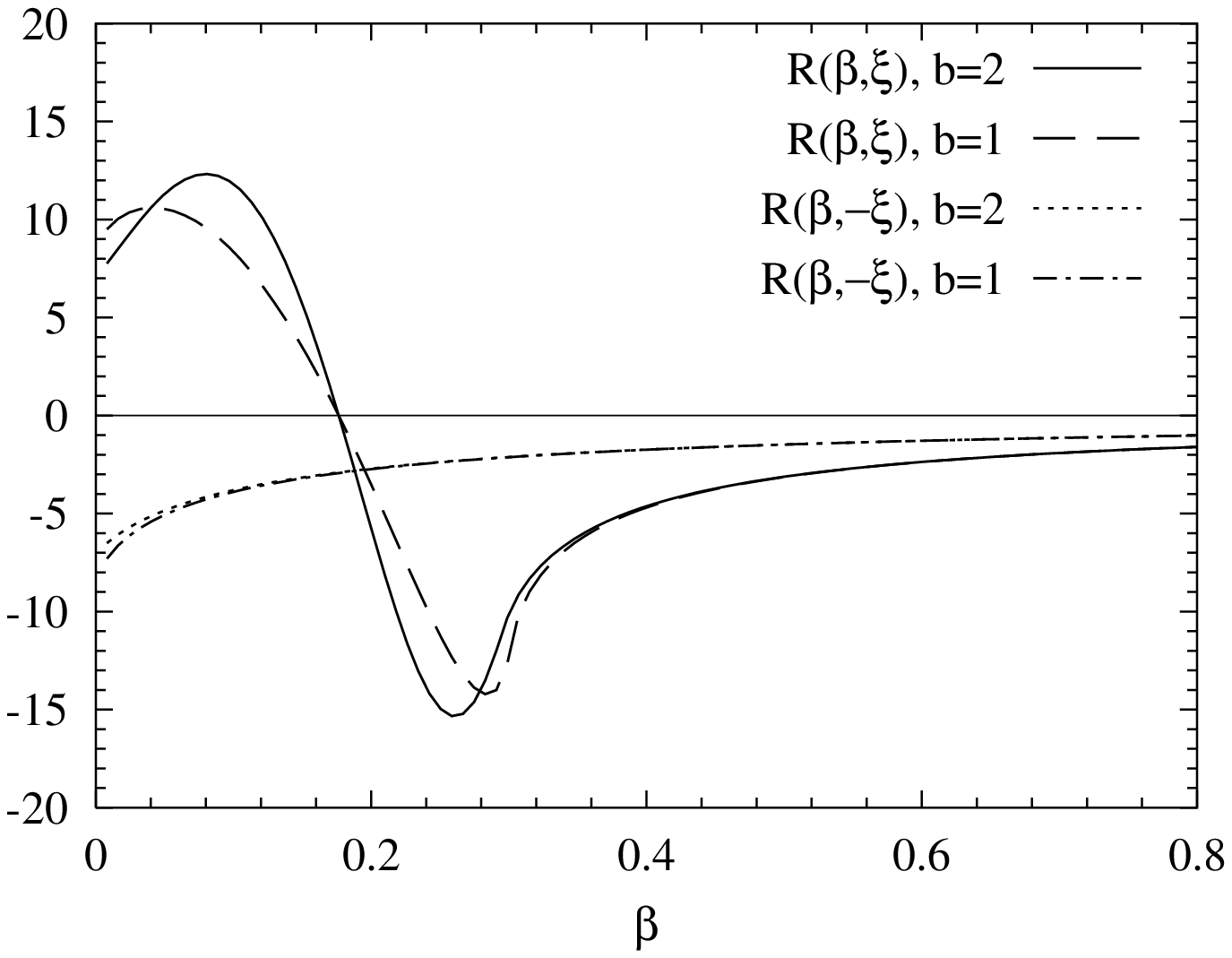}
\caption{\label{fig:dd-integrands} The functions appearing in the
  integrals (\protect\ref{im-part}) and (\protect\ref{re-part}) for
  the meson production amplitude for $\xi= 0.053$ (top) and $\xi=
  0.18$ (bottom), which respectively correspond to $\xb =0.1$ and $\xb
  = 0.3$.  Note the different ranges for $\beta$ in the plots.}
\end{figure}
The real parts of the amplitudes involve principal value integrals,
whose numerical evaluation requires some care, especially for small
$\xi$.  For our choices of profile parameters $b=1$ and $b=2$ one can
explicitly perform the $\alpha$ integral after inserting
(\ref{dd-models}) into (\ref{amp-integrals}) and
(\ref{amp-g-integrals}).  The result is
\begin{eqnarray}
  \label{re-part}
\re I^q(\xi) &=& \int_0^1 d\beta\, 
  \Big[ R(\beta,\xi)\, q(\beta) + R(\beta, -\xi)\, \bar{q}(\beta) \Big]
\nonumber \\
 &=& \int_0^1 d\beta\, 
\Bigg\{ R(\beta,\xi) \Big[q(\beta) - \bar{q}(\beta)\Big] 
    + \Big[R(\beta,\xi) + R(\beta, -\xi)\Big]\, \bar{q}(\beta) \Bigg\}
  \,, 
\nonumber \\
\re I^g(\xi) &=& \int_0^1 d\beta\,
  \Big[ R(\beta,\xi) + R(\beta, -\xi) \Big] \beta g(\beta) \,,
\nonumber \\[0.3em]
\re \tilde{I}^q(\xi) &=& \int_0^1 d\beta\, 
  \Big[ R(\beta,\xi)\, \Delta q(\beta) 
      - R(\beta, -\xi)\, \Delta\bar{q}(\beta)
  \Big] \,,
\end{eqnarray}
with
\begin{eqnarray}
R(\beta,\xi) & \stackrel{b=1}{=} & 
  - \frac{3}{4 \xi^3 (1-\beta)^3}\, 
  \Bigg(\, 2 \xi (1-\beta) (\beta-\xi)
\nonumber \\
 && \hspace{6.1em}
  {}+ \beta (1-\xi) \Big[ \beta (1+\xi) - 2\xi \Big]
  \log\frac{|\beta (1+\xi) - 2\xi|}{\beta(1-\xi)} \,\Bigg) ,
\nonumber \\
R(\beta,\xi) & \stackrel{b=2}{=} & 
  \frac{5}{16 \xi^5 (1-\beta)^5}\, 
  \Bigg(\, 2 \xi (1-\beta) (\beta-\xi)
           \Big[ 3 (\beta-\xi)^2 - 5\xi^2 (1-\beta)^2 \Big]
\nonumber \\
 && \hspace{5.8em}
  {}+ 3 \beta^2 (1-\xi)^2 \Big[ \beta (1+\xi) - 2\xi \Big]^2
  \log\frac{|\beta (1+\xi) - 2\xi|}{\beta(1-\xi)} \,\Bigg) .
\end{eqnarray}
For both $b=1$ and $b=2$, the function $R(\beta,\xi)$ is continuous in
the full interval of integration, with finite limits at $\beta=0$ and
$\beta=1$.  If $\xi>0$ it is positive for $\beta< \xi$ and negative
for $\beta> \xi$, and if $\xi<0$ it is negative in the entire
interval.  Convergence of the integral for polarized quark
distributions requires that $\Delta q(\beta)$ and
$\Delta\bar{q}(\beta)$ have integrable singularities at $\beta=0$,
which is the case for the parton densities we use in this study.  The
unpolarized quark distributions have a steeper behavior at small
$\beta$, but since $R(\beta,\xi) + R(\beta,-\xi) \sim \beta$ for
$\beta\to 0$ it is sufficient to have integrable singularities for
$q(\beta)-\bar{q}(\beta)$ and for $\beta \bar{q}(\beta)$.

In Fig.~\ref{fig:dd-integrands} we illustrate the behavior of the
functions multiplying the parton distributions in the integrals
(\ref{im-part}) and (\ref{re-part}).  The imaginary part of the
amplitude involves momentum fractions in the parton densities between
$0$ and $2\xi /(1+\xi) = \xb$, with a maximum of the shape function
$I(\beta,\xi)$ for $\beta$ around $\xi$.  In contrast, the real part
is sensitive to higher momentum fractions, with a partial cancellation
from values above and below $\xi$.  One also clearly sees the stronger
sensitivity to small $\beta$ if $b=1$.  Note that the functions shown
in the figure will be multiplied in the amplitude with functions
showing a strong rise towards $\beta=0$.


\section{Distribution of pions or kaons from vector meson decay}
\label{app:decay}

\begin{figure}[t]
\centering 
\includegraphics[width=16.5cm]{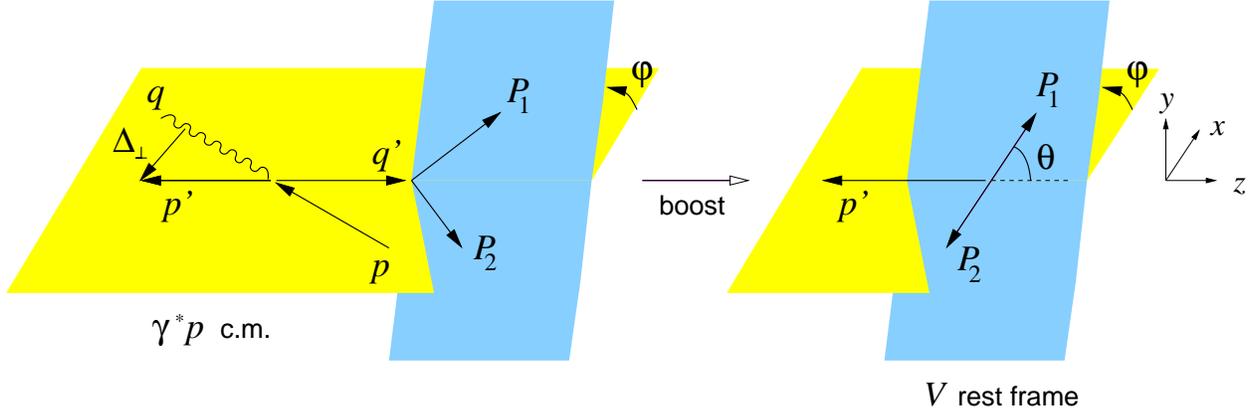}
\caption{\label{fig:decay-kin} Kinematic variables for $\gamma^*(q)
  + p(p) \to V(q') + B(p')$ followed by the decay $V \to P_1 P_2$,
  shown in the $\gamma^*\,p$ center-of-mass and in the rest frame of
  $V$.  Here $\theta$ and $\varphi$ are the spherical coordinates of
  the momentum of $P_1$ in the depicted coordinate system.}
\end{figure}

In this appendix we discuss the decay of a vector meson into two
pseudoscalar mesons and derive the $z$ distribution given in
(\ref{dsig-dz-sim}).  Consider the contribution of $ep\to V B$ with
subsequent decay $V\to P_1 P_2$ to semi-inclusive production $ep\to
P_1 + X$.  A useful set of variables to describe the decay of the
vector meson are the polar and azimuthal angles $\theta$ and $\varphi$
of $P_1$ in the vector meson center-of-mass, as shown in
Fig.~\ref{fig:decay-kin}.  The distribution in these angles is
connected in a straightforward way with the spin density matrix of the
produced vector meson \cite{Schilling:1973ag}, and the phase space
element has a factorized form in the variables $Q^2$, $\xb$, $t$ and
$\theta$, $\varphi$.  The variable $z$ used for semi-inclusive
production of $P_1$ is then given by
\begin{equation}
  \label{z-expression}
z = a + b \cos\theta + c \sin\theta \cos\varphi
\end{equation}
with
\begin{eqnarray}
  \label{abc}
a &=& \frac{E_{P1}}{m_{V}} \;
      \frac{r_2\, (1+2\xb\sms m_p^2/Q^2)
          + r_3 \sqrt{1+4\xb^2\sms m_p^2/Q^2}}{2r_1}\;
      \Big[ 1 + O(\xb \Delta_T^2 /Q^2) \Big]
\:\approx\:  \frac{E_{P1}}{m_{V}} \,,
\nonumber \\[0.5em]
b &=& \frac{|\sms\mbox{\boldmath{$q$}}_{P1}|}{m_{V}} \;
      \frac{r_3\, (1+2\xb\sms m_p^2/Q^2)
          + r_2 \sqrt{1+4\xb^2\sms m_p^2/Q^2}}{2r_1}\;
      \Big[ 1 + O(\xb \Delta_T^2 /Q^2) \Big]
\:\approx\:  \frac{|\sms\mbox{\boldmath{$q$}}_{P1}|}{m_{V}} \,,
\nonumber \\[0.3em]
c &=& {}- \frac{|\sms\mbox{\boldmath{$q$}}_{P1}| \Delta_T}{Q^2} \,
        \frac{2\xb}{1-\xb}\, 
        \frac{\sqrt{1+4\xb^2\sms m_p^2/Q^2}}{r_3} \,,
\end{eqnarray}
where we abbreviated
\begin{eqnarray}
  \label{r1r2}
r_1 &=&   1 + \frac{\xb}{1-\xb}\, \frac{m_p^2}{Q^2} \,,
\qquad\qquad
r_2 \:=\: 1 + \frac{\xb}{1-\xb}\, \frac{m_V^2-m_B^2+m_p^2}{Q^2} \,,
\nonumber \\[0.3em]
r_3 &=&  \Bigg[ \Bigg( 1 - \frac{\xb}{1-\xb}\, 
         \frac{m_V^2+m_B^2-m_p^2}{Q^2} \Bigg)^2
         - \Bigg( \frac{\xb}{1-\xb}\,
           \frac{2 m_V m_B}{Q^2} \Bigg)^2 \,\Bigg]^{1/2} \,.
\end{eqnarray}
The energy and momentum $E_{P1}$ and
$|\sms\mbox{\boldmath{$q$}}_{P1}|$ of $P_1$ in the rest frame of $V$
have already been given in (\ref{meson-masses}), and $\Delta_T$ is the
transverse momentum of the scattered baryon with respect to the
initial proton in the $\gamma^* p$ center-of-mass (see
Fig.~\ref{fig:decay-kin}).  The approximate expressions in (\ref{abc})
are valid up to relative corrections of order $\xb\sms m_p^2 /Q^2$ and
$\xb\sms \Delta_T^2 /Q^2$, and to the same accuracy one has
$\Delta_T^2 = (1-\xb) (t_0-t)$.  Changing variables from $\theta$ to
$z$ gives for the cross section
\begin{equation}
  \label{dsig-dz-dphi}
\frac{d\sigma(ep\to P_1 + P_2\sms B)}{dQ^2\, d\xb\, dt\, d\varphi\, dz}
 = \frac{1}{b - c \cot\theta \cos\varphi}\,
   \frac{d\sigma(ep\to P_1 + P_2\sms B)}{dQ^2\, d\xb\, dt\, d\varphi\;
                                    d\smb\cos\theta} .
\end{equation}
In Bjorken kinematics one has $c \ll b$ and can replace the Jacobian
in (\ref{dsig-dz-dphi}) by $1/b$ (except in the small region where
$\sin\theta \sim c/b$, which is not relevant for our purposes).
Neglecting $\Delta_T$ we get $z = a + b \cos\theta$ with $a$ and $b$
evaluated at $\Delta_T=0$, and integration over $t$ and $\varphi$
gives
\begin{equation}
\label{dsig-dz}
\frac{d\sigma(ep\to P_1 + P_2\sms B)}{dQ^2\, d\xb\, dz} = 
\frac{3}{4 b^3} \Bigg[ 
  2 (z-a)^2\, \frac{d\sigma(ep\to V_L\sms B)}{dQ^2\, d\xb} 
+ (z-a+b) (a+b-z)
  \frac{d\sigma(ep\to V_T B)}{dQ^2\, d\xb} \Bigg]
\end{equation}
in terms of the cross sections for the production of longitudinally or
transversely polarized vector mesons.  Using $s$-channel helicity
conservation, which is experimentally seen to hold at the few 10\%
level in $\rho^0$ and $\phi$ production
\cite{Borissov:2001fq,Tytgat:2000th,Rakness:2000th}, these cross
sections respectively correspond to the production from longitudinally
or transversely polarized virtual photons, and we finally obtain
(\ref{dsig-dz-sim}).  In our numerical applications we have used the
exact expressions from (\ref{abc}) and (\ref{r1r2}) at $\Delta_T=0$,
and thus in particular neglected $c$.  Since the integrated cross
sections are dominated by small $\Delta_T$, this should be a very good
approximation for the values of $Q^2$ and $\xb$ we focus on in the
present study.  The inclusion of finite $\Delta_T$ effects in the
kinematics would considerably complicate any analysis.

\end{appendix}


\end{document}